\documentclass[10pt]{article}
\usepackage{amsmath,amsfonts}
\usepackage{url}
\usepackage{algorithmicx}
\usepackage[ruled]{algorithm}
\usepackage{algpseudocode}
\usepackage{algpascal}
\usepackage{algc}
\usepackage{algorithm}
\usepackage{caption}
\usepackage{graphicx}
\usepackage{blindtext}
\usepackage{times}

\newcommand{\cO}{\mathcal{O}}

\newcommand{\F}{\mathbb{F}}

\newcounter{lastnote}

\title{{A Survey on Hardware Implementations of Elliptic Curve Cryptosystems}}

\author
{Bahram Rashidi\\
\\
\normalsize{Dept. of Elec. Eng., University of Ayatollah ozma Boroujerdi}\\
\normalsize{Boroujerd, 69199-69411, Iran}\\
\normalsize{E-mail: b.rashidi@ec.iut.ac.ir, b.rashidi@abru.ac.ir}
}

\date{}

\begin{document} 


%\baselineskip24pt


\maketitle


\begin{abstract}

In the past two decades, Elliptic Curve Cryptography (ECC) have become increasingly advanced. ECC, with much smaller key sizes, offers equivalent security when compared to other asymmetric cryptosystems. In this survey, an comprehensive overview of hardware implementations of ECC is provided. We first discuss different elliptic curves, point multiplication algorithms and underling finite field operations over binary fields $\F_{2{^m}}$ and prime fields $\F_p$ which are used in the literature for hardware implementation. Then methods, steps and considerations of ECC implementation are presented. The implementations of the ECC are categorized in two main groups based on implementation technologies consist of field programmable gate array (FPGA) based implementations and application specific integrated circuit (ASIC) implementations. Therefore, in these categories to have a better presentation and comparison, the implementations are presented and distinguished based on type of finite fields. The best and newest structures in the literature are described in more details for overall presentation of architectures and approaches in each group of implementations. High-speed implementation is an important factor in the ECC applications such as network servers. Also in smart cards, Wireless Sensor Networks (WSN) and Radio Frequency Identification (RFID) tags require to low-cost and lightweight implementations. Therefore, implementation methods related to these applications are explored. In addition, a classification of the previous works in terms of scalability, flexibility, performance and cost effectiveness is provided. Finally, some words and techniques about future works that should be considered are provided.

\end{abstract}

\textit{keywords}:\textit{Elliptic Curve Cryptography, FPGA, ASIC, Finite fields, point multiplication.}

\section{Introduction}\label{sec intro}

Elliptic Curve Cryptography (ECC) was proposed independently by Victor Miller \cite{Miller} and Neal Koblitz \cite{Koblitz} in the mid 1980's. It is a public key cryptography, which is based on the Elliptic Curve Discrete Logarithm Problem (ECDLP) of the elliptic curve over a finite fields \cite{Guide}. ECC provides various security applications such as key exchange, digital signatures, data encryption and authentication. The main advantages of ECC, when compared to other public key cryptosystems such as RSA is smaller key size and better performance with equivalent security level. In the two last decades, the application of the elliptic curves in cryptography has been considered and is attractive for many scientists. 
ECC has different applications in public key cryptography, e.g., banking transactions, mobile security, digital right management, Wireless Sensor Networks (WSN) and other security applications. Also it is applicable in many internet protocols and network applications such as SSL (Secure Sockets Layer), TLS (Transport Layer Security) \cite{TLS}, WAP WTLS (Wireless Transport Layer Security) \cite{WTLS} (for elliptic curves over prime) and IPsec which are commonly used today in over-the-web transactions and secure document transfers.
The ECC has been adopted by many standards such as ANSI \cite{ANSI}, IEEE \cite{IEEE}, ISO \cite{ISO} and NIST \cite{NIST}. In December 2010, Chinese State Cryptography Administration (SCA) published the national public key cryptographic algorithm based on ECC in \cite{SCA}, known as SM2. The industry has taken extreme interest in the ECC for internet protocols, smart cards, Radio Frequency Identification (RFID) tages and cell phones. Manufacturing companies related to ECC are consist of Sun Microsystems, Certicom, MasterCard, Fujitsu, MIPS Technologies, Digital Signature Trust Co and DataKey. The small key size, low area consumption and fast implementation make ECC one of the best choices for hardware implementation. The hardware-based implementations can provide significant security improvements by protecting secret keys and other parameters compared to software solutions. There is a growing need for hardware implementation of the ECC. Hardware implementations have better performance and better power efficiency than that of software implementations based on a microcontrollers. In the ECC, main and backbone operation is point multiplication (scalar multiplication) which is based on field operations. The efficiency of the ECC implementation depends on point multiplication. Efficient hardware implementation of field operations have direct impact on speed and performance of the ECC applications. In computation of elliptic curve point multiplication, the main operations are field multiplication and field inversion. Therefore, for implementation of the ECC these two field operations have more complexity. Some of the time critical applications, such as network servers where millions of heterogeneous client devices need to be connected. Therefore, high-speed hardware implementation of the ECC is an important factor. It could be only feasible and acceptable solution to reach a performance. Also in low-cost and low-area applications such as smart cards, WSN and RFID tags low-power and lightweight hardware implementations are only solutions for realization of these applications.

\begin{figure}[H]
	\centering
	%	\captionsetup{justification=centering}
	\includegraphics[scale=0.6]{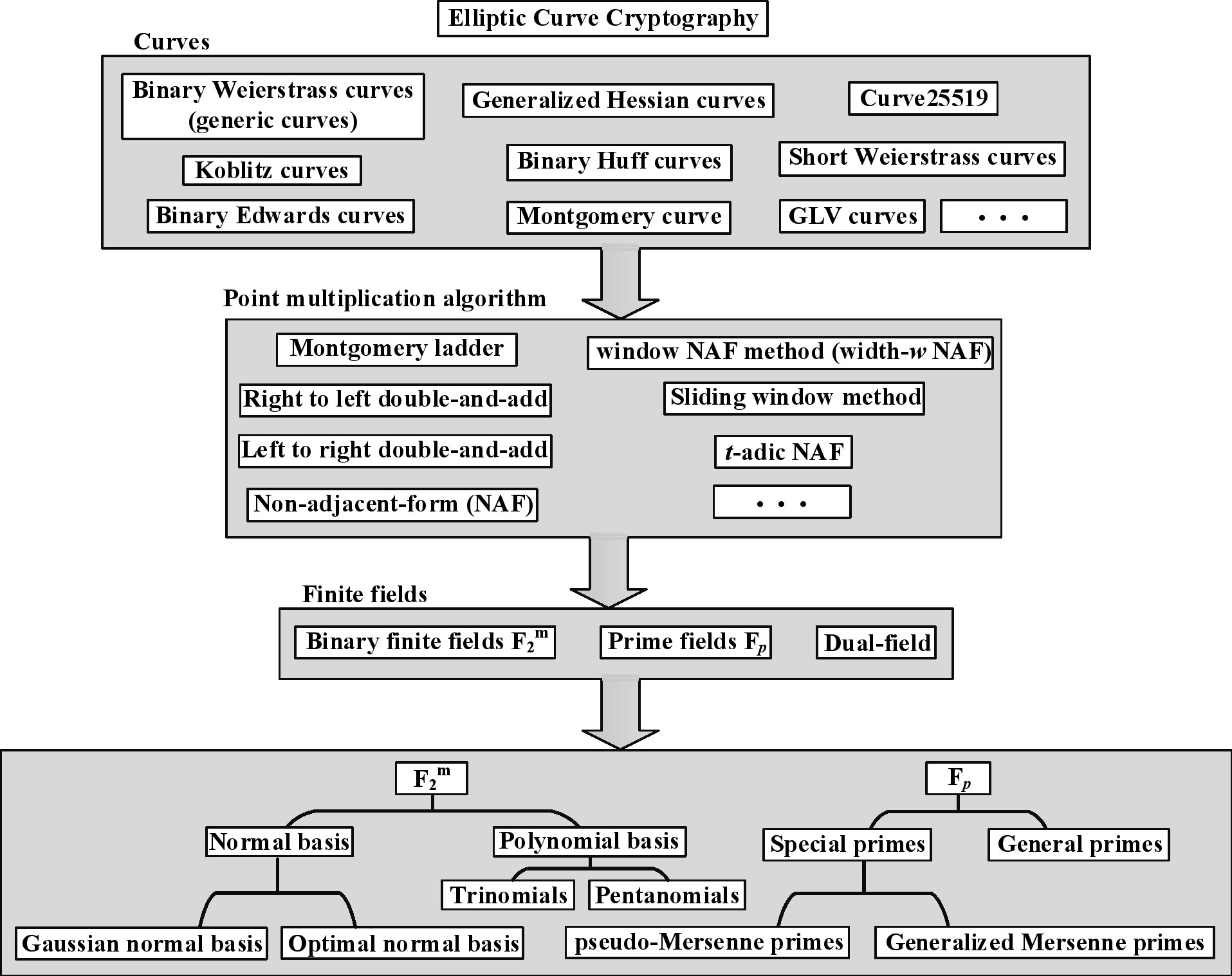}
	\caption{Various options in implementation of the ECC.}
	\label{fig:FI1}
\end{figure}

Implementation of an elliptic curve cryptosystem, like many other systems, follows a hierarchical approach, in which the performance of the top implemented layers is greatly influenced by the performance of the underlying layers. Therefore, it is important to have efficient implementations of the finite field operations in the underlying layer. To that end, three main steps have been performed. The first step is related to design and implementation of finite field operations such as field multiplication, field addition and field inversion. In the second step, design and implementation of point addition and point doubling operations are performed. In the third step, based on point multiplication algorithm, implementation are performed. Different option for ECC implementation are shown in Fig.\ref{fig:FI1}.
In the first step of the figure type of the elliptic curve is selected. In the second step, the point multiplication algorithm based on elliptic curve can be selected. Also in the third step, the type of finite field, size and other properties are selected. Two applicable fields are binary field $\F_{2{^m}}$ and prime field $\F_p$. In addition, different works for flexibility are implemented based on two fields that are called dual-field implementations. 
The binary finite field operations are suitable for hardware implementations due to these structures are carry free. Therefore, the field addition is implemented by a simple bit-wise XOR operation without carry bit. Moreover, the efficiency of the field multiplication depends on representation of elements in $\F_{2{^m}}$. There are two main practical bases called polynomial basis (PB) and normal basis (NB). In PB by using irreducible Trinomials and Pentanomials the field multiplication and squaring can be implemented efficiently. Also special types of NB representation called Gaussian normal basis representation (GNB) where the field multiplication is implemented efficiently. On the other hand, for prime fields, special primes are highly suited for efficient reduction techniques and implementation, the most simple form of such primes being the Mersenne primes and other are Generalizations Mersenne primes and pseudo-Mersenne primes. In next sections, we explain these options in more details.

This paper focuses on survey of techniques for implementing ECC at a high-speed and low-area and the existing hardware implementations of the ECC. The presentation is organized based on implementation technologies of the ECC such as field programmable gate array (FPGA) and application specific integrated circuit (ASIC). The survey starts by defining different elliptic curves, point multiplication algorithms and finite field arithmetics and also discussing their impact on implementations. Also, methods, steps and considerations in the implementation of the ECC are discussed. Then, the implementations of the ECC found in the literature are presented in related sections also a categorized and comprehensive comparison of the existing works is performed. For fair comparison and better analysis, the works are categorized and presented based on used finite field, type of elliptic curves, representation basis, implementation technology and platforms. 
We study implementations in terms of (1) hardware consumption (area) which is important for any cost-sensitive application, (2) execution time which is important for many applications especially in high-speed application and is related to speed processing of implementation, (3) maximum operation frequency that has a direct impact on the computation time, power consumption which is important for low-power, low-energy and low-area ASIC implementations.

The rest of the paper is organized as follows. Section \ref{sec-BM} describes the mathematical background. The point multiplication algorithms are discussed in Section \ref{sec-PMA}. Section \ref{sec-FFA} presents finite field arithmetics. In Section \ref{sec-ECC1} methods, steps and considerations of ECC implementation are presented. Elliptic curve implementations are presented in Section \ref{sec-ECC2}. Finally, the paper is concluded in Section \ref{sec-end}.

\section{\textbf{Mathematical background}}
\label{sec-BM}
In this section, we will briefly introduce the mathematical background relevant to the present survey. We start with a short review of defining different elliptic curves, point multiplication algorithms and finite field arithmetics and also discussing their effect on implementations. The elliptic curves can be defined over any field such as field of rational numbers, real numbers and complex numbers. For cryptographic application, elliptic curves are defined over finite fields. Two important finite fields for hardware implementation are binary fields and prime fields. Elliptic curves are traditional represented by the so called Weierstrass equations. For cryptographic applications, many other forms of elliptic curves have been proposed and investigated to improve high-speed and efficient implementations. In following, we briefly recall elliptic curves used in implementations.

\subsection{\textit{Elliptic curves over $\F_{2^{m}}$}}

Various binary elliptic curves over $\F_{2^{m}}$ and their properties are discussed and reviewed in this subsection.

\subsubsection{Binary Weierstrass curves}

An elliptic curve over a field $\F$ can be defined by

\begin{equation}
E: y^2+a_1 xy+a_3 y=x^3+a_2 x^2+a_4 x+a_6
\end{equation}

This equation is called \textit{long Weierstrass} equation, where $a_1,a_2,a_3,a_4$ and $a_6$ are in $\F$. The discriminant of the field is given by 

\begin{equation}
\Delta=-d_2^2 d_8-8d_4^3-27d_6^2+9d_2 d_4 d_6
\end{equation}
where $d_2=a_1^2+4a_2, d_4=2a_4+a_1 a_3, d_6=a_3^2+4a_6$, and $d_8=a_1^2 a_6+4a_2 a_6-a_1 a_3 a_4+a_2 a_3^2-a_4^2$. $\Delta \neq 0$, since the elliptic curve is nonsingular. The set of affine points $(x,y)$ satisfying the curve equation with the point at infinity denoted by $\cO$ construct a group \cite{Guide}. The set of $\F$-rational points on $E$ is defined as follows:

\begin{equation}
E(\F)=\{(x,y)\in \F \times \F: y^2+a_1 xy+a_3 y-x^3-a_2 x^2-a_4 x-a_6=0\}\cup\{\cO\}
\end{equation}

Based on a group of points defined over an elliptic curve, group law operation for two points $P_1, P_2$, where $P_1 \neq P_2$, defines the point addition (PA) $P_3=P_1+P_2$ using the tangent and chord rule as the primary group operation. For $P_1=P_2$ we have point doubling (PD) $P_3=2P_1$. Basically a point $ P $ over the curve can generate all the other point by PA.

The binary elliptic curves defined over a binary field $\F_{2^{m}}$. \textit{Binary Weierstrass curves} (BWCs) is defined by following equation

\begin{equation}
W: y^2+xy=x^3+ax^2+b 
\end{equation}
where $a,b \in \F_{2^{m}}$ and $b\neq 0$. This equation is called \textit{non-super singular} which is suitable for cryptographic applications. For this family of curves, NIST recommended standard elliptic curves over $\F_{2^{m}}$ fields consist of \{B-163, B-233, B-283, B-409 and B-571\}.

In following point addition and point doubling on BWCs in affine coordinate are presented. Let $P_1=(x_1,y_1)$ and $P_2=(x_2,y_2)$ be two points on the BWCs with $P_1\neq \pm P_2$ where $-P_2=(x_2,x_2+y_2)$. 
Then the addition of points $P_1,P_2$ is the point $P_3$ denoted by $P_3=P_1+P_2=(x_3,y_3)$, where $x_3=\lambda^2+\lambda+x_1+x_2+a$, and $y_3=\lambda(x_1+x_3)+x_3+y_1,$ where, $\lambda=\dfrac{y_1+ y_2} {x_1+ x_2}$. Also for the point doubling we have $P_3=2P_1=(x_3,y_3)$, where $x_3=\lambda^2+\lambda+a$, and $y_3=\lambda(x_1+x_3)+x_3+y_1$, where, $\lambda =x_1+\dfrac {y_1} {x_1}$. In this case, point addition and point doubling are computed by 1$ \textbf{I} $+2$ \textbf{M} $+1$ \textbf{S} $, where $ \textbf{I} $, $ \textbf{M} $ and $ \textbf{S} $ are cost of computation field inversion, field multiplication and field squaring respectively. Inversion is the most time-consuming operation in among other field operations. Therefore, the projective coordinate system (each point is represented by three coordinates $(X,Y,Z)$) is used to reduce the complexity of the point addition and point doubling computation. The more details of the projective coordinates are presented in point multiplication subsection.

\subsubsection{Koblitz Curves}

In the binary Weierstrass curves if $a \in \{0, 1\}$ and $b=1$, it is called \textit{Koblitz curves} or \textit{anomalous binary curves} \cite{Koblitz2}. Therefore, the Koblitz curves are defined over $\F_{2^{m}}$ by following equation:

\begin{equation}
K: y^2+xy=x^3+ax^2+1 
\end{equation}

Koblitz curves offer considerable computational advantages compared to the binary Weierstrass curves, because can be used to computation of the point multiplication without the need for point doubling \cite{Guide}. NIST recommended standard Koblitz curves over $\F_{2^{m}}$ consist of \{K-163, K-233, K-283, K-409 and K-571\}.

\subsubsection{Binary Edwards curves}

\textit{Binary Edwards curves} (BECs) are the first family of the binary elliptic curves with complete group law operation \cite{BLF8}. Let $d_1$, $d_2$ be elements of $\F_{2^m}$ such that $d_1\ne 0$ and $d_2\ne d_1(d_1+1)$. The binary Edwards curve with parameters $d_1$ and $d_2$ is given by the equation

\begin{equation}
	E~:~d_1(x+y)+d_2(x+y)^2=xy(x+1)(y+1).
\end{equation}

The equation of the binary Edwards curve $E$ is symmetric in $x, y$ and the negation of the point $(x,y)$ is $(y,x)$. 
The point $\cO=(0,0)$ is the neutral element of the addition law and the point $(1,1)$ has order 2. The addition, doubling and differential addition formulas for the binary Edwards curves are presented in~\cite{BLF8}. The addition group law is complete if 
$Tr(d_2)=1$, where $Tr$ is the trace function from $\F_{2^m}$ to $\F_2$.

\subsubsection{Generalized Hessian curves}

The Hessian curve is a symmetric curve shape representing an elliptic curve \cite{FJ}. The arithmetic in this curve is faster than that of Weierstrass form. Therefore, use of Hessian curve in cryptography has been studied. The family of generalized Hessian curves over a finite field covers more isomorphism classes of elliptic curves and it is equivalent to the family of all elliptic curves with a point of order 3 \cite{FJ}. Generalized Hessian curves provide efficient unified addition formulas which is resist against side-channel attacks. They also have complete addition formulas with suitably chosen parameters. A \textit{generalized Hessian curve} (GHC) over $\F_{2^m}$ is defined by

\begin{equation}
H: x^3+y^3+c=dxy
\end{equation}
where $c, d$ are elements of $\F_{2^m}$, $c\neq 0$ and $d^3\neq 27c$. This equation a symmetric cubic equation. The Hessian addition formulas, called the Sylvester formulas. In \cite{FJ} a suitable modification of the Sylvester formulas for fast and efficient unified addition formulas on generalized Hessian curves is presented. For the point $ P=(x,y) $ on $H$ the additive inverse is given by $-P=(y,x)$. 

\subsubsection{Binary Huff curves}

The affine model of \textit{binary Huff curve} (BHC) \cite{Huff} given by

\begin{equation}
Huff: ax(y^2+y+1)=by(x^2+x+1)
\end{equation}
where $a,b \in \F^{*}_{2^{m}}$ and $a\neq b$. Also this curve is birationally equivalent to the Weierstrass elliptic curve \cite{Huff}

\begin{equation}
v(v+(a+b)u)=u(u+a^2)(u+b^2)
\end{equation}
under the inverse maps

\begin{equation}
(x,y)\leftarrow (\dfrac {b(u+a^2)} {v}, \dfrac {a(u+b^2)} {v+(a+b)u}) ~~~~~and~~~~~ (u,v)\leftarrow (\dfrac {ab} {xy}, \dfrac {ab(axy+b)} {x^2y})
\end{equation}

The set of points on a BHC forms a group. The identity element is $\cO=(0,0)$. While the above maps are not line-preserving, the group law on a BHC satisfies the tangent and chord rule \cite{Huff}. Binary Huff curves are curves with unified point addition and point doubling formula with resistance against power attacks.

\subsection{\textit{Elliptic curves over $\F_p$}}

Let $p$ be a prime with $p > 3$, an elliptic curve over $\F_p$ is defined the so-called \textit{short Weierstrass} equation as:

\begin{equation}
y^2=x^3+ax^2+b
\end{equation}
where $a, b, x, y \in \F_p$ and $4a^3+27b^2 \neq 0$. In this curves the characteristic is not equal 2 and 3 or $p > 3$ . Also group operations on elliptic curves over $\F_p$ is defined. NIST recommended standard elliptic curves over prime fields consist of \{$ p $-192, $ p $-224, $ p $-256, $ p $-384 and $ p $-521\}.

In recent years in \cite{Curve25519} a special elliptic curve called \textit{Curve25519} over $\F_p$ is defined, where $ p=2^{255}-19$. The curve Curve25519 is defined as follows:

\begin{equation}
y^2=x^3+486662x^2+x
\end{equation}

Other traditional elliptic curve over prime fields $\F_p$ with $p\geq 5$ is called \textit{Montgomery curve} \cite{MontCur} and defined by following equation:
 
\begin{equation}
M: By^2=x^3+Ax^2+x
\end{equation}
where $A, B$ are elements in $\F_p$ and $B(A^2-4) \neq 0$.

Also, the so-called \textit{Gallant-Lambert-Vanstone curves} for simply GLV curves, are elliptic curves over $\F_p$ which possess an efficiently computable endomorphism $ \varphi $ whose characteristic polynomial has small coefficients. The three family of GLV curves that can be defined, over a prime field $\F_p$, by a Weierstrass equation based on \cite{GLV} are as follows:

In the first case we have

\begin{equation}
GLV1: y^2=x^3+ax
\end{equation}
In this equation, let $\alpha \in \F_p$ be an element of order 4. Then the map $ \varphi: GLV1\rightarrow GLV1$ defined by $(x, y)\rightarrow(−x,\alpha y)$ and $\cO \rightarrow \cO$ is an endomorphism defined over $\F_p$. The second family is

\begin{equation}
GLV2: y^2=x^3+b
\end{equation}
where $p \equiv$ 1 mod 3. Let $\beta \in \F_p$ be an element of order 3. Then the map $ \varphi: GLV2\rightarrow GLV2$ defined by $(x,y)\rightarrow(\beta x,y)$ and $\cO \rightarrow \cO$ is an endomorphism defined over $\F_p$. Let $p > 3$ be a prime such that −7 is a perfect square in $\F_p$ , and let $\omega$=(1+$\sqrt{-7}$)/2, and let $ a $=($\omega$-3)/4. And also for third family of the elliptic curve defined over $\F_p$ we have

\begin{equation}
GLV3: y^2=x^3-\dfrac {3} {4}x^2-2x-1
\end{equation}

Then the map $ \varphi: GLV3\rightarrow GLV3$ defined by $(x,y)\rightarrow (\omega^{-2}\dfrac {x^2-\omega} {x-a},\omega^{-3}y\dfrac {x^2-2ax+\omega} {(x-a)^2})$ and $\cO \rightarrow \cO$ is an endomorphism defined over $\F_p$. Computing the endomorphism is a little harder than doubling a point. Galbraith, Lin, and Scott (GLS) \cite{GLS} generalized the GLV technique to a broader class of elliptic curves defined over $\F_{p^2}$. The GLS curves were generalized for binary curves over $\F_{2^{2m}}$ in \cite{Hankerson}.

Also \textit{Jacobian curve} \cite{Jacobian} can be used in cryptography instead of the Weierstrass form because it can provide a robustness against simple and differential power analysis attacks. In addition, this curve has faster arithmetic compared to the Weierstrass curve.

\section{\textbf{Point multiplication algorithms}}
\label{sec-PMA}
The most important operation and dominates the execution time of elliptic curve cryptography is called \textit{point multiplication} or \textit{scalar multiplication}. In this operation we have, $kP=P+P+...+P$, where $k$ is a positive integer and $ P $ is a point on the curve. Therefore, in a straightforward way the point multiplication can be computed by $ k $ times addition of point $ P $ by self using PA and PD operations. Here, we presented different point multiplication methods. There are several ways to implement point multiplication \cite{Guide}: Right to left double-and-add, Left to right double-and-add, Non-adjacent-form (NAF) method, window NAF method (width-$w$ NAF), Sliding window method, $ \tau $-adic NAF ($\tau$NAF) method and Montgomery ladder method.
Possible coordinates are affine and projective. In the projective coordinate, projective point $(X,Y,Z),Z \neq 0$ is corresponds to the affine point $(\dfrac {X} {Z^c}, \dfrac {Y} {Z^d})$. Applicable and the most popular projective coordinates are consist of Standard $(c=1,d=1)$, Jacobeans $(c=2,d=3)$ and Lopez-Dahab $(c=1,d=2)$. For example, in Lopez-Dahab (LD) coordinate \cite{LD} projective version of the BWCs in Eq.(2) is obtained by replacing $x$ and $y$ with $\dfrac {X} {Z}$ and $\dfrac {Y} {Z^2}$ as:

\begin{equation}
E: Y^2+XYZ=X^3 Z+aX^2 Z^2+bZ^4
\end{equation}

\begin{figure}[H]
	\centering
	%	\captionsetup{justification=centering}
	\includegraphics[scale=0.7]{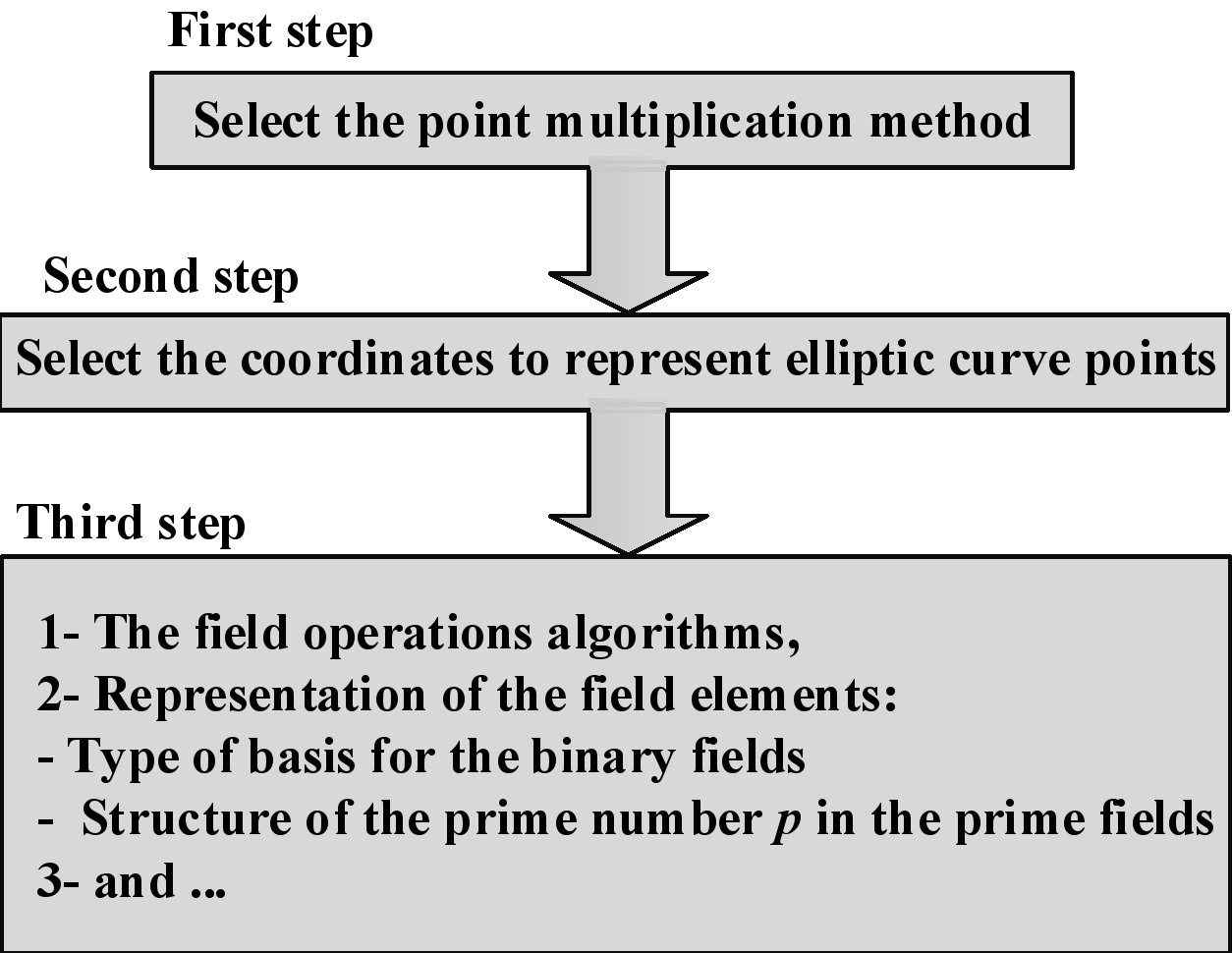}
	\caption{Three main steps for compute of the point multiplication.}
	\label{fig:PM_1}
\end{figure}

In the point multiplication algorithm, for use with most cryptographic protocols, it is required to convert the output result with projective coordinates to affine coordinates. A point multiplication is performed in three main steps. In the first step, the point multiplication algorithm must be selected. In the second step, the coordinates to represent elliptic curve points must be defined. Finally in the last step, the field operations algorithms, representation of the field elements (type of basis for the binary fields and structure of the prime number $p$ in the prime fields) are defined and selected. Fig.\ref{fig:PM_1} shows three main steps for compute of the point multiplication.

In following traditional point multiplication algorithm are presented. Algorithm \ref{alg:A1} and Algorithm \ref{alg:A2} show right to left, left to right point multiplication algorithms respectively. Algorithm \ref{alg:A1} processes the bits of $ k $ from right to left and Algorithm \ref{alg:A2} processes the bits from left to right. In these algorithms every bit of scalar $ k $ is scanned, then based on value of each bit, '0' or '1', a PD or both PD and PA operations are performed. In the right to left algorithm, the PD and PA operations can be performed in parallel form.

\begin{minipage}[h]{7cm}
  \vspace{0pt}  
  \begin{algorithm}[H]
    \caption{Right-to-left point multiplication algorithm}
    \label{alg:A1}
    \textbf{Input:} $k =(k_{l-1},k_{l-2},...,k_{2},k_{1},k_{0}), P \in \F_q$\\
    \textbf{Output:} $kP$\\
1. $Q\leftarrow\cO$\\
2. \textbf{for} $ i $ \textbf{from} 0 \textbf{downto} $ l-1 $ \textbf{do}\\
3. ~~\textbf{if} $ k_i=1 $ \textbf{then} $Q\leftarrow P+Q$\\
4. ~~$P\leftarrow 2P$\\
5. ~~\textbf{end} \textbf{if}\\
6. \textbf{end} \textbf{for}\\
7. \textbf{Return} $Q$
\end{algorithm}
\end{minipage}
\begin{minipage}[h]{7cm}
  \vspace{0pt}
  \begin{algorithm}[H]
    \caption{Left-to-right point multiplication algorithm}
        \label{alg:A2}
    \textbf{Input:} $k =(k_{l-1},k_{l-2},...,k_{2},k_{1},k_{0}), P \in \F_q$\\
    \textbf{Output:} $kP$\\
1. $Q\leftarrow\cO$\\
2. \textbf{for} $ i $ \textbf{from} $ l-1 $ \textbf{downto} 0 \textbf{do}\\
3. ~~$Q\leftarrow 2Q$\\
4. ~~\textbf{if} $ k_i=1 $ \textbf{then} $Q\leftarrow P+Q$\\
5. ~~\textbf{end} \textbf{if}\\
6. \textbf{end} \textbf{for}\\
7. \textbf{Return} $Q$
  \end{algorithm}
\end{minipage}\\
\\

The Hamming weight (HW) of $ k $, i.e., $ H(k) $, is the number of nonzero terms in the representation of $ k $. The PA is required in the algorithm only when $ k_i=1 $. When a signed-bit representation in NAF, i.e., $ k_i \in \{0, \pm 1\} $ so that $ k_ik_{i+1}=0 $, for all $ i $, is used, $ H(k)=m/3 $ on average. Therefore, it is of interest to reduce $ H(k) $. Hamming weight of $ k $ can be further reduced with windowing methods, but then certain points need to be precomputed \cite{K10}. In Koblitz curves, PD operation can be replaced efficiently by Frobenius endomorphism \cite{Koblitz2}. An algorithm similar to Algorithm \ref{alg:A1} and Algorithm \ref{alg:A2}, can be devised so that PDs are replaced by Frobenius endomorphisms. Frobenius map $ \varphi $ is an endomorphism that raises every element to its power of two, i.e., $ \varphi: x \rightarrow x^2 $. Koblitz curves have the appealing feature that if the point $ P=(x,y) $ is on curve, so is the point $ (x^2,y^2) $. Frobenius map for a point $ P=(x,y) $ can be defined as $ \varphi:(x,y)\rightarrow(x^2,y^2) $. Frobenius endomorphism can be carried out efficiently if the field elements are represented in normal basis.

Algorithm\ref {alg:A3} shows point multiplication algorithm by using NAF$(k)$ instead of the binary representation of $ k $. In this algorithm, if processes $ w $ digits of $ k $ at a time by using a window method we have Algorithm \ref{alg:A4}.

\begin{minipage}[h]{7cm}
  \vspace{0pt}  
  \begin{algorithm}[H]
    \caption{Binary NAF method for point multiplication}
    \label{alg:A3}
    \textbf{Input:} $k =(k_{l-1},k_{l-2},...,k_{2},k_{1},k_{0}), P \in \F_q$\\
    \textbf{Output:} $kP$\\
1. Compute NAF$(k)=\sum_{i=0}^{l-1} k_i2^i$ \\
2. $Q\leftarrow\cO$\\
3. \textbf{for} $ i $ \textbf{from} $ l $-1 \textbf{downto} $ 0 $ \textbf{do}\\
4. ~~$Q\leftarrow 2Q$\\
5. ~~\textbf{if} $ k_i=1 $ \textbf{then} $Q\leftarrow Q+P$\\
6. ~~\textbf{end} \textbf{if}\\
7. ~~\textbf{if} $ k_i=-1 $ \textbf{then} $Q\leftarrow Q-P$\\
8. ~~\textbf{end} \textbf{if}\\
9. \textbf{end} \textbf{for}\\
10. \textbf{Return} $Q$
\end{algorithm}
\end{minipage}
\begin{minipage}[h]{7.5cm}
  \vspace{0pt}
  \begin{algorithm}[H]
    \caption{Window NAF method for point multiplication}
        \label{alg:A4}
    \textbf{Input:} $k =(k_{l-1},k_{l-2},...,k_{2},k_{1},k_{0}), P \in \F_q$\\
    \textbf{Output:} $kP$\\
1. Compute NAF$_w(k)=\sum_{i=0}^{l-1} k_i2^i$ \\
2. Compute $P_i = iP$ for $i\in \{1, 3, 5, ..., 2^{w-1}-1\}$\\
3. $Q\leftarrow\cO$\\
4. \textbf{for} $ i $ \textbf{from} $ l $-1 \textbf{downto} $ 0 $ \textbf{do}\\
5. ~~$Q\leftarrow 2Q$\\
6. ~~\textbf{if} $ k_i\neq 0 $ \textbf{then}\\ 
7. ~~~~\textbf{if} $ k_i > 0 $ \textbf{then} $Q\leftarrow Q+P_{k_i}$\\
8. ~~~~\textbf{else} $Q\leftarrow Q-P_{-k_i}$\\
9. ~~~~\textbf{end} \textbf{if}\\
10. ~\textbf{end} \textbf{if}\\
11. \textbf{end} \textbf{for}\\
12. \textbf{Return} $Q$  \end{algorithm}
\end{minipage}\\

In order to utilize fast Frobenius endomorphisms, $k$ must be converted into an optimize NAF representation such as $\tau$NAF. In \cite{Solinas} efficient algorithms for finding $\tau$NAF are presented. $\tau$NAF is analogous with the binary NAF as it has on average the same length and HW. In the Koblitz curves, the point multiplication can be further improved by converting the scalar $ k $ into $\tau $-adic nonadjacent form $\tau $NAF which rewrites $ k $ into the form $\sum_{i=0}^{l-1} u_i\tau^i$, where $u_i \in \{0, \pm1\}$ and $\tau=((-1)^{1-a}+\sqrt{-7})/2$. $\tau$NAF point multiplication algorithm is shown in Algorithm \ref{alg:A5}. In the point multiplication based on NAF representation of the scalar $ k $, it is needs to be converted in NAF expansion. Therefore, this conversion is one of the important steps in the hardware implementation of the point multiplication. One popular point multiplication algorithm is Montgomery ladder \cite{MLP} which is shown in Algorithm \ref{alg:A6}. In this algorithm point multiplication computes based on the $x$ and $z$ and the $y$-coordinate is recovered in the end. Both PD and PA operations are computed very efficiently for every $k_i$, where $k_i$ is $i^{th}$ bit of scalar $k$. The point multiplication is computed recursively by projective PA and PD operations without using the $Y$ coordinate. Therefore, the numbers of the field multiplications are reduced. This is an important property of Algorithm \ref{alg:A6}. In the final step of the algorithm, the projective coordinate is converted to the affine coordinate. In Montgomery ladder algorithm, both PA and PD are done simultaneously for each bit of scalar $k$, so the power trace has regular and unified form.

\begin{minipage}[h]{7cm}
  \vspace{0pt}  
  \begin{algorithm}[H]
    \caption{$\tau$NAF method for point multiplication}
    \label{alg:A5}
    \textbf{Input:} $k =(k_{l-1},k_{l-2},...,k_{2},k_{1},k_{0}), P \in \F_q$\\
    \textbf{Output:} $kP$\\
1. Compute $\tau$NAF$(k)= \sum_{i=0}^{l-1} u_i\tau^i$ \\
2. ~\textbf{if} $ u_{\l-1} = 1 $ \textbf{then} $Q \leftarrow P$ \textbf{else} $Q \leftarrow -P$\\
3. ~\textbf{end} \textbf{if}\\
4. \textbf{for} $ i $ \textbf{from} $ l $-2 \textbf{downto} $ 0 $ \textbf{do}\\
5. ~~~$Q\leftarrow \varphi(Q) $\\
6. ~~\textbf{if} $ u_i=1 $ \textbf{then} $Q\leftarrow Q+P$\\
7. ~~\textbf{end} \textbf{if}\\
8. ~~\textbf{if} $ u_i=-1 $ \textbf{then} $Q\leftarrow Q-P$\\
9. ~~\textbf{end} \textbf{if}\\
10. \textbf{end} \textbf{for}\\
11. \textbf{Return} $Q$
\end{algorithm}
\end{minipage}
\begin{minipage}[h]{7.5cm}
  \vspace{0pt}
  \begin{algorithm}[H]
    \caption{Montgomery ladder algorithm for point multiplication}
        \label{alg:A6}
    \textbf{Input:} $k =(k_{l-1},k_{l-2},...,k_{2},k_{1},k_{0})$, with $ k_{l-1}=1 $,  $P \in \F_q$\\
    \textbf{Output:} $kP$\\
1. $Q_1\leftarrow P, Q_2\leftarrow 2P$ \\
2. \textbf{for} $ i $ \textbf{from} $ l $-2 \textbf{downto} $ 0 $ \textbf{do}\\
3. ~~\textbf{if} $ k_i=1 $ \textbf{then}\\ 
4. ~~~~$Q_1\leftarrow Q_1+Q_2, Q_2\leftarrow 2Q_2$  \\ 
5. ~~\textbf{else}\\
6. ~~~~$Q_2\leftarrow Q_1+Q_2, Q_1\leftarrow 2Q_1$\\
7. ~~\textbf{end} \textbf{if}\\
8. \textbf{end} \textbf{for}\\
9. \textbf{Return} $Q_1$ 
\end{algorithm}
\end{minipage}\\\\

For GLV curves, the cost for the computation of a point multiplication can be significantly reduced by efficiently-computable endomorphism. This endomorphism allows one to accomplish an $m$-bit point multiplication $kP$ by a computation of the form $ k_1P+k_2Q $, where $k_1, k_2$ have only half the length of $k$ \cite{GLV}. The two half-length point multiplications can be carried out simultaneously via Shamir’s trick, which takes $m$/2 point doubling and roughly $m$/4 additions when $k_1, k_2$ are represented in Joint Sparse Form (JSF). Therefore, $ k_1P+k_2Q $ can be computed by a \textit{Simultaneous multiple point multiplication} type algorithms.

Comparisons between the implementations of point multiplication in Montgomery ladder by Lopez and Dahab coordinate system and binary Edwards curves shows the traditional Weierstrass curves perform faster than the recently proposed binary Edwards. But, the main advantage of using binary Edwards curves compared to other forms of elliptic curves is their complete formulas, that is providing hardware implementation which works for any inputs.

\section{\textbf{Finite field arithmetics}}
\label{sec-FFA}

Elliptic curves can be defined over any field such as field of rational numbers, real numbers, and complex numbers. For cryptographic application, elliptic curves are defined over finite fields. There are different finite fields and representations for use in ECC. Two important finite fields for hardware implementation are binary fields $\F_{2^m}$ and prime fields $\F_p$ \cite{W18}. The field operations required to implement the elliptic curve point multiplication are field multiplication, field addition, field squaring and field inversion.

\subsection{\textbf{Field arithmetic over $\F_{2^m}$}}

Binary finite fields $\F_{2^m}$ are defined as a vector space with dimension equal $ m $ over $\F_2$ rather a basis.
A basis can be represented by set of elements $\{e_0,e_1,...,e_{m-1}\}$ in $\F_{2^m}$. In this case, each element in the field, i.e. $A \in \F_{2^m}$ can be represented as $ A=\sum_{i=0}^{m-1} a_ie_i$, where $ a_i \in \F_2$. There are different basis for $\F_{2^m}$, \textit{polynomial basis} (PB) and \textit{normal basis} (NB) are the most used basis in cryptographic applications over binary finite fields. In the normal basis representation field squaring is cost free, it is implemented by a simple cyclic shift, but in polynomial basis it is based on an array of XOR gates if irreducible polynomial of the field is an irreducible pentanomial or trinomial. In the literature polynomial basis is the most used, due to the field multiplier in this basis is more efficient than that of the normal basis. In following polynomial basis and normal basis representation are explained briefly. 

\subsubsection{\textit{Polynomial basis representation}}

The irreducible polynomial $ P(x)=x^m+p_{m-1} x^{m-1}+...+p_1 x+p_0 $ with degree $ m $ and $ p_i \in \F_2 $ is called reduction polynomial. If $ \alpha $ be one roots of the $ P(x) $, i.e. $P(\alpha)=0$, therefore, set $ \{\alpha^{m-1},\alpha^{m-2},...,\alpha^2,\alpha,1\} $ is a polynomial basis (or canonical basis). In this case, elements of the field $\F_{2^m}$ are presented based on set of the polynomials with degrees $ 0 \leq d \leq m-1$ such as $A(x)=a_{m-1} x^{m-1}+a_{m-2} x^{m-2}+...+a_1 x+a_0$, where $ a_i \in \F_2 $. Also, the polynomial $A(x)$ is simply given by its coefficients in $\F_2$ as the $m$-bit number $(a_{m-1},a_{m-2},...,a_1,a_0)$, that is the binary representation of the corresponding element in $\F_{2^m}$. In the polynomial basis, numbers 0 and 1 are represented by $0=(0,0,...,0,0)$ and $1=(0,0,...,0,1)$ respectively.
As know, the number of nonzero terms in the irreducible polynomial $P(x)$ must be an odd number. On the other hand, complexity of the field arithmetics is depends on the number of nonzero terms in the irreducible polynomial, so that lower number of nonzero terms is better for an efficient implementation because sparse polynomials offer considerable computational advantages. Therefore, first candidate for $P(x)$ is irreducible trinomials (three nonzero terms) and other candidate is irreducible pentanomials (five nonzero terms). In practice, these two irreducible polynomials have the most widely used in the implementations. \\

\begin{itemize}
  \item \textbf{Trinomial basis representation}
\end{itemize}

For binary finite fields which are generated by irreducible trinomials, the reduction polynomial is defined as $T_{m,k} (x)=x^m+x^k+1$, where $0 \leq k \leq m-1$. A trinomial is irreducible if only if it's Reciprocal polynomial, i.e. $ T_{m,m-k} (x)=x^m T_{m,k} (\dfrac {1} {x})=x^m+x^{m-k}+1 $ be irreducible. Hence, we should be interested in trinomials of the form only $T_{m,k} (x)=x^m+x^k+1$, where $1 \leq k \leq m/2$ \cite{S1}. Such trinomials exist for certain values of $m$ only. If they exist, we should choose the reduction polynomial with the smallest $k$. Such a trinomial are the most efficient for the point multiplication implementation.\\

\begin{itemize}
  \item \textbf{Pentanomial basis representation}
\end{itemize}

The reduction polynomial in this case is an irreducible pentanomial as follows:

\begin{equation*}
 P(x)=x^m+x^{k_3}+x^{k_2}+x^{k_1}+1,~~~~ 1\leq k_1\leq k_2\leq k_3\leq m-1 
\end{equation*}

Irreducible pentanomials $P(x)$ always exist for $m\geq 4$. In practice, it is recommended to use pentanomials whose coefficient triples $k_1, k_2, k_3$ or $k_3, k_2, k_1$ will have the first coefficient as small as possible and next coefficients are kept as small as possible after fixing the previous one or ones in the triple order \cite{S1}. These irreducible pentanomials are efficient for computations of the field operations and reduce hardware and time complexity.

\subsubsection{\textit{Normal basis representation}}

Normal basis representation can be defined for any finite field $\F_{q^m}$ where $q$ is power of a prime number. A normal basis for $\F_{2^m}$ over $\F_2$ is shown as follows:

\begin{equation*}
\textbf{B}=\{\beta^{2^{m-1}},\beta^{2^{m-2}},...,\beta^{2^2},\beta^{2^1},\beta^{2^0}\}
\end{equation*}
where $\beta \in \F_{2^m}$ and it is called generator of base \textbf{B}. For any binary field, always exists a normal basis \textbf{B}. Each normal element such as $ A $ is represented as follows:

 \begin{equation*}
 A=\sum_{i=0}^{m-1} a_i\beta^{2^i}=a_{m-1}\beta^{2^{m-1}}+a_{m-2}\beta^{2^{m-2}}+...+ a_{2}\beta^{2^2}+a_1\beta^2+a_0\beta
 \end{equation*}
where $a_i\in \F_2$. In the vector representation, similar to polynomial basis, for the element $A$ we have the $m$-bit number $(a_{m-1},a_{m-2},...,a_1,a_0)$. In this basis, zero element and multiplicative identity can be represented by $(0,0,...,0,0)$ and $(1,1,...,1,1)$ respectively.\\

\begin{itemize}
  \item \textbf{Gaussian normal basis}
\end{itemize}

The \textit{Gaussian normal basis} (GNB), is a special class of normal basis. The field multiplication is performed simpler and more efficient in GNB \cite{Ash} and \cite{IEEE}. The complexity of the GNB multiplication is measured by its type, which is a positive integer related to the number of nonzero entries of the multiplication matrix. Therefore, a more efficient multiplier has a smaller type. The GNB is considered in several standards such as IEEE P1363 \cite{IEEE} and NIST \cite{NIST}. For example these two standards recommended even types $T=\{4,2,6,4$ and $10\}$, corresponded to the fields \{$\F_{2^{163}}$, $\F_{2^{233}}$, $\F_{2^{283}}$, $\F_{2^{409}}$ and $\F_{2^{571}}$\}. For each field $\F_{2^m}$ where $ m $ is not divisible by 8, a GNB of some type exists, and also for each positive integer $T$ at most one GNB of type $T$ exists. In more details, for the given positive integers $m$ and $T$, if $p=mT+1$ be a prime number such that $gcd(h,m)=1$, where $h= \dfrac {mT} {k}$ and $k$ is the multiplicative order of 2 module $p$, then a GNB over $\F_{2^{m}}$ of type $T$ exists. The GNBs with odd values of $m$ are applicable for cryptography, that implies $T$ is an even number.\\

\begin{itemize}
  \item \textbf{Optimal normal basis}
\end{itemize}

The \textit{optimal normal basis} (ONB) is a GNB of type 1 or 2 that provide the most efficient multiplication algorithm among all other normal bases \cite{Mullin}. For ONB the number of nonzero entries of the multiplication matrices is minimum and is equal 2$m$-1. The ONB of type 1 or 2 are defined as follows \cite{Mullin}:

\begin{enumerate}
\item ONB of type-1 exists if $m+1$ is a prime number and '2' is a primitive element of the prime field $\F_{m+1}$.
\item In $\F_{2^m}$ ONB of type-2 exists if $m+1$ is a prime number and either '2' is a primitive element in $\F_{2m+1}$ or $2m+1\equiv 3(mod~4)$ and the order of '2' in $\F_{m+1}$ is $m$.
\end{enumerate}

\begin{itemize}
  \item \textbf{\textit{Addition in $\F_{2^m}$}}
\end{itemize}

Addition is the simplest field operation in among other field operations over $\F_{2^m}$. It is a bit-wise addition in $\F_2$ which is implemented by XOR gates in hardware. For example, for addition of two field elements $ A, B $ in binary fields we have:

 \begin{equation*}
A(x)+B(x)= \sum_{i=0}^{m-1} ((a_i+b_i)~ mod~2)x^i=\sum_{i=0}^{m-1} (a_i\oplus b_i)x^i=(a_{m-1}\oplus b_{m-1}, a_{m-2}\oplus b_{m-2}, ..., a_{1}\oplus b_{1}, a_{0}\oplus b_{0})
 \end{equation*}

\begin{itemize}
  \item \textbf{\textit{Squaring in $\F_{2^m}$}}
\end{itemize}

Squaring of the element $A(x)=a_{m-1} x^{m-1}+a_{m-2} x^{m-2}+...+a_1 x+a_0$ in polynomial basis is a linear operation, such that

 \begin{equation*}
C(x)=\sum_{i=0}^{m-1} c_ix^{i}=A^2(x)=\sum_{i=0}^{m-1} a_ix^{2i}=a_{m-1}x^{2(m-1)}+a_{m-2}x^{2(m-2)}+...+a_{\lceil \dfrac {m} {2} \rceil}x^{2\lceil\dfrac {m} {2} \rceil}+...+ a_{1}x^{2}+a_{0}
 \end{equation*}
 
The vector representation of $A^2(x)$ is obtained by inserting a '0' bit between consecutive bits of the vector representation of $A(x)$, i.e. $(a_{m-1},a_{m-2},...,a_1,a_0)^2$=$(a_{m-1},0, a_{m-2}, 0,...,a_1, 0, a_0)$. The result must be reduced by modulo $ P(x)$. In this case, in $A^2(x)$ terms with degree equal $m$ and greater by reduction matrix $R_s$ which is a $m \times m-\lceil \dfrac {m} {2} \rceil$ dimension matrix are transformed to terms with degree lower than $m$ as follows \cite{Huapeng_Wu}:

 \begin{equation*}
R_s(1, x, ..., x^{m-1})^t=(x^{2\lceil \dfrac {m} {2} \rceil}, x^{2\lceil \dfrac {m} {2} \rceil+1}, ..., x^{2m-2} )^t~ mod~ P(x)
 \end{equation*}

\begin{equation*}
\mathsf{\textbf{B}} =
\begin{pmatrix}
q_{0,0} & q_{0,1} & \ldots
& q_{0,m-1} \\
q_{1,0} & q_{1,1} & \ldots
& q_{1,m-1} \\
\vdots & \vdots & \ddots
& \vdots\\
q_{m-\lceil \dfrac {m} {2} \rceil-1,0} & q_{m-\lceil \dfrac {m} {2} \rceil-1,1} & \ldots
& q_{m-\lceil \dfrac {m} {2} \rceil-1,m-1}

\end{pmatrix}
\begin{pmatrix}
1 \\
x \\
\vdots\\
x^{m-1}
\end{pmatrix}
=\begin{pmatrix}
x^{2\lceil \dfrac {m} {2} \rceil} \\
x^{2\lceil \dfrac {m} {2} \rceil+1}\\
\vdots\\
x^{2m-2}

\end{pmatrix}
~ mod~ P(x)
\end{equation*}
where $q_{i,j} \in \{0,1\}$. So coefficients of $C(x)$ can be computed by

$ c_i=\begin{cases}
a_{\dfrac {i} {2}}+a_{\lceil \dfrac {m} {2} \rceil}q_{0,i}+a_{\lceil \dfrac {m} {2} \rceil+1}q_{1,i}+...+a_{m-1}q_{\lfloor \dfrac {m} {2} \rfloor,i} & i~ even\\
a_{\lceil \dfrac {m} {2} \rceil}q_{0,i}+a_{\lceil \dfrac {m} {2} \rceil+1}q_{1,i}+...+a_{m-1}q_{\lfloor \dfrac {m} {2} \rfloor,i} & ~ otherwise
\end{cases}$\\

\begin{figure}[H]
	\centering
	\includegraphics[scale=0.7]{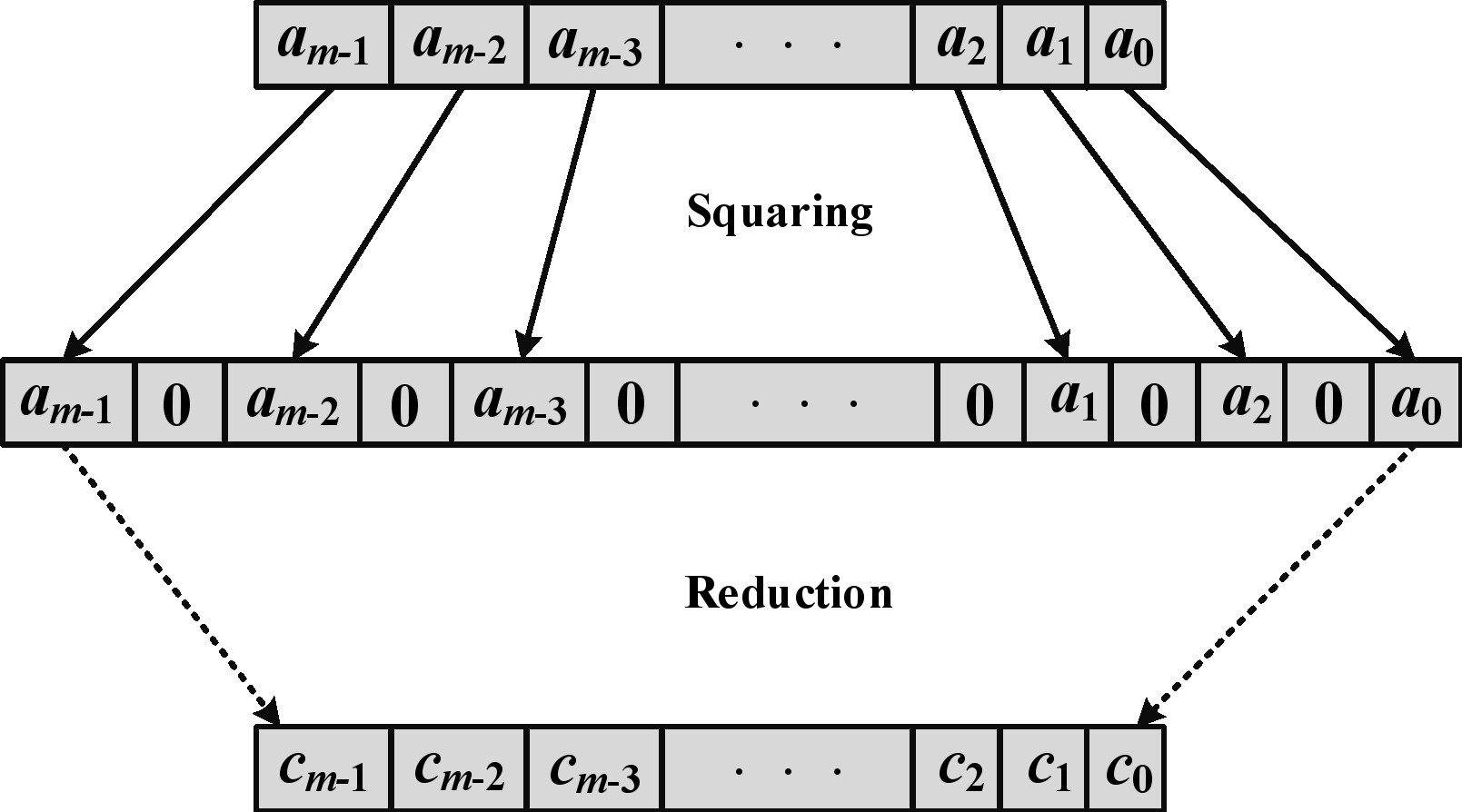}
	\caption{Squaring in polynomial basis.}
	\label{fig:Sq}
\end{figure}

Therefore, in polynomial basis squaring operation is implemented by array of XOR gates \cite{Inv_B}, \cite{Fault_B}. Fig.\ref{fig:Sq} shows the squaring in polynomial basis.  

Also in normal basis the squaring of element $A$ is expressed as:

\begin{equation*}
A^2=(\sum_{i=0}^{m-1} a_i\beta^{2^i})^2=\sum_{i=0}^{m-1} a_i\beta^{2^{i+1}}=a_{m-1}\beta+\sum_{i=1}^{m-1} a_{i-1}\beta^{2^{i}}
\end{equation*}

This means $A^2$ is represented by $(a_{m-2},a_{m-3},...,a_2,a_1,a_0,a_{m-1})$. So, one important property of the normal basis representation is that the squaring  is performed very efficiently by a simple one-bit cyclic shift.

\begin{itemize}
  \item \textbf{\textit{Multiplication in $\F_{2^m}$}}
\end{itemize}

There are three main architectures for implementation of finite field multipliers consist of bit-serial, digit-serial (or word-level) and bit-parallel. The bit-serial structures need very low hardware resources but on the other hand, the complete output bits are computed after $m$ clock cycles because one bit of the output is processed at each clock cycle. These architectures are suitable for lightweight cryptosystems. In digit-serial architectures, there is a trade-off between number of clock cycles and area; therefore digit-serial architecture can be a better choice for hardware implementation of the elliptic curves point multiplication. The third architectures are bit-parallel multipliers in which the output is computed in one clock cycle. However, the critical path delay and area are increased. The area consumption in the digit-serial architectures is higher than that of the bit-serial, but it is much lower than the bit-parallel architectures.

If finite field $\F_{2^m}$ is generated by irreducible polynomial $P(x)=x^m+\sum_{i=0}^{m-1} p_ix^i$ for multiplication of two elements $A(x)=\sum_{i=0}^{m-1} a_ix^i$ and $B(x)=\sum_{i=0}^{m-1} b_ix^i$ we have following steps \cite{Huapeng_Wu}:

\begin{equation*}
S(x)=A(x)\times B(x)=\sum_{k=0}^{2m-2} s_kx^k
\end{equation*}
where for coefficients of $s_k$ we have

\begin{equation*}
s_k=\sum_{i+j=k ~\\ 0\leq i,j\leq m-1} a_ib_j ~~~ k=0, 1, 2, ..., 2m-2
\end{equation*} 
The $S(x)$ must be reduced with module $P(x)$ 

\begin{equation*}
C(x)=S(x)~mod~P(x) \Rightarrow C(x)=\sum_{i=0}^{m-1} c_ix^i,~c_i\in \F_2
\end{equation*}

There are different methods for polynomial basis multiplication such as Mastrovito method, Karatsuba-Ofman algorithm, bit-serial MSB method, bit-serial LSB method, Least Significant Digit-serial (LSD) method, Interleaving method, Two-Step method, Matrix-Vector method and Montgomery method for more details see \cite{FF_Rodrigez}, \cite{FF_Book}, \cite{Gorge}, \cite{Conf_B}, \cite{Isecure_B} and \cite{Mont_B}.

Also for multiplication two normal elements $ A={\sum_{i=0}^{m-1} a_i\beta^{2^i}} $ and $ B={\sum_{j=0}^{m-1} b_j\beta^{2^j}} $ 
 
\begin{equation*}
C=A\times B=\sum_{i=0}^{m-1} a_i\beta^{2^i} \times \sum_{j=0}^{m-1} b_j\beta^{2^j}= \sum_{i=0}^{m-1} \sum_{j=0}^{m-1} a_ib_j \beta^{2^i} \beta^{2^j}=\sum_{k=0}^{m-1} c_k\beta^{2^k}
\end{equation*}
for $\beta^{2^i} \beta^{2^j}$ we have 

\begin{equation*}
\beta^{2^i} \beta^{2^j}=\sum_{k=0}^{m-1} c_k\lambda_{i,j}^{(k)}\beta^{2^k}, ~~\lambda_{i,j}^{(k)}=0~or 1
\end{equation*}
which $\lambda^{(k)}$ is $m$ dimensional matrix called multiplication matrix with entries $\lambda_{i,j}^{(k)}$. So, coordinates of $C$, i.e., $c_k$ are as follows:

\begin{equation*}
c_k=\sum_{i=0}^{m-1} \sum_{j=0}^{m-1} a_ib_j\lambda_{i,j}^{(k)}
\end{equation*}
More details for computation of $ \lambda_{i,j}^{(k)} $ and normal basis multiplication is presented in \cite{HandBook} and \cite{VLSIONB_B}.   

The GNB multiplication of $C=A\times B$ can also be computed by the following approach \cite{IEEE}. Let $\F_{2^m}$ has a GNB of type $T$. Also let $u$ be an integer of order $T~mod~p$, where $p=mT+1$ is a prime number. Then, the set 

\begin{equation*}
Z=\{z_{i,j}:z_{i,j}=2^iu^j|i\in \{0, 1, ..., m-1\},~j\in\{0, 1, ..., T-1\} \}
\end{equation*}
is a reduced residue system modulo $p$. Therefore, each positive integer $x$ less than $p$ can be uniquely represented as $x=z_{i,j}~mod~p$. Let $F$ be a function given by:

\begin{equation*}
F(x)=i,~x=z_{i,j}~mod~p,~~~~and~~~F:\{ 1,2,...,p-1\}\rightarrow\{0,1,...,m-1\}
\end{equation*}
for an even type $T$, the first coordinate $c_0$ of product $C$ is computed by:

\begin{equation*}
c_0=\sum_{k=1}^{p-2} a_{F(k+1)}b_{F(p-k)}
\end{equation*}

Also, other coordinates $c_i $, $ 1\leq i\leq m-1$, are computed similarly by one bit right cyclic shift of inputs \cite{GNB_FPGA_B}-\cite{GNB_VLSI_B}.
Normal basis multiplication approaches and comprehensive comparisons for GNB and ONB multiplier structures can be found in \cite{VLSIONB_B}-\cite{GNB_VLSI_B}.

\begin{itemize}
  \item \textbf{\textit{Inversion in $\F_{2^m}$}}
\end{itemize}

There are mainly two ways of computing field inversion. The first is  to  use  the  extended  Euclidean  algorithm. The second method is to use the \textit{Fermat's little} theorem in the multiplicative group of the finite field. If $ A $ be a nonzero element in $\F_{2^m}$, then one element such as $ C $ exist which satisfies in following equation:

\begin{equation*}
A\times C=1
\end{equation*}
here the $ C $ element is called invert of $ A $.
  
In general, for computation of inversion in any finite field one can use the multiplicative structure of the group of the nonzero elements with \textit{Lagrange's theorem}. In particular, for the binary field $\F_{2^m}$ and for any nonzero element $ A $ we have:

\begin{equation*}
A^{2^{m-1}}=A.A^{2^m-2}=1  
\end{equation*}   

Therefore, for binary fields the computation of the inversion of $A$ can be performed by the modular exponentiation $A^{2^m-2}$. This computation can be done using the basic \textit{Square-Multiplication algorithm} \cite{SM}. In this algorithm inversion is computed with $m-1$ squaring and $m-2$ multiplication operations. An efficient method to compute the inversion in $\F_{2^m}$ based on the Fermat's little theorem is proposed by Itoh and Tsujii in \cite{ITA}. In their algorithm the number of multiplication is reduced to $log_2^{m-1}+HW(m-1)-1$. The hardware implementations of the inversion operation in the literature often use the \textit{Itoh-Tsujii algorithm} (ITA) for the point multiplication. In \cite{Inv_B}, \cite{Inv_GNB1} and \cite{Inv_GNB2} recent hardware implementations of ITA on polynomial basis and normal basis are presented. Also Euclidean-based inversion algorithms which are presented in \cite{SoftPM} are consist of \textit{Extended Euclidean Algorithm} (EEA), \textit{Almost Inverse Algorithm} (AIA), \textit{Modified Almost Inverse Algorithm} (MAIA).

\subsection{\textbf{Field arithmetic in $\F_{p}$}}

One of the popular fields which are choice for ECC are prime fields with large prime characteristic. To hardware implementation of the ECC over $\F_{p}$ we have various choices for prime number $p$. In following, we present these choices based on \cite{smart}. \\

\begin{itemize}
  \item \textbf{General Primes}
\end{itemize}

In general prime fields the number $p$ has not a special pattern. Therefore, implementation of modular field arithmetics are not very efficient. The most hardware implementations for these prime numbers are based on Montgomery method \cite{MM}. In this method uses a special representation to perform efficient arithmetic, the division and remaindering are essentially implemented by shift operation. \\

\begin{itemize}
  \item \textbf{Generalized Mersenne Primes}
\end{itemize}

Special primes are highly suited for efficient reduction techniques, the most simple form of such primes being the Mersenne primes, which are primes of the form $p=2^k-1$. Generalization on the Mersenne primes is considered in literature because in practice the number of Mersenne primes of the correct size for cryptography is limited. In \cite{Crandall} the use of primes of the form $p=2^k-c$, where $c$ is a small integer is proposed. Primes of the form $p=2^k\pm c$ for a small value of $c$ are called pseudo-Mersenne primes. NIST recommended five efficient prime fields, with prime numbers: $p_{192}=2^{192}-2^{64}-1$, $p_{224}=2^{224}-2^{96}+1$, $p_{256}=2^{256}-2^{224}+2^{192}+2^{96}-1$, $p_{384}=2^{384}-2^{128}-2^{96}+2^{32}-1$ and $p_{521}=2^{521}-1$.

\begin{itemize}
  \item \textbf{\textit{Addition and Subtraction in $\F_{p}$}}
\end{itemize}

Addition in $\F_p$ has carry bit which is propagates in the structure. This leads to long critical path delay and reduce operating frequency for fields with a large prime $p$ in a hardware implementation. The modular addition is defined as the computation of $S=A+B~(mod~p)$ given the integers $A, B$ and $p$, where $A=\sum_{i=0}^{m-1} a_i2^i$ and $B=\sum_{i=0}^{m-1} b_i2^i$ are $m$-bit positive integers with $0<A, B<p$. One of the efficient method for computation of the modular addition and subtraction is \textit{Omura's method} \cite{Omura}. The modular addition is computed as follows:

\begin{equation*}
S=A+B~(mod~p)=\begin{cases}
 A+B-p, &~A+B \geq 2^m~or~A+B-p \geq 0\\
A+B & ~ otherwise
\end{cases}\\
\end{equation*}
and for modular subtraction we have

\begin{equation*}
S=A-B~(mod~p) =\begin{cases}
 A-B+p, &~if~A-B < 0\\
A-B & ~ otherwise
\end{cases}\\
\end{equation*}

\begin{figure}[H]
	\centering
	\includegraphics[scale=0.45]{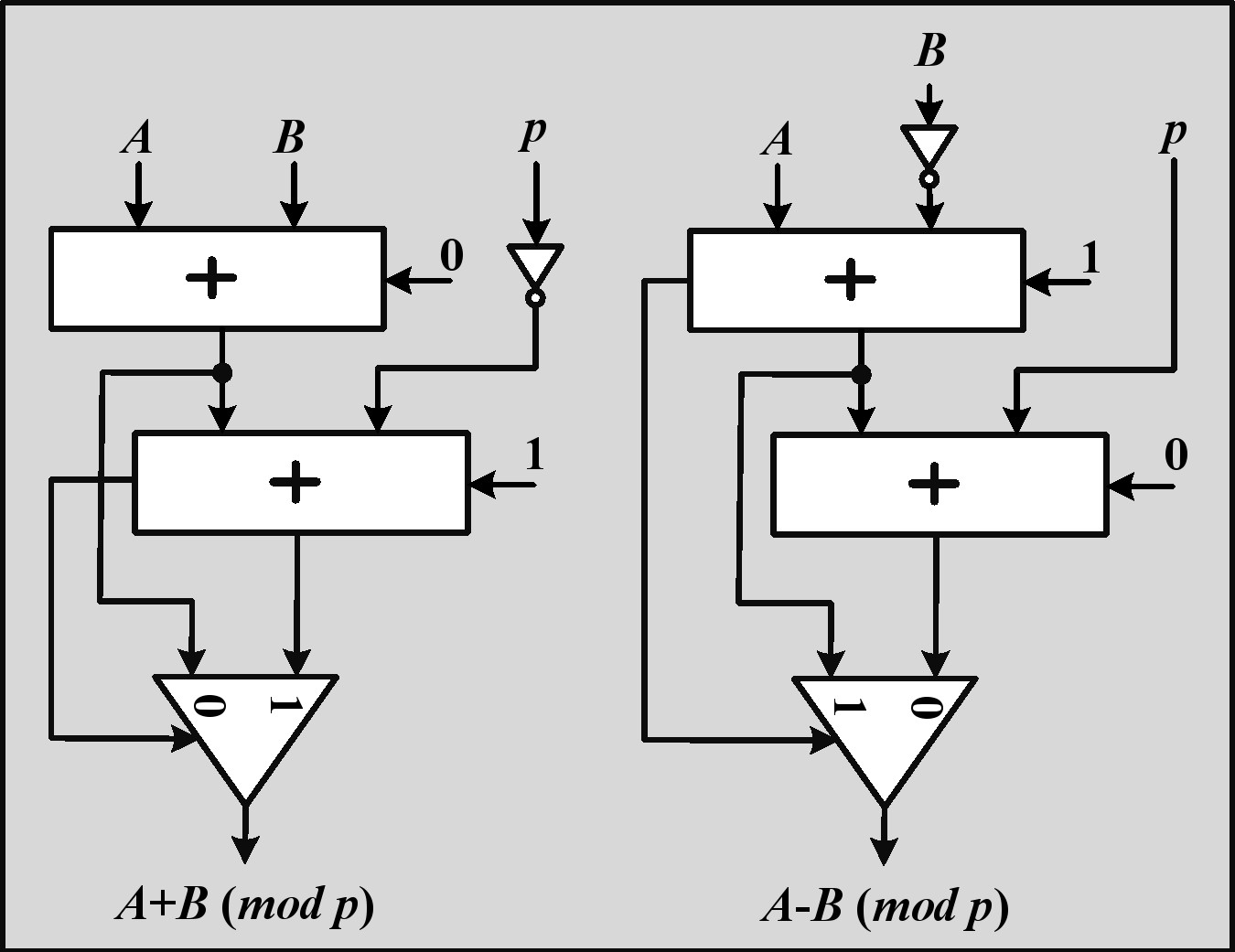}
	\caption{Modular addition and subtraction.}
	\label{fig:MA}
\end{figure}

Fig.\ref{fig:MA} shows circuits for modular addition and subtraction.

A modular addition over $\F_p$ is implemented by using two adders. The two elements $A$ and $B$ are added together by the first adder. Then, the second one subtracts the modulus $p$ based on carry bit of the first adder. To reduce the critical path delay, the implemented adders for $\F_p$ are based on carry propagate adders such as \textit{carry skip adder}, \textit{carry select adder}, \textit{carry lookahead adder}, \textit{carry delayed adder} and \textit{carry save adder}. See \cite{FF_Rodrigez}-\cite{Gorge} for algorithms, structures and more details.

\begin{itemize}
  \item \textbf{\textit{Multiplication in $\F_{p}$}}
\end{itemize}

The modular multiplication of two elements $A$ and $B$ in $\F_p$ is defined as product of these two field elements modulo $p$

\begin{equation*}
C=A\times B~(mod~p)
\end{equation*}
where $ A, B, C, p \in \F_p $ and $ A, B < p $. There are many different methods for computation of the \textit{modular multiplication consist of multiply and then divide}, \textit{Interleaved modular multiplication}, \textit{Brickell's method}, \textit{Montgomery modular multiplication}, \textit{Bipartite modular multiplication} and \textit{Tripartite modular multiplication} \cite{FF_Rodrigez}, \cite{Gorge} and \cite{Marcelo}-\cite{Kazuo}. The Montgomery multiplication is the most used in between other modular multiplication methods. Montgomery multiplication of $A$ and $B~(mod~p)$, is defined as $A\times B\times 2^{−k}~(mod~p)$ for some fixed integer $k$. Algorithm \ref{alg:A7} shows the Montgomery modular multiplication.

\begin{algorithm}[H]
\caption{Montgomery Modular Multiplication}
\label{alg:A7}
\textbf{Input:} $A ,B < p < 2^k$, with $ 2^{k-1} < p < 2^k$ and $ p=2t+1 $, with $t \in N$\\
\textbf{Output:} $u = A\times B\times 2 ^{−k}$ mod $ p $.\\
1. $u=0$ \\ 
2. \textbf{for} $ i $ \textbf{from} 0 \textbf{to} $ k $-1 \textbf{do}\\
3.~~$ u=u+a_i.B $\\ 
4.~~\textbf{if} $ u_0=1 $ \textbf{then}  \\ 
5.~~ $ u=u+p $ \\
6.~~\textbf{end} \textbf{if}\\
7.  $ u=u~div~2 $ \\
8.~~\textbf{end} \textbf{for}\\
9.~~\textbf{if} $ u \geq p $ \textbf{then}  \\ 
10.~~~ $ u=u-p $ \\
11.~~\textbf{end} \textbf{if}
\end{algorithm}

In Montgomery method, the operations are computed based on shift (division by 2) and addition which leads to efficient implementation. Therefore, this is the most important property of the algorithm.

\begin{itemize}
  \item \textbf{\textit{Inversion in $\F_{p}$}}
\end{itemize}
If for nonzero element $A\in\F_p$ we have $A\times B=1~mod~p$ or $B=A^{-1}~mod~p$, where $B\in\F_p$ then $B$ is inverse of $A$ in $\F_p$. There are two general approaches, Similar to binary fields, for find $B$. The first is based on Fermat's little theorem which can be implemented by modular exponentiation and the second is based on the extended Euclidean (GCD) algorithm.\\
The Fermat's little theorem states that for the prime number $p$ and for any integer $A$ not divisible by $p$, we have, 

\begin{equation*}
A^{p-1}\equiv 1~(mod~p)
\end{equation*}

We write $A\times A^{p-2}\equiv 1~(mod~p)$, that means, $A^{-1}\equiv A^{p-2}~(mod~p)$. So, the inversion of $A$ can be performed by the modular exponentiation $A^{p-2}$. The disadvantage of this approach is high execution time \cite{Gorge}. In another approach to inversion is implemented based on extended Euclidean algorithm. The more efficient methods in this type of approaches are \textit{Kaliski Inversion for Montgomery Domain} and \textit{Almost Montgomery Inverse} (AMI) \cite{Gorge}. In \cite{Gorge} and \cite{HandBook} more detailed information about modular inversion in $\F_p$ are presented.

\section{\textbf{Implementation of the elliptic curve cryptography:Methods, Steps and Considerations}}
\label{sec-ECC1}

\begin{figure}[H]
	\centering
	\includegraphics[scale=0.75]{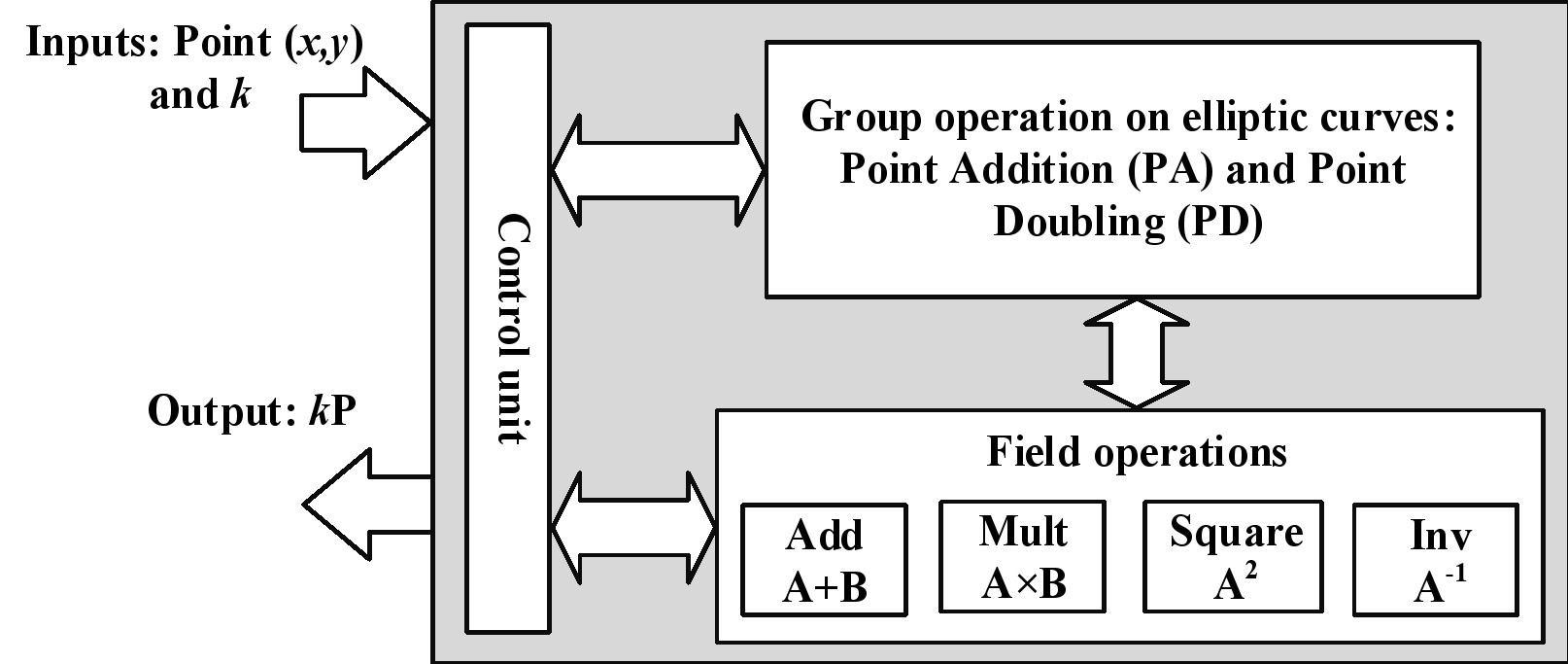}
	\caption{Overall architecture for computation of the elliptic curve point multiplication.}
	\label{fig:FFF1}
\end{figure}

The hardware implementation of ECC has two important stages that both have direct impact on the efficiency of the implementation. In the first stage, type of elliptic curve, finite field, point multiplication algorithm and structures related to field operations are selected. And in the second stage, the hardware architecture of the circuit in two levels of time scheduling of field operations, based on available resources and design of microarchitecture of field operations is implemented. Overall architecture for computation of the elliptic curve point multiplication is shown in Fig.\ref{fig:FFF1}.

\begin{figure}[H]
	\centering
	\includegraphics[scale=0.65]{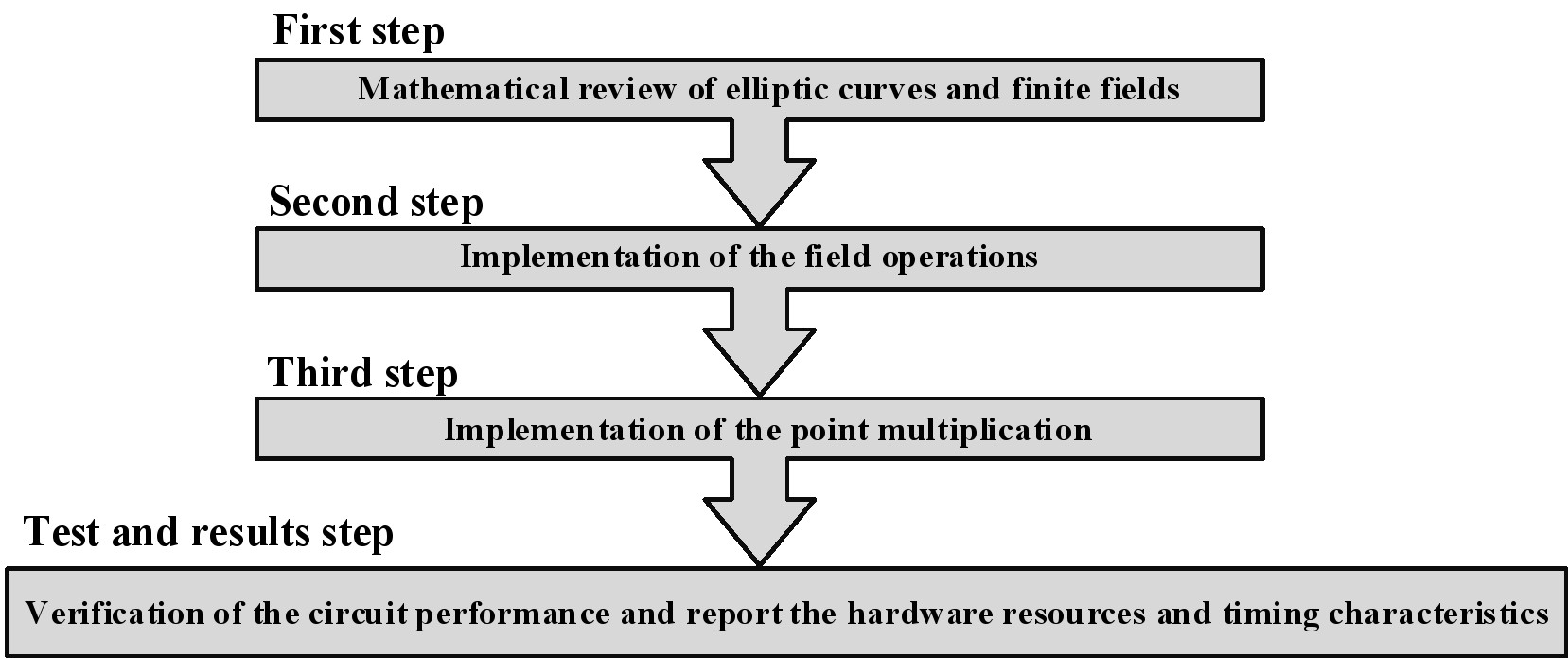}
	\caption{Steps of the hardware implementation of the elliptic curve point multiplication with aims and related works in each step.}
	\label{fig:HPM_1}
\end{figure}

\begin{figure}[H]
	\centering
	\includegraphics[scale=0.65]{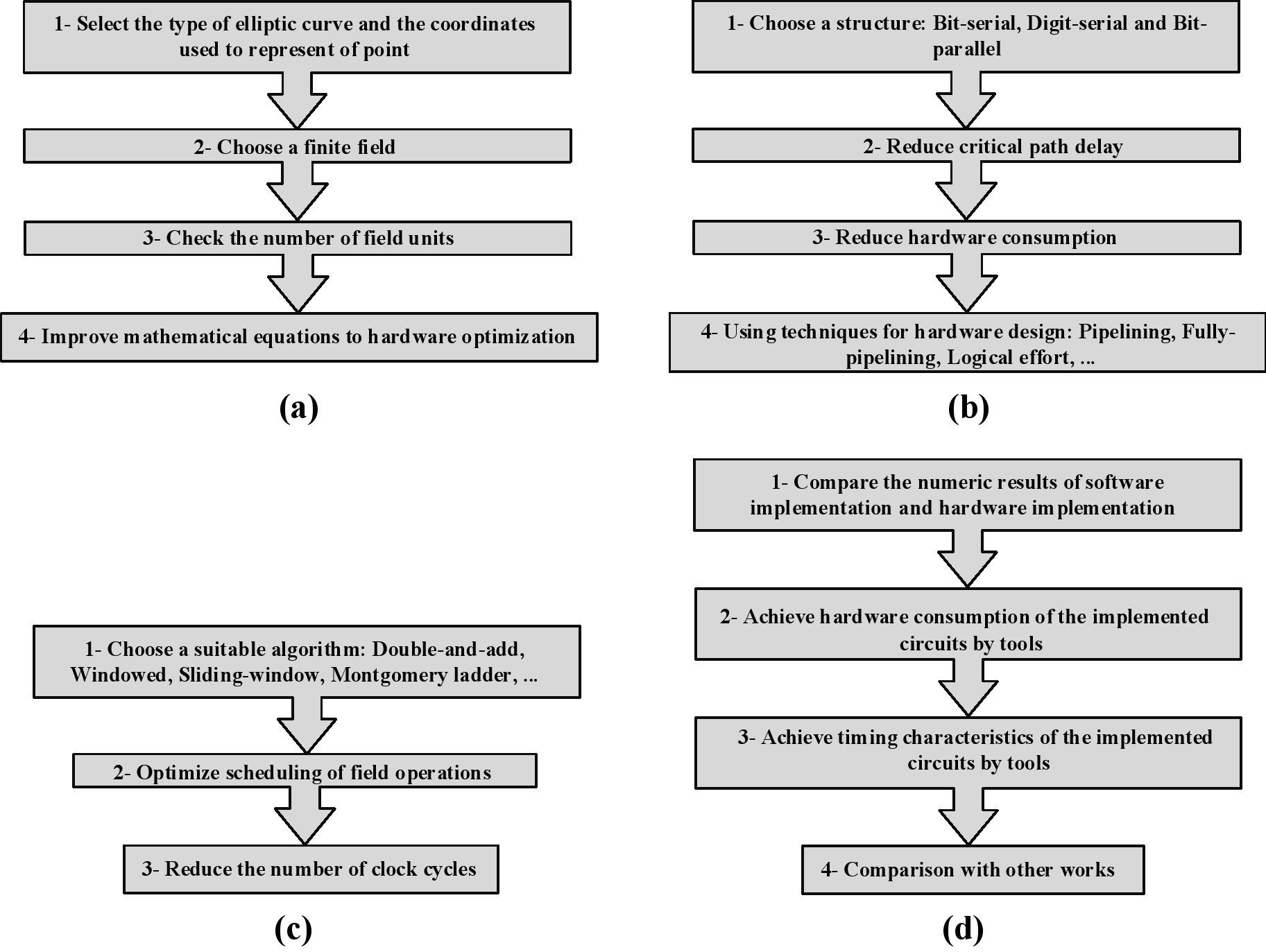}
	\caption{Main details of the first step (a), second step (b), third step (c) and fourth step (d) of the hardware implementation of the elliptic curve point multiplication}
	\label{fig:S_1}
\end{figure}

Fig.\ref{fig:HPM_1} shows more details of the hardware implementation of the elliptic curve point multiplication with aims and related works in each step. As seen form this figure, implementation is split into four steps: (1) Mathematical review (study) of elliptic curves and finite fields. (2) Implementation of the field operations. (3) Implementation of the point multiplication algorithm. (4) Verification of the circuit performance and report the hardware resources and timing characteristics. Main details of the first step, second step, third step and fourth step of the hardware implementation of the elliptic curve point multiplication are shown in Figs.\ref{fig:S_1} (a), (b), (c) and (c) respectively.

\subsection{\textbf{Design Considerations for the hardware implementation of the elliptic curve cryptography}}

Different hardware implementation of the elliptic curve point multiplication are proposed. In the following main design considerations for the hardware implementation of the ECC in the different level of speed, efficiency, reconfigurability and hardware consumption are present.

\begin{itemize}
  \item \textbf{\textit{Flexibility}}
\end{itemize}

\begin{figure}[H]
	\centering
	\includegraphics[scale=0.6]{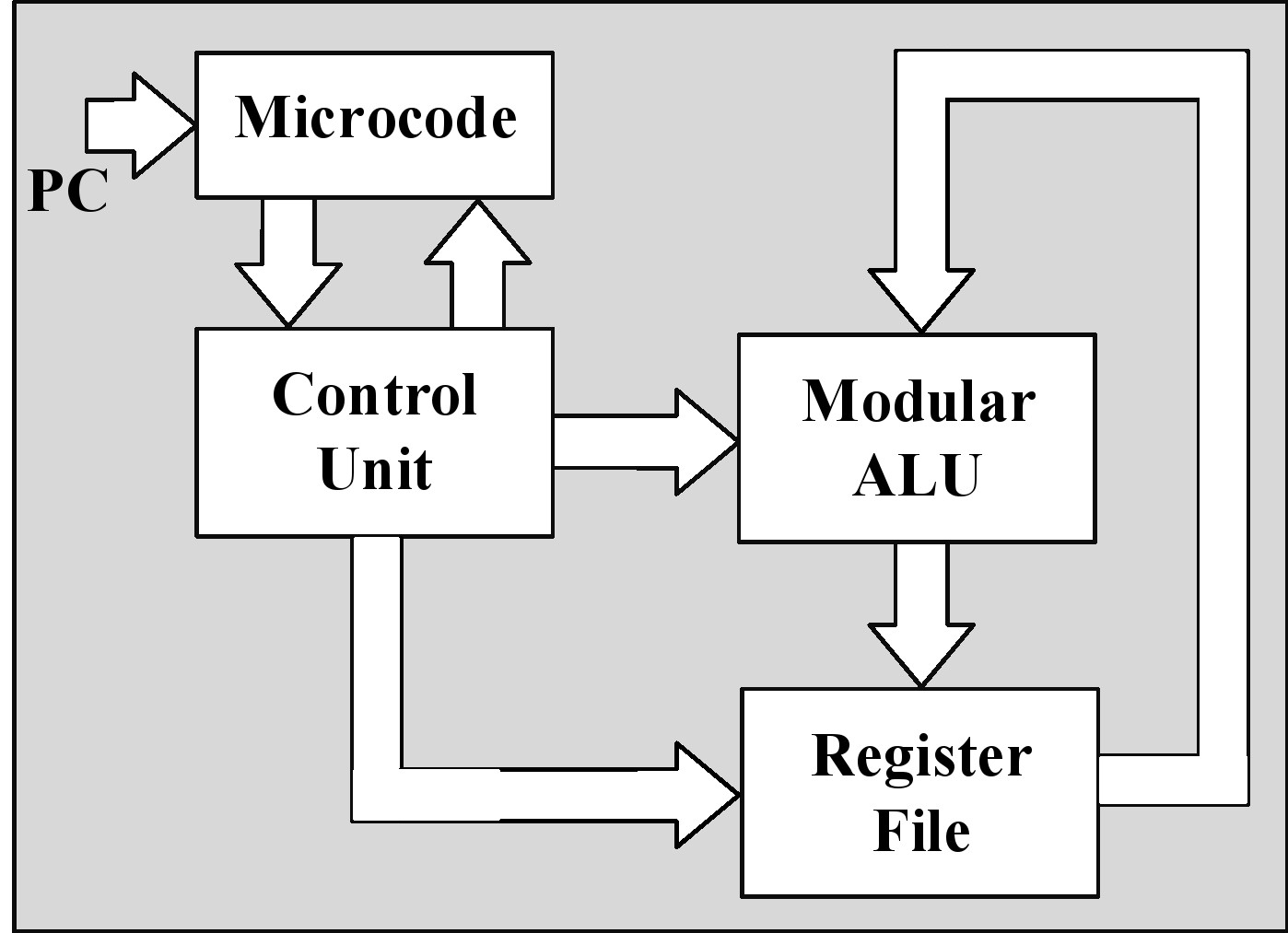}
	\caption{Overall block diagram of architecture flexible processor.}
	\label{fig:FL}
\end{figure}

Flexibility and reconfigurability of ECC implementations make them one of the best choice for high-performance applications. It allows designing ECC implementations that are optimized for support several parameters such as arbitrary elliptic curve, different standards, fields, algorithms and coordinate choices can be achieved through reconfiguration. In this case, implementation of the cryptographic algorithms in hardware are preformed without losing flexibility. But in specific structures, ECC implementation are optimized based on only specific parameters such as one field size, one irreducible polynomial in binary field, one prime number as modulo in prime fields. The flexible ECC implementation are usually based on hardware-software codesign approach. The structure is implemented by a parameterized module generator, which can accommodate arbitrary parameters. The control part of the processor is microcoded, enabling curve operations to be incorporated into the processor. The microcoded approach also has a shorter development time, algorithmic optimization and is more flexible. Fig.\ref{fig:FL} shows overall block diagram of flexible processor.

\begin{figure}[H]
	\centering
	\includegraphics[scale=0.9]{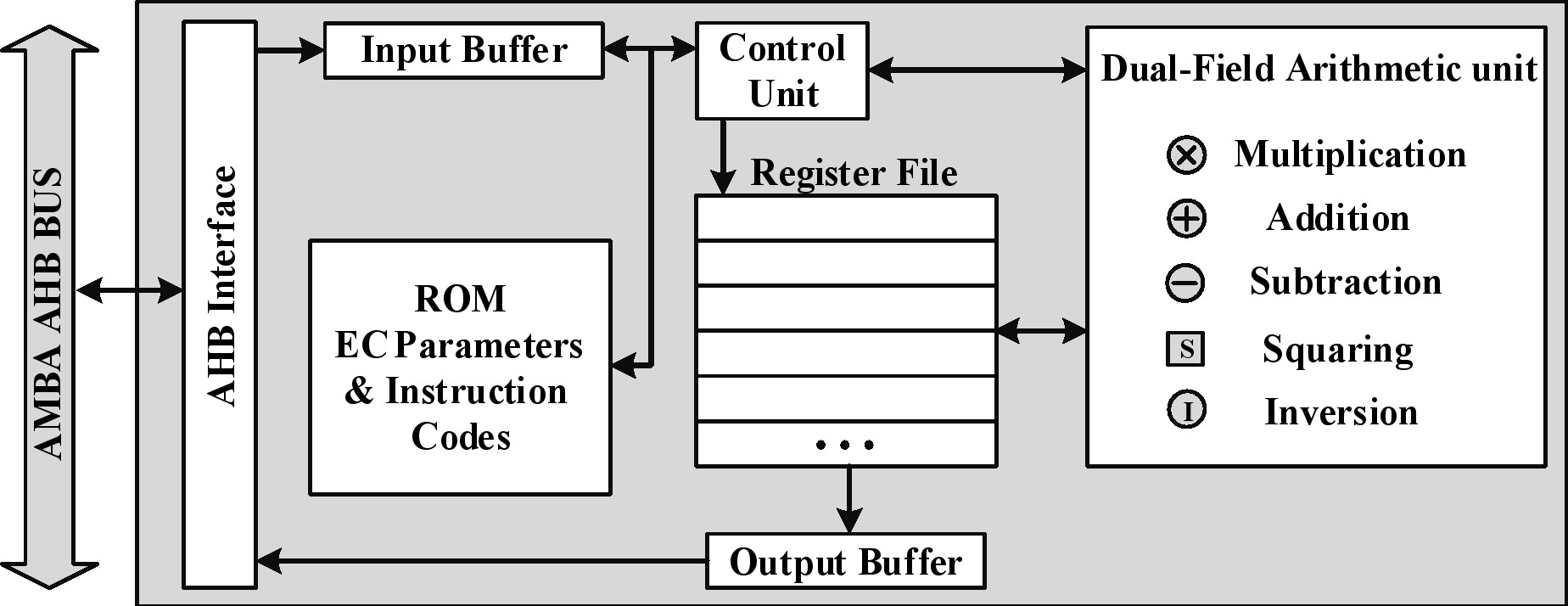}
	\caption{General block diagram of dual-field ECC processor.}
	\label{fig:DF}
\end{figure}

Dual-field ECC processor are traditional flexible implementation. General block diagram of Dual field ECC processor is shown in Fig.\ref{fig:DF}. In this structure, arithmetic unit support both field operations on $\F_{p}$ and $\F_{2^m}$. It is consists of a control unit, dual-field ALU, ROM memory, register file and standard advanced microcontroller bus architecture (AMBA) advanced high-performance bus (AHB) interface. By initializing memory with curve parameters and instruction codes, the processor can flexibly perform arbitrary elliptic curve operations over dual-field and different point multiplication algorithms.

\begin{itemize}
  \item \textbf{\textit{Scalability}}
\end{itemize}

One of the important subject when designing crypto-processor is scalability. It is the ability to perform cryptographic operations with support various field sizes (for various level of security) and irreducible polynomials (or modulo in prime field) without reconfiguration. In this case, the performance and power consumption of crypto-processor can be controllable. There are two primary approaches for realization of scalable crypto-processor. The first approach is over-designing. It is based on consideration of require hardware for performing the full field operation but the total clock cycles are decreased when performing cryptographic operations with minimum level of security. The second approach is the hardware-software codesign method. In this method the minimum bit size for operations (proper to minimum level of security) implements in hardware and uses software to perform the extension to maximum bit size.

\begin{itemize}
  \item \textbf{\textit{Security}}
\end{itemize}

The security is one of the most issues in the ECC implementation. Countermeasures to attacks consist of power, timing and electromagnetic radiation must be considered until leaks no information about the bit pattern of the secret key. In a side-channel attack the implementation is under attack by power consumption and execution time of operations. Bit pattern of the scalar $ k $ in the point multiplication can be approximated based on power consumption and picks on power trace in a simple power attack (SPA). Montgomery ladder algorithm is highly regular for each bit of scalar $ k $. In other words, for any bit of $ k $ the point addition and point doubling are computed simultaneously. Therefore, the power trace has a unified and bit pattern of $ k $ is not visible from the power trace. In timing analysis attack which is a side-channel attack, the time taken to execute cryptographic algorithms is analyzed for compromise a cryptosystem. To resistant against this attack, the implementation must be reduces data dependent timing information. So, in the implementation of the point multiplication execution time for each point multiplication must be fixed and independent from inputs and scalar number $ k $.

\begin{itemize}
  \item \textbf{\textit{Implementation platforms}}
\end{itemize}

The the most of the hardware implementations of ECC are realized based on FPGA design and ASIC design. FPGAs are reconfigurable platforms, so the functionality of the implementation can be modified through reconfiguration. For ASIC implementations, the circuit is specialized forever. It should be noted that low number of papers about full-custom implementation of the point multiplication in chip-level have been found.

\subsection{\textbf{Different used techniques and the proposed ideas for hardware implementations of the elliptic curve cryptography}}

\begin{itemize}
  \item \textbf{\textit{Pipelining and Fine-Grain pipelining}}
\end{itemize}

\begin{figure}[H]
 	\centering
 	\includegraphics[scale=0.5]{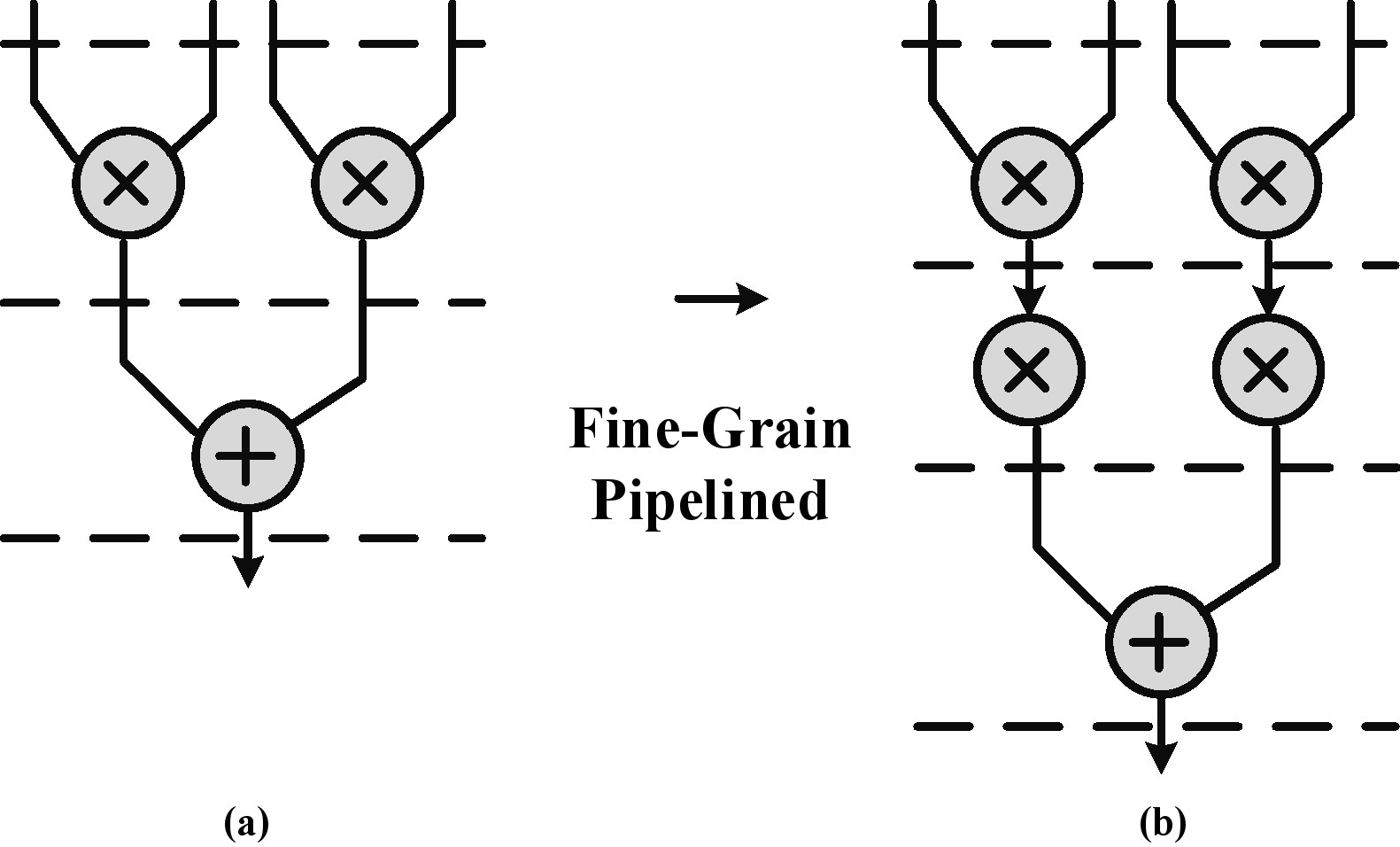}
 	\caption{(a) Pipelined structure and (b) implementation of the Fine-Grain pipelining technique for field multipliers.}
 	\label{fig:FF2}
\end{figure}

In pipelining technique we use registers between field operations in the data path to reduce critical path delay. Therefor, pipelining of data path of the point multiplication can increase operation frequency and processing speed. Besides  using  pipelining of data path, to further increase speed and throughput in implementations the Fine-Grain pipelining technique is employed. In this case, in addition data path, field operations are pipelined. Fig.\ref{fig:FF2} shows implementation of the Fine-Grain pipelining technique for field multipliers. In this figure 2 field multipliers shown at left are broken into 4 smaller and faster parts in right. Breaking and replacing slower parts with some faster units in pipeline architecture will increase clock frequency and throughput of the circuit. As shown in the figure for desired operation frequency, the multiplier is broken into two smaller units with lower critical path delay than that of original structure.

\begin{figure}[H]
 	\centering
 	\includegraphics[scale=0.6]{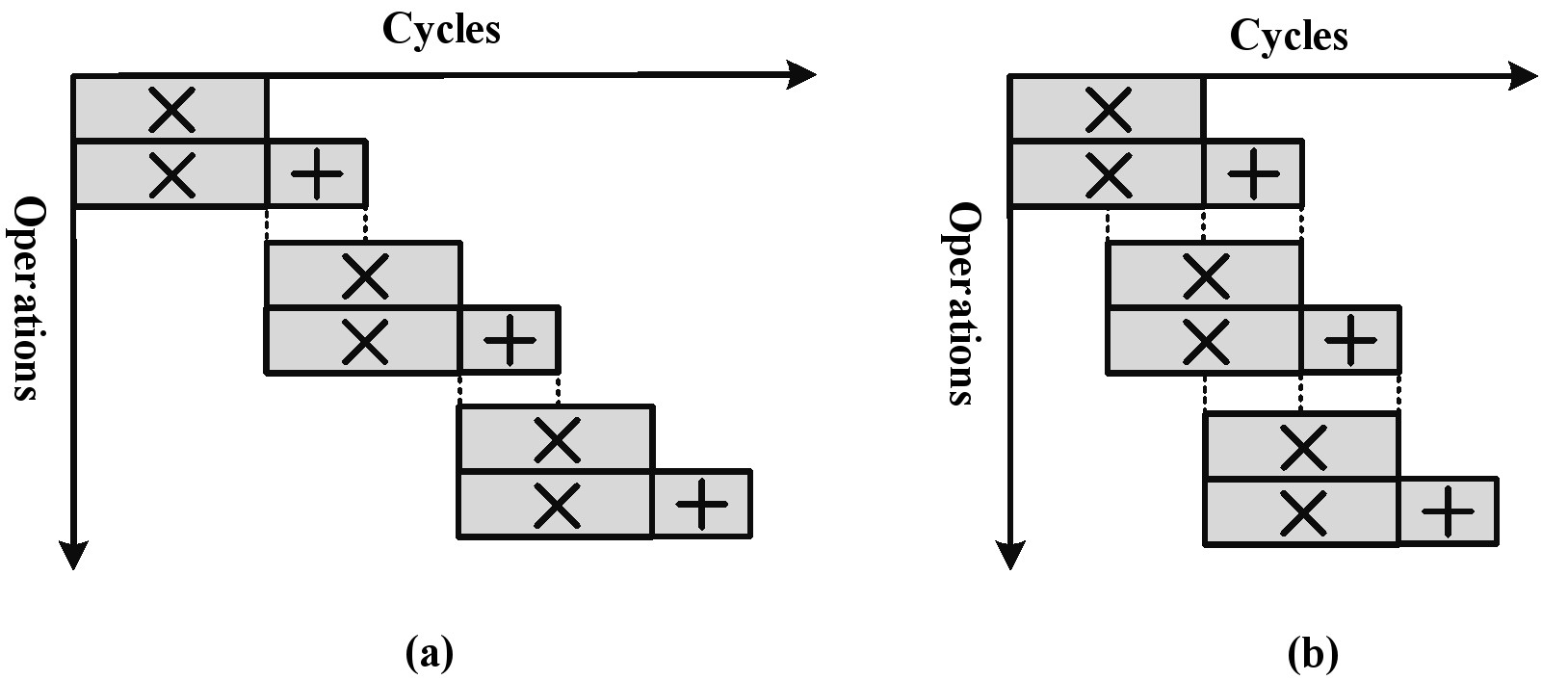}
 	\caption{(a) The scheduling of Fig.\ref{fig:FF2} (a) and (b) the scheduling of Fig.\ref{fig:FF2} (b).}
 	\label{fig:FF3}
\end{figure}

In Figs.\ref{fig:FF3} (a) and (b) show the scheduling of Fig.\ref{fig:FF2} (a) (pipelined) and Fig.\ref{fig:FF2} (b) (Fine-Grain pipelined) respectively. It can be seen that the number of field operations processed in the structure with Fine-Grain pipelined is more than that of in pipelined structure.

\begin{itemize}
  \item \textbf{\textit{Retiming: minimize the clock  period and the number of registers in the circuit}}
\end{itemize}

\begin{figure}[H]
 	\centering
 	\includegraphics[scale=0.5]{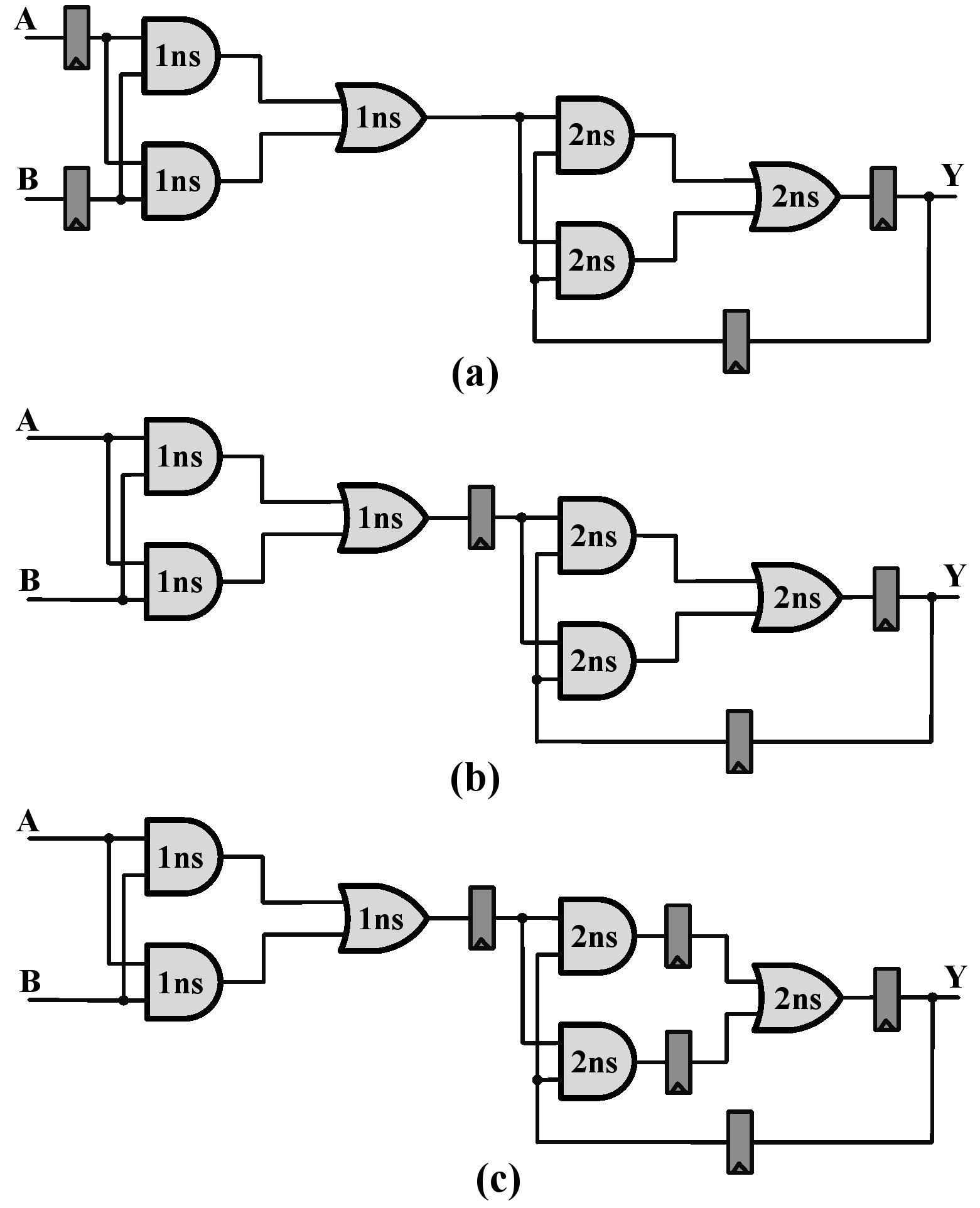}
 	\caption{A simple circuit (a), retiming for minimum registers (b) and retiming for minimum period (c).}
 	\label{fig:Re}
 \end{figure}

Retiming is a technique for optimizing sequential digital circuits. It repositions the registers between the combinational parts of digital circuits. The main aim of retiming is to find a digital circuit with the minimum number of registers for a specified clock period. There are two general approaches; minimizing the clock period of the circuit without regard to the number of registers and minimizing the number of registers in the circuit with no constraints on the clock period \cite{Re}. For explain the concept of retiming, consider a simple circuit in Fig.\ref{fig:Re} (a), where delay of each gate is shown inside it. The typical clock period for this circuit is given by the maximum delay of critical path of gates. So, in Fig.\ref{fig:Re} (a) the clock period is 6ns. In Fig.\ref{fig:Re} (b) an equivalent circuit with three D flip-flops and clock period of 4ns can be obtained by repositioning D flip-flops. This circuit has the minimum number of D flip-flops. On the other hand, the minimum clock period achievable by moving D flip-flops is 2ns at a cost of 4 D flip-flops as shown in Fig.\ref{fig:Re} (c).  

Therefore, a simple reconfiguration of D flip-flops product designs with differing area costs (number of D flip-flops) and performance (clock period). The retiming handles a trade-off between area and performance to provide solutions for varying clock periods. This technique can be proposed as an efficient method for hardware implementation of ECC.
 
\begin{itemize}
  \item \textbf{\textit{Scheduling of the underling field arithmetics:Parallel processing and Resource sharing}}
\end{itemize}

To efficient implementation of the point addition and point doubling based on selected coordinate and also management of hardware consumption and computation time, we can use proper scheduling. The scheduling of point multiplication operations must be carefully performed. A suitable parallelization of the operations can be employed based on the scheduling of the underling field arithmetics in the point addition and point doubling computation. It can be implemented by performing point addition and point doubling, in each loop iteration of the point multiplication, concurrently. The limitations of the parallel processing method is hardware resources. In the ECC processors, computation time is one of the most important factor that is considered in all previous works (specially in FPGA-based implementations). In the binary Weierstrass curves, the point addition and point doubling formulas are performed in parallel by using two levels of multiplications. In more details, for example computation of PA and PD in LD coordinate requires 6 field multipliers as follows:

The point addition $(Z_a,X_a)=Add(X_1, Z_1, X_2, Z_2,x)$ is given by:

\begin{equation*}
Z_a=(X_1 \times Z_2+X_2 \times Z_1)^2, X_a=xZ_a+(X_1 \times Z_2)\times(X_2 \times Z_1)
\end{equation*}
and for point doubling $(X_d, Z_d)=Double(X_i, Z_i, b)$, we have: 

\begin{equation*}
Z_d=X_i^2 \times Z_i^2,  X_d=X_i^4+bZ_i^4.
\end{equation*}

The parallel point addition and point doubling operations are computed in at least tho steps due to the data dependency of the formulas. And, in each step at most three field multiplier are used. In the first step the three multiplications $A=X_1\times Z_2$, $B=X_2\times Z_1$ and $X_1^2\times Z_1^2$ are computed in parallel by multipliers $M_1,M_2$ and $M_3$ respectively. In the second step $A\times B$, $x\times Z_a$ and $b\times Z_1^4$ are performed similarly. Now if we have restriction on hardware resources these computation based on two field multipliers can be implemented in three steps.

\begin{equation*}
Step-1 =\begin{cases}
 t_1=X_1 \times Z_2 \overset{by}\longrightarrow M_1,\\
 t_2=X_2 \times Z_1 \overset{by}\longrightarrow M_2
\end{cases}\\
\end{equation*}

\begin{equation*}
Step-2 =\begin{cases}
 x \times Z_a \overset{by}\longrightarrow M_1,\\
 t_1 \times t_2 \overset{by}\longrightarrow M_2
\end{cases}\\
\end{equation*}

\begin{equation*}
Step-3 =\begin{cases}
 X_1^2 \times Z_1^2 \overset{by}\longrightarrow M_1,\\
 b \times Z_1^4 \overset{by}\longrightarrow M_2
\end{cases}\\
\end{equation*} 

\begin{figure}[H]
 	\centering
 	\includegraphics[scale=0.8]{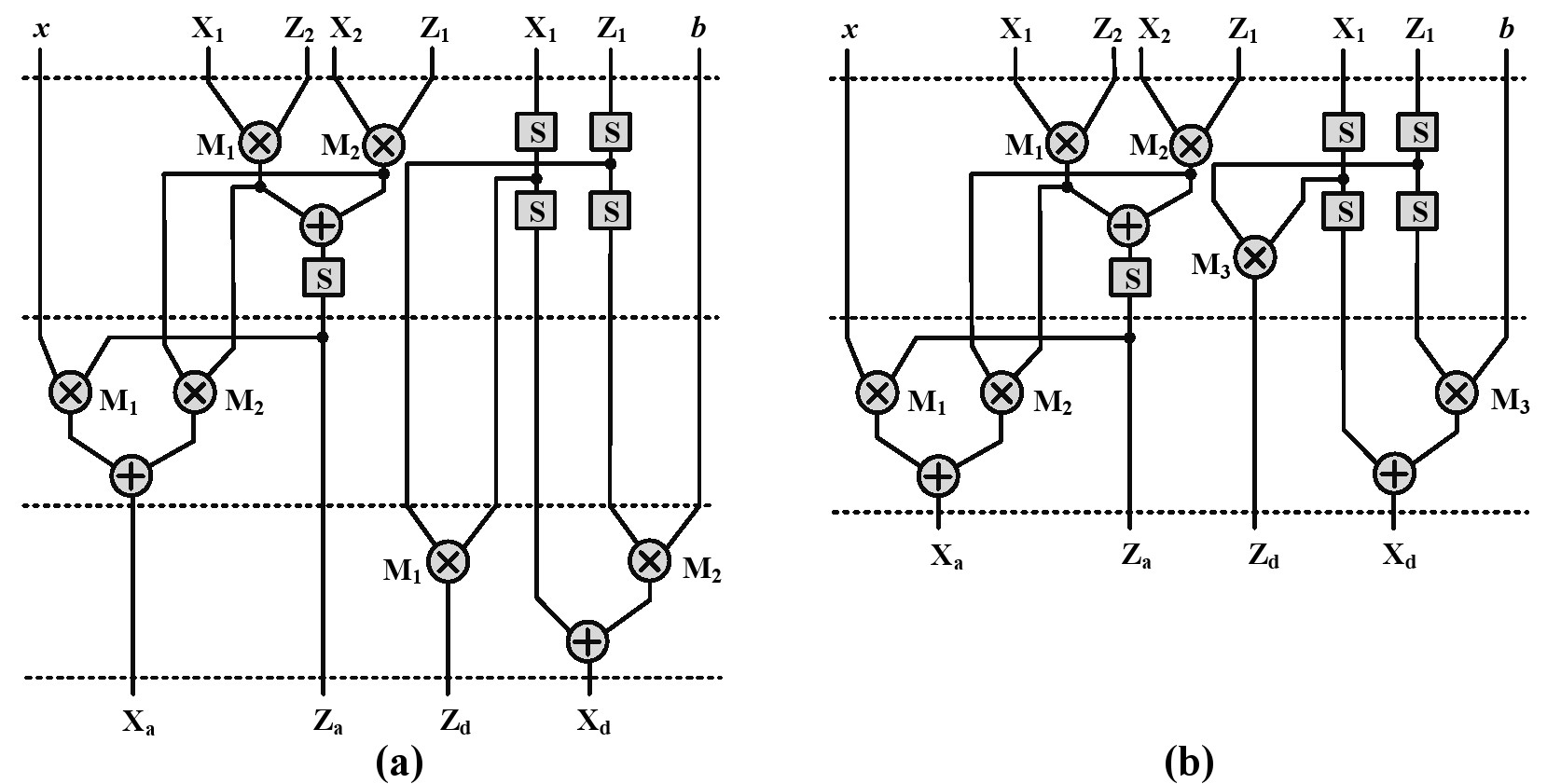}
 	\caption{Scheduling with restriction on hardware resources based on two field multipliers (a), the resource allocation to reduce the number of clock cycles based on three field multipliers (b).}
 	\label{fig:FF_1}
 \end{figure}

The scheduling of parallel computation of the point addition and point doubling of binary Weierstrass curves is shown in Fig.\ref{fig:FF_1}. The scheduling with restriction on hardware resources based on two field multipliers is shown in Fig.\ref{fig:FF_1} (a). In Fig.\ref{fig:FF_1} (b) the scheduling of the resources is performed to reduce the number of clock cycles. In Fig.\ref{fig:FF_1} 6 field multiplication operations for the point addition and point doubling are implemented by three and two multipliers by resource sharing in separate steps.

\begin{figure}[H]
 	\centering
 	\includegraphics[scale=0.45]{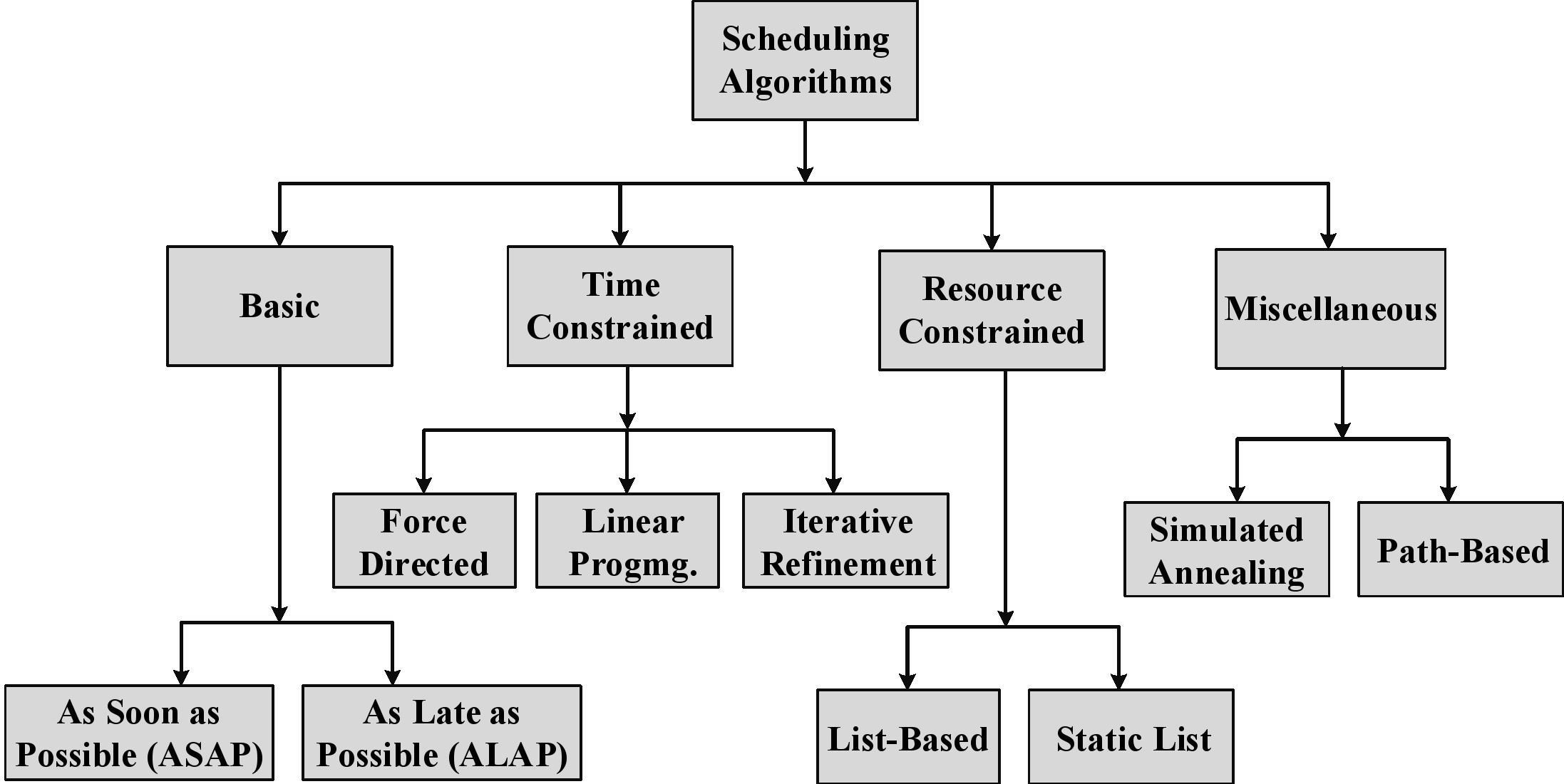}
 	\caption{Scheduling algorithms.}
 	\label{fig:Sch}
 \end{figure}

There are two different application scenarios: resource constrained and fast execution time. For resource constrained architectures, reduce area and power has more priority than execution time parameter. The fast execution time scenario is specifically for servers that involve key exchange and signatures. In this application scenario there will be thousands of point multiplication requests simultaneously, and hence the server should be fast enough to satisfy the requests. Therefore, an accurate scheduling of the field operations in the point multiplication algorithm could be an efficient solution for high speed and resource constrained design. Category of the scheduling algorithms for hardware implementations is shown in Fig.\ref{fig:Sch}. Future implementations could be done based on advantages of these algorithms. In \cite{Sch} there are more details about scheduling algorithms. 
 
\begin{itemize}
  \item \textbf{\textit{Analysis of mathematical topics related to elliptic curves:Finite field operations and Group law operations}}
\end{itemize}

\begin{figure}[H]
	\centering
	\includegraphics[scale=0.8]{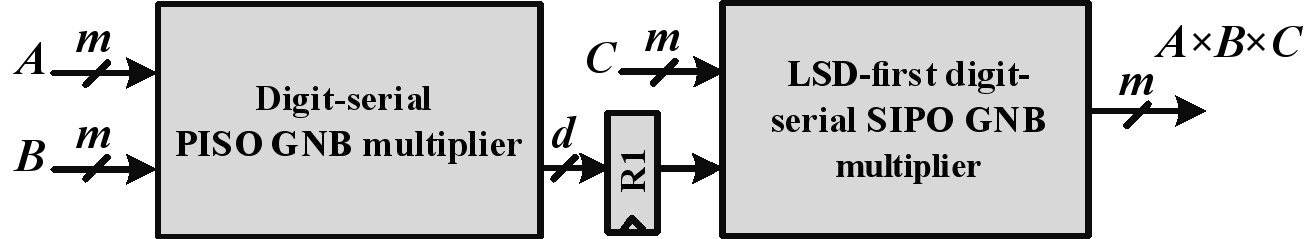}
	\caption{Structure of the hybrid-double GNB multiplier over $\F_{2^m}$.}
	\label{fig:FF4}
\end{figure}

Analysis of point addition and point doubling formulates, coordinate and optimization of related mathematics and algorithms of the field operations are one of the methods for efficient implementation of the ECC. In this case aim is reduce the number of field operation especially field multiplier. Also, new and modified hardware structures of the field multiplier in polynomial basis, normal basis and and prime field are introduced. This mathematical optimization can be useful for point addition and point doubling in differential addition coordinate in binary Edwards and generalized Hessian curves. One of the contributions in analysis of mathematical topics is the modify finite field operations for efficient implementation. For example, in \cite{HD} a hybrid-double multiplier is proposed. This structure performs double multiplications with a latency of $\lceil\dfrac {m}{d}\rceil+1$ (where $ m $ is field size and $ d $ is digit size) clock cycles assuming that one clock cycle is required to load the output of the first multiplier to the input of the second multiplier. A hybrid-double multiplier is composed based on a digit-serial parallel-input serial-output (PISO) GNB multiplier, a LSD-first digit-serial serial-input parallel-output (SIPO) multiplier and a register for loading and saving intermediate results. The structure of the hybrid-double multiplier is shown in Fig.\ref{fig:FF4}.

\begin{itemize}
  \item \textbf{\textit{Logical effort}}
\end{itemize}

The logical effort technique is a method for achieving the least delay for a given load in a logic circuit \cite{LE}. In the design of the field multiplier in the ECC structure, to balance the delay among the stages and to obtain a minimum over all delay, the logical effort technique can be applied. A hardware implementation of the logical effort in the cryptographic applications can be found in \cite{VLSIONB_B} and \cite{L_BECs}. This technique is suitable for high-speed hardware implementation of the ECC. We can design an algorithmic and automatics approach based on logical effort for compute size of transistors in the critical path delay for different loads. Also, in this case the best trade-offs between area and speed can be achieved.

\section{\textbf{Elliptic curve cryptography implementations}}
\label{sec-ECC2}
During the last decade, many papers about hardware implementation of ECC have been published in the literature. The most previous works have similar selection in the implementation, i.e., in selecting type of finite field, elliptic curve, point multiplication algorithm, and algorithm of field operations. For example, the Montgomery ladder and the Itoh-Tsujii algorithm are widely used for point multiplication and field inversion respectively. The aim of this section is review of the these implementations and architectures. In this survey, hardware structures of the ECC are categorized based on implementation technologies as follows:

\begin{enumerate}
\item Hardware implementations of the elliptic curve cryptosystems on FPGAs
\item ASIC Hardware Implementations of the elliptic curve cryptosystems
\end{enumerate}

\subsection{\textbf{Hardware implementations of the elliptic curve cryptosystems on FPGAs}}

In this subsection, we review hardware architectures for ECC on FPGA. FPGA implementation leads to faster architectures which have more parallelism in performing field operations. The main parts of cryptographic applications are increasingly implemented in FPGA platforms according to the recent advancements in these applications \cite{AES}. In particular, parallel and pipelined architectures and also low-power and low-cost designs are implemented on FPGAs, such that can operate at very high data rates. Therefore, these properties and the reconfigurability of the FPGAs make them one of the best devices for high-performance and low-power reconfigurable implementation. To implement cryptographic algorithms in hardware without losing flexibility the FPGA platforms are the best choose. Reconfigurability advantage of FPGAs, allows designing ECC implementations that are optimized for specific parameters, because support for other parameters such as different fields, algorithm and coordinate choices can be achieved through reconfiguration. Here, the FPGA-based implementations are categorized based on type of finite fields into three groups. The first and second groups are implementations on binary fields and prime fields respectively. Also the third group is the FPGA-based implementations with dual-field property. The elliptic curve point multiplication with field operations over dual-fields, required for the ECC schemes such as signature, authentication and key exchange.

\subsubsection{FPGA implementations of the point multiplication on binary fields}

In this subsection, FPGA implementations of the point multiplication on binary fields are presented. These works are consist of \cite{K10}, \cite{W18}, \cite{GLS1}-\cite{K15} and \cite{E1}-\cite{E6}. Two works \cite{GLS1}-\cite{GLS2} are based on GLS curves. To better comparison, the implementations are categorized based on type of the curve. The efficient and popular curves are include binary Weierstrass curves, Koblitz curves, binary Edwards curves, generalized Hessian curves and binary Huff curves.

\begin{itemize}
  \item \textbf{\textit{FPGA implementations of the binary Weierstrass curves:}}
\end{itemize}

FPGA-based implementations of the point multiplication on binary Weierstrass curves are presented in \cite{K10}, \cite{W18} and \cite{W1}-\cite{W33}. The most of these implementations are implemented by using polynomial basis representation. The main special techniques, which are used in this category of implementations are summarized as follows:   

\begin{enumerate}

\item In \cite{K10} \textbf{\textit{tools for evaluating the use of parallelism}} and shows where it should be used in order to maximize efficiency are provided.

\item In \cite{W18} a \textbf{\textit{clock switch circuit}} is used to manage the clock signal so that the circuit operates at its maximum clock frequency at different steps of the Montgomery ladder algorithm.

\item In \cite{W1} a parallel version of the \textbf{\textit{half-and-add method}} using the mixed-coordinate representation for PA, PD and point halving are implemented.

\item In \cite{W3} a \textbf{\textit{parameterized generator}}, which can produce field multipliers with different speed and area trade-offs. The curve operations to be incorporated into the processor based on microcoded control unit. 

\item In \cite{A21} three finite field RISC cores and a \textbf{\textit{main controller to achieve instruction-level parallelism (ILP)}} for elliptic curve point multiplication based on the analysis of both data dependency and critical path is proposed.

\end{enumerate}

\begin{figure}[H]
	\centering
	%	\captionsetup{justification=centering}
	\includegraphics[scale=0.45]{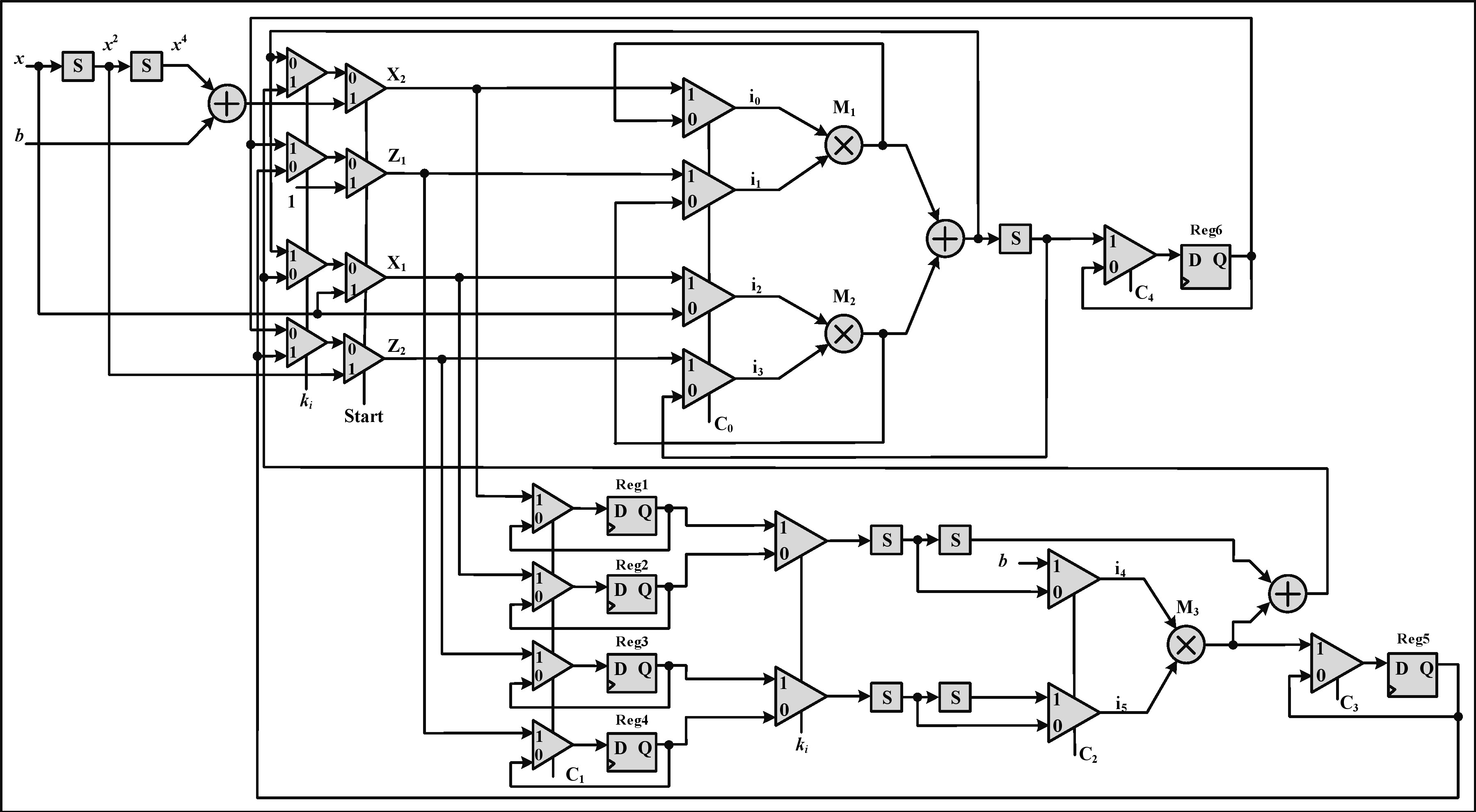}
	\caption{Structure for implementation of loop iterations in the Montgomery ladder algorithm proposed in \cite{W18}.}
	\label{fig:W3_1}
\end{figure}

In following, we present recent works in this category in more details. In \cite{W18} a hardware structure of the point multiplication based on Montgomery ladder algorithm for binary Weierstrass curves is presented. In this work, the PA and PD are performed concurrently in parallel by three pipelined digit-serial multipliers in polynomial basis. The field multiplier is based on a parallel and independent computation of multiplication by power of the variable polynomial. An efficient architecture of the Itoh-Tsujii inversion algorithm is implemented for field inversion more details of this architecture is presented in \cite{Inv_B}. A clock switch circuit is used to manage the clock signal so that the circuit operates at its maximum clock frequency at different steps of the Montgomery ladder algorithm. The proposed structure for implementation of the Montgomery ladder loop iterations in \cite{W18} is shown in Fig.\ref{fig:W3_1}. As seen in this figure the point addition and point doubling are computed independently in parallel by three multipliers.

As know in the point multiplication algorithm, computations of coordinate conversion from projective to affine is start at the end of loop iterations. In \cite{W18} the critical path delay of the proposed circuit in loop iterations mode is less than that of in coordinate conversion mode. Therefore, to increase the speed processing, multi-frequency clock technique is used. In this way, the structure can switch between two different fast and slow clock frequencies, that determined by the different critical path delays of loop iterations part and coordinate conversion part. The structure of the clock switch circuit and its performance is explained in \cite{W18}.

In \cite{W14} two high-speed ECC implementations for point multiplication is proposed. A pipelined full-precision field multiplier is used to reduce the latency, and the Lopez-Dahab Montgomery ladder algorithm is modified for accurate scheduling to avoid data dependency. It the first proposed high-performance architecture includes a 2-stage pipelined full-precision $m$ bit field multiplier, one field squaring, one quad-squaring, and two field addition units in order to perform point operations within 6 clock cycles. To performing operations in 6 clock cycles, squarer block or quad-square block or both blocks in parallel along with the multiplication is applied. In this structure one of the adders is placed in the common data path. The second adder is used to add the two outputs of the multiplier. In addition, the circuit can save some intermediate results of the operations in the registers (accumulator) to avoid loading/unloading to the main memory.

The second ECC implementation for point multiplication in \cite{W14} is based on three full-precision field multipliers called $Mul1$, $Mul2$ and $Mul3$ to achieve the lowest latency high-speed ECC. The one field multiplication is pipelined by one stage so output is ready in one clock cycle. Also, the field square and field adder are performed in the same clock cycle based on combinational logic. The field operation circuits are cascaded, therefore different operations in the same clock cycle can achieve by tapping the results.

\begin{figure}[H]
	\centering
	\includegraphics[scale=0.65]{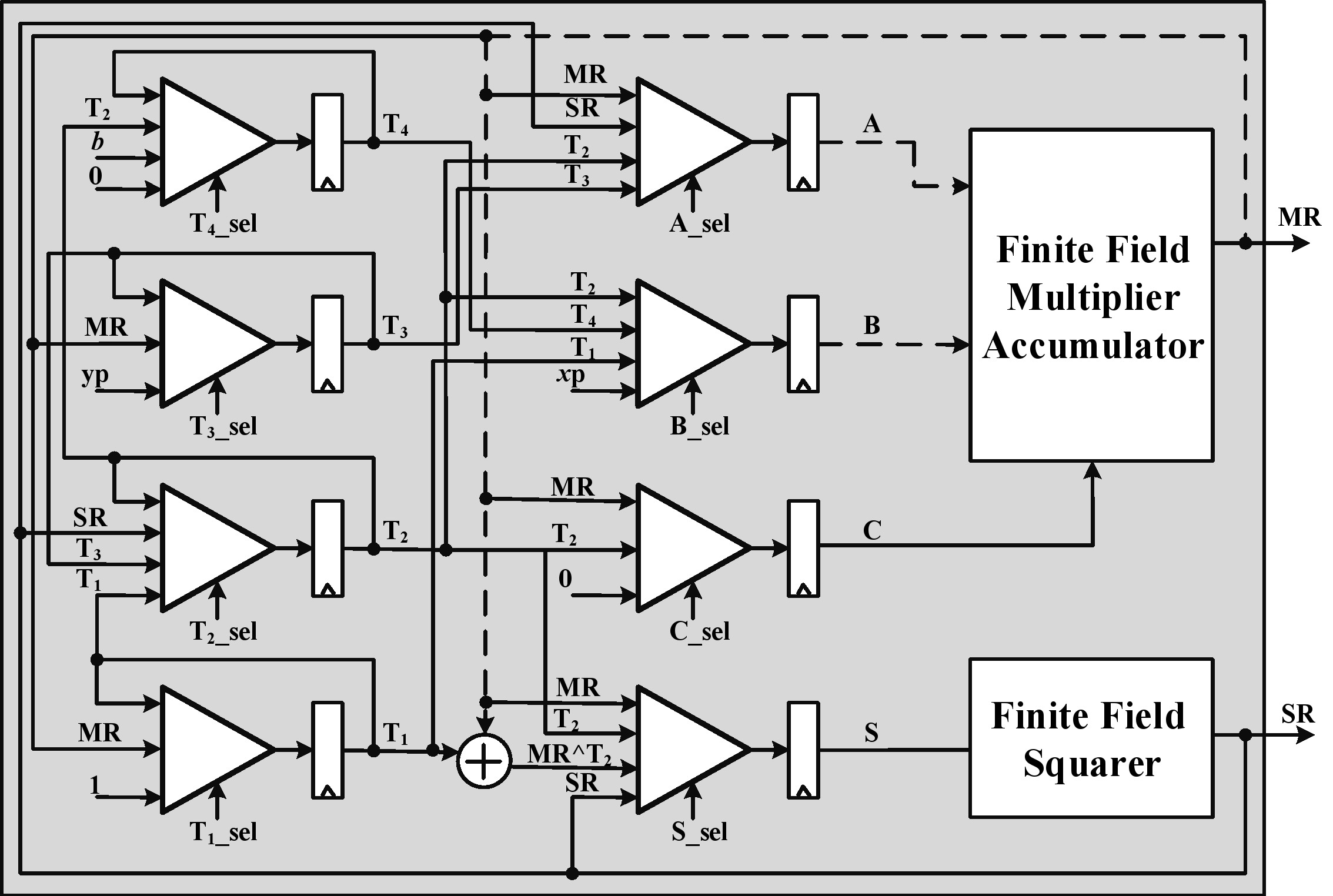}
	\caption{Data-path of elliptic curve scalar multiplication using a three stages pipelined MAC in \cite{W16}.}
	\label{fig:W3}
\end{figure}

In \cite{W16} an efficient pipelined architecture of the point multiplication over $\F_{2^{m}}$ is proposed. The architecture uses a multiplier accumulator (MAC) by bit-parallel field multiplier based on the Karatsuba-Ofman algorithm. In this work, for better sharing of execution paths the Montgomery ladder algorithm is modified. The data path in the architecture is well designed, so that the critical path contains few extra logic primitives apart from the MAC. To find the optimal number of pipeline stages, placement of pipeline registers is analyzed. Therefore, scheduling schemes with different pipeline stages are proposed. The data path of implemented ECC using a three stages pipelined MAC is shown in Fig.\ref{fig:W3}. The proposed architecture consists of one bit-parallel MAC, one field squarer, a register file, a finite state machine (FSM) and a 6$\times$18 control ROM. The inputs to field squarer and MAC are all registered. For data caching, 4 registers T$_1$ to T$_4$ are used in the data path. A multiplexer is before each register. The control signals T1\_sel, T2\_sel, T3\_sel and T4\_sel are given at each clock cycle to select different operations in the point multiplication implementation. Therefore, the input delay for registers is only the delay of a 4 to 1 multiplexer. In the Fig.\ref{fig:W3}, the critical path of the 3-stage pipelined architecture is shown by the bold dashed line. The critical path is consists of a pipelined MAC, a field adder and one 4 to 1 multiplexer.

Other recent work is presented in \cite{W21}. In this work, a theoretical model to approximate the delay of different field operations used in a point multiplication structure is implemented on $k$ input lookup table (LUT) based FPGAs. Also a suitable scheduling for performing PA and PD in a pipelined data path of the point multiplication is implemented. 

The point multiplication architecture presented in \cite{W21} uses the left to right double-and-add algorithm with binary signed digit representation. The used coordinate in the processor is Lopez-Dahab projective coordinate. The inputs of the arithmetic unit are provided by the register bank, at each clock cycle, through six buses. At the end of the clock cycle, the results of the computation are stored in the registers through buses. Control signals are generated at every clock based on the state of the FSM and key digit.

In \cite{W27} a high-speed elliptic curve point multiplication using FPGA is presented. To find out an optimal digit size different levels of digit-serial computation are applied to the data-path of field multipliers and dividers. Results for the five NIST recommended curves are provided in \cite{W27}. 

The point multiplication architecture presented in \cite{W27} is constructed based on three field multipliers, three field squarers, nine field adders and one field divider. In this structure, for increase speed processing the multipliers are parallel. The four output registers are used for storing of the output parameters in the point multiplication algorithm, in addition, they are employed for loading of initial values in the start of the algorithm. 
Table \ref{table:T1} shows the results of the FPGA implementations of the point multiplication on binary Weierstrass curves.

\begin{table}[H]
	\centering
	\captionsetup{justification=centering}
{\scriptsize	
\caption{Results of the FPGA implementations of the point multiplication on binary Weierstrass curves.}
	\centering 
	%\vspace*{-1}
	\label{table:T1}
	\begin{tabular}{|c|c|c|c|c|c|c|}
		\hline
		\bf Works(Year) & \bf Field & \bf Device & \bf Area & \bf F$_{max}$(MHz) & \bf Time($\mu s$) \\
		\hline
		\cite{W1}, PB, (2005) & 163 & VE (XCV3200) & 11616 Slices & 41 & 25 \\
		\hline		
		\cite{W2}, PB, (2015) & 233 & Kintex-7 (XC7K325T) & 3016 Slices & 255.66 & 2660 \\		
		\hline
		\cite{W2}, PB, (2015) & 283 & Kintex-7 (XC7K325T) & 4625 Slices & 251.98 & 5540 \\		
		\hline		
		\cite{W4}, PB, (2004) & 191 & V2 (XCV2600E) & 17630 Slices & --- & 63 \\		
		\hline			
		\cite{W6}, PB, (2008) & 163 & Spartan-3 (XC3S2000) & 10379 Slices & 44 & 325 \\		
		\hline	
		\cite{W6}, PB, (2008) & 163 & V4 (XC4VSX35) & 10488 Slices & 99 & 144 \\		
		\hline		
		\cite{W8}, $D=16$, PB, (2000) & 167 & VE (XCV400E) &  \shortstack{3002 LUTs +  \\ 1769 FFs + 10 BRAMs}  & 76.7 & 210 \\		
		\hline			
		\cite{W10}, PB, (2009) & 163 & V2 (XC2V6000) & 15527 LUTs + 3994 FFs & 98.3 & 31.17 \\		
		\hline		
		\cite{W11}, PB, (2007) & 163 & V4 (XC4VFX100) & 3568 Slices & 253 & 9 \\		
		\hline			
		\cite{W11}, PB, (2007) & 283 & V4 (XC4VFX100) & 6128 Slices & 157 & 23 \\		
		\hline	
		\cite{W12}, PB, (2009) & 163 & V2 & 1842 Slices & 234.9 & 852.5 \\		
		\hline		
		\cite{W14}, Design 1, PB, (2016) & 163 & V5 (XC5VLX50) & 4393 Slices & 228 & 4.91 \\		
		\hline		
		\cite{W14}, Design 2, PB, (2016) & 163 & V5 (XC5VLX110) & 11777 Slices & 113 & 3.99 \\		
		\hline	
		\cite{W14}, Design 1, PB, (2016) & 163 & V7 (XC7V330T) & 4150 Slices & 352 & 3.18 \\		
		\hline		
		\cite{W14}, Design 2, PB, (2016) & 163 & V7 (XC7V690T) & 11657 Slices & 159 & 2.83 \\		
		\hline		
		\cite{W15}, PB, (2014) & 163 & V4 (XC4VLX200) & 10417 Slices & 121 & 9 \\		
		\hline
		\cite{W16}, PB, (2016) & 163 & V4 (XC4VLX200) & 7354 Slices & 222 & 6.1 \\		
		\hline
		\cite{W16}, PB, (2016) & 163 & V5 (XC5VLX110) & 3041 Slices & 294 & 4.6 \\		
		\hline		
		\cite{W16}, PB, (2016) & 233 & V4 (XC4VLX200) & 11708 Slices & 194 & 9.9 \\		
		\hline
		\cite{W16}, PB, (2016) & 233 & V5 (XC5VLX110) & 4762 Slices & 244 & 7.9 \\		
		\hline			
		\cite{W16}, PB, (2016) & 283 & V4 (XC4VLX200) & 15169 Slices & 179 & 13 \\		
		\hline
		\cite{W16}, PB, (2016) & 283 & V5 (XC5VLX110) & 6286 Slices & 213 & 10.9 \\		
		\hline			
		\cite{W17}, $D=55$, PB, (2013) & 163 & V4 (XC4VLX200) & 17929 Slices & 250 & 9.6 \\		
		\hline					
		\cite{W18}, $D=41$, PB, (2016) & 163 & V4 (XC4VLX100) & 17144 Slices & 280.348 & 6.2 \\		
		\hline
		\cite{W18}, $D=41$, PB, (2016) & 163 & V5 (XC5VLX110) & 5768 Slices & 343.300 & 5.08 \\		
		\hline		
		\cite{W18}, $D=41$, PB, (2016) & 163 & V7 (XC7VX485T) & 5575 Slices & 437.062 & 3.97 \\		
		\hline
		\cite{W18}, $D=58$, PB, (2016) & 233 & V4 (XC4VLX100) & 30141 Slices & 312.305 & 7.84 \\		
		\hline			
		\cite{W18}, $D=58$, PB, (2016) & 233 & V5 (XC5VLX110) & 10601 Slices & 359.661 & 6.84 \\		
		\hline
		\cite{W18}, $D=58$, PB, (2016) & 233 & V7 (XC7VX485T) & 10528 Slices & 496.672 & 4.913 \\		
		\hline
		\cite{W21}, PB, (2013) & 163 & V4 (XC4VLX200) & 8095 Slices & 131 & 10.7 \\		
		\hline
		\cite{W22}, PB, (2004) & 191 & VE (XCV3200E) & 18314 Slices & 9.99 & 56 \\		
		\hline
		\cite{W23}, PB, (2004) & 167 & VE (XCV400E) & 4245 LUTs + 1393 FFs & 82.3 & 1300 \\		
		\hline	
		\cite{W24}, PB, (2009) & 191 & V2 (XC2V6000) & 25963 Slices & 30.1 & 72.939 \\		
		\hline
		\cite{W26}, $D=41$, PB, (2008) & 163 & V2 (XC2V2000) & 4192 Slices & 128 & 41 \\		
		\hline					
		\cite{W27}, $D=41$, PB, (2013) & 163 & V5 (XC5VLX110) & 6150 Slices & 250 & 5.48 \\		
		\hline	
		\cite{W27}, $D=39$, PB, (2013) & 233 & V5 (XC5VLX110) & 6487 Slices & 192.3 & 19.89 \\		
		\hline					
		\cite{W28}, PB, (2008) & 163 & V4 (XC4VLX200) & 16209 Slices & 153.9 & 19.55 \\			
		\hline			
		\cite{W29}, GNB, (2008) & 163 & V4 (XC4VLX80) & 24363 Slices & 143 & 10 \\		
		\hline			
        \cite{A21}, GNB, (2010) & 163 & V4 (XC4VLX80) & 20807 Slices & 185 & 7.7 \\		
		\hline   			
	    \cite{W32}, PB, (2015) & 163 & V4 (XC4VLX25) & 3536 Slices & 290 & 14.39 \\		
		\hline
	    \cite{W32}, PB, (2015) & 163 & V5 (XC5VLX50) & 1089 Slices & 296 & 14.06 \\		
		\hline
	    \cite{W32}, PB, (2015) & 163 & V7 (XC7VX550T) & 1476 Slices & 397 & 10.51 \\		
		\hline											
	    \cite{W33}, PB, (2012) & 163 & V4 (XC4VLX80) & 8070 Slices & 147 & 9.7 \\		
		\hline	
	    \cite{W33}, PB, (2012) & 163 & V5 (XC5VLX85T) & 3446 Slices & 167 & 8.6 \\
		\hline	    	
	    \cite{W33}, PB, (2012) & 233 & V4 (XC4VLX100) & 13620 Slices & 154 & 12.5 \\		
		\hline	
	    \cite{W33}, PB, (2012) & 233 & V5 (XC5VLX85T) & 5644 Slices & 156 & 12.3 \\		    	    	
		\hline						
		\end{tabular}\\
		~\\ $D$: Digit size; PB: Polynomial basis; GNB: Gaussian normal basis; S-II:Stratix II; S-V:Stratix V; S-IV:Stratix IV; V4:Virtex-4; V5:Virtex-5; V7:Virtex-7.
}
\end{table}

\begin{figure}[H]
	\centering
	\includegraphics[scale=0.45]{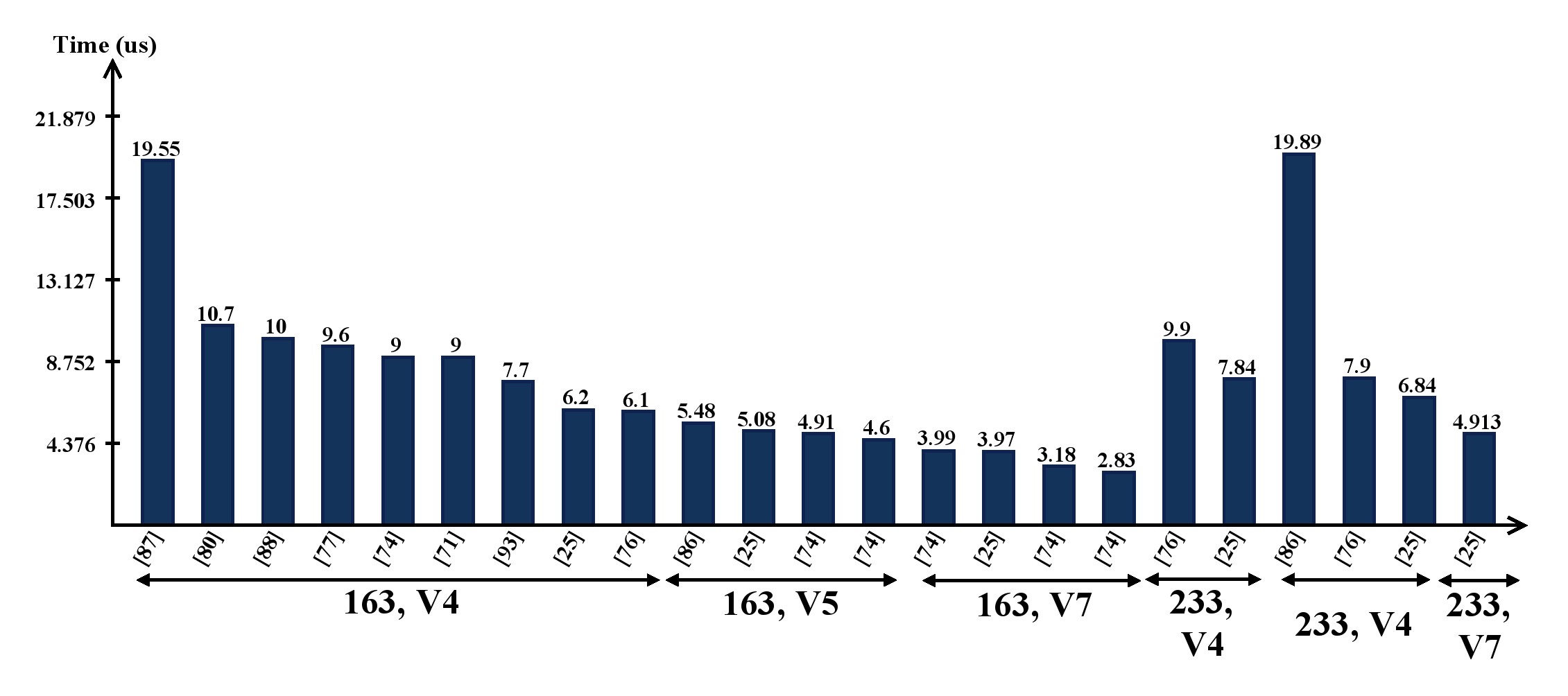}
	\caption{Graphical representation of the execution time in the some previous FPGA-based works for binary Weierstrass curves.}
	\label{fig:W4}
\end{figure}

\begin{figure}[H]
	\centering
	\includegraphics[scale=0.45]{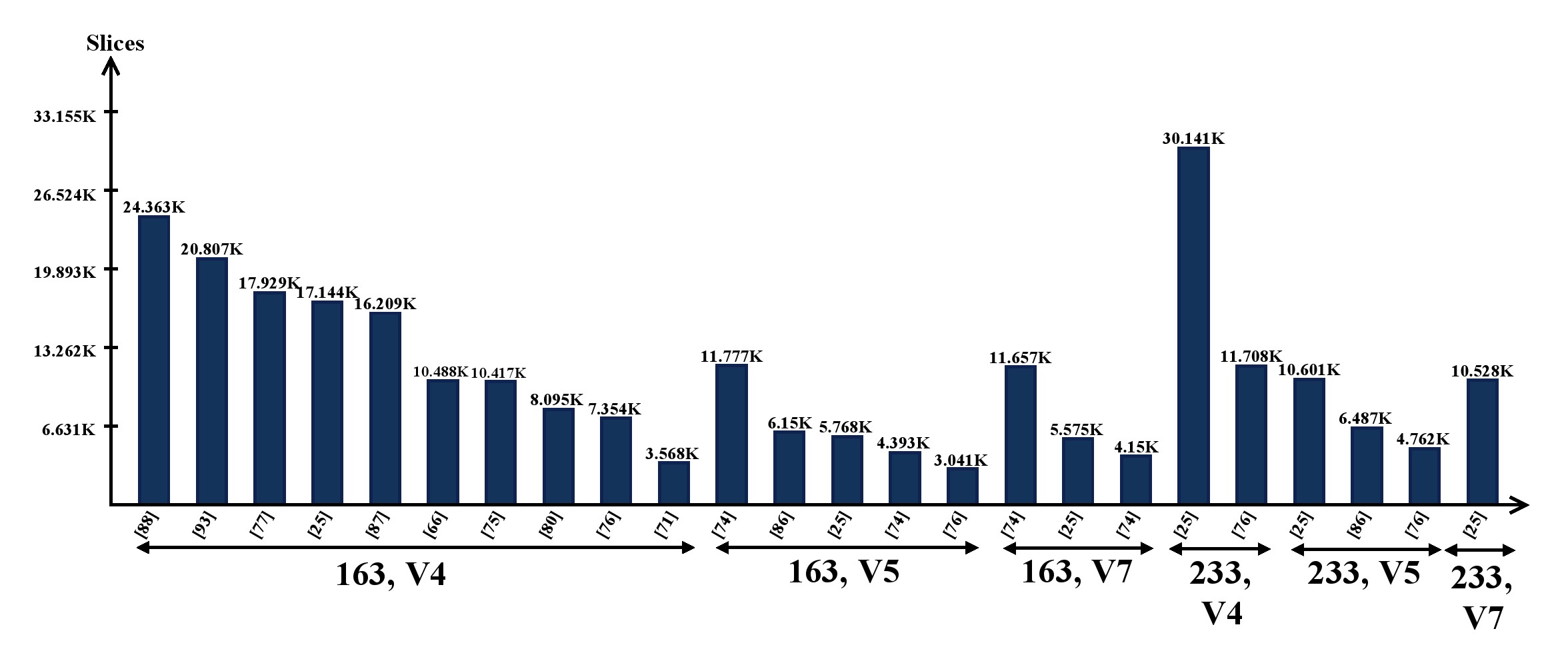}
	\caption{Graphical representation of the number of Slices in the some previous FPGA-based works for binary Weierstrass curves.}
	\label{fig:W5}
\end{figure}

The recent works \cite{W14}, \cite{W16}, \cite{W18} and \cite{W27} are the best time efficient implementations of the point multiplication on binary Weierstrass curves. High-throughput design presented in \cite{W32} is the best reported work in terms of area$\times$time metric. Work presented in \cite{W14} over $\F_{2^{163}}$ on Virtex-7 achieves a better metric value. Also it is the fastest FPGA design to date on Virtex-7. Execution time in work \cite{W18} over $\F_{2^{233}}$ is 4.913$\mu s$, which is outperforms compared to other works. For Virtex-4 over $\F_{2^{163}}$, the previous highest speed 3-stage pipelined implementation is presented in \cite{W16} and consumed 7354 slices to achieve 6.1$\mu s$. For Virtex-4 over $\F_{2^{233}}$, the highest speed work is \cite{W18} with 7.84$\mu s$. For Virtex-5, the best reported performance result over $\F_{2^{163}}$ and $\F_{2^{233}}$ are 4.6$\mu s$ and 6.84$\mu s$ presented in \cite{W16} and \cite{W18} respectively. Point multiplication implemented in \cite{W14} consumes only 4393 slices to compute a point multiplication in 4.91$\mu s$, which is 10\% and 29\% better in both speed and area than that recent work \cite{W27}. The HPECC architecture in \cite{W14} over $\F_{2^{571}}$ is the first reported implementation based on full-precision multiplier and sets a new time record equal 37.5$\mu s$ for the point multiplication on Virtex-7.\\
Fig.\ref{fig:W4} and Fig.\ref{fig:W5} show a graphical representation of the execution time and the number of Slices for the previous structures respectively. To have a better comparison, in these graphs the results are distinguished based on field size and type of FPGA platform. As seen in Fig.\ref{fig:W4} and Fig.\ref{fig:W5} the works \cite{W14}, \cite{W16}, \cite{W18} and \cite{W27} have acceptable performance in terms of speed and hardware resources.

\begin{itemize}
  \item \textbf{\textit{FPGA implementations of the Koblitz Curves:}}
\end{itemize}

The first FPGA-based implementation of the point multiplication for NIST Koblitz curve K-163 is presented in \cite{A10} with computation time of 45.6$ms$ on Altera Flex 10K FPGA. Other previous FPGA-based implementations of the Koblitz curves are presented in \cite{K1}-\cite{K15}. The main special contributions in this category of implementations are summarized as follows:   

\begin{enumerate}

\item In \cite{K2} a \textbf{\textit{parallelization method utilizing point operation interleaving}} is provided.

\item In \cite{K6} algorithms for point multiplication on Koblitz curves using \textbf{\textit{multiple-base expansions of the form $k=\sum_{}^{} \pm \tau^a(\tau-1)^b$ and $k=\sum_{}^{} \pm \tau^a(\tau-1)^b(\tau^2-\tau-1)^c$}} are described. Also, the first rigorously-proven sub-linear point multiplication using complex bases is presented.

\item In \cite{K13} presents \textbf{\textit{parallelization of scalable point multiplication}} that can support all 5 NIST Koblitz curves without reconfiguring structure.

\end{enumerate}

In following these works are discussed. The focus of the works are reducing the computation time of the point multiplication and flexibility for support all five NIST Koblitz curves. With the increase in hardware resources in recent FPGAs, designers are enabled for maximum parallelism of the several field operations in the ECC hardware implementation. For example, in \cite{K2} discuss implementation of the point multiplication on Koblitz curves with parallel field multipliers. In this work, a novel parallelization method by using interleaving of point operation is presented. The effects of field basis selection is studied in \cite{K2} and conclude that polynomial basis has faster results than normal basis.

In \cite{K3} a scalable ECC processor is presented. This ECC processor supports all five NIST Koblitz curves K-163, K-233, K-283, K-491 and K-571 without the need to reconfigure the FPGA. A finite field arithmetic unit (FFAU) that reduces the number of clock cycles is proposed. Also an improved PA algorithm to take advantage of the FFAU structure is presented. The structure computes the point multiplication after the $\tau$NAF(\textit{k}) computation, therefore it gets as inputs the point $P$ with two affine coordinates $x_1$ and $y_1$ and $\tau$NAF converted value of \textit{k}. The outputs of the scalable structure are the two affine coordinates, $x_3$ and $y_3$, of the output point $Q=kP$.

\begin{figure}[H]
	\centering
	\includegraphics[scale=0.7]{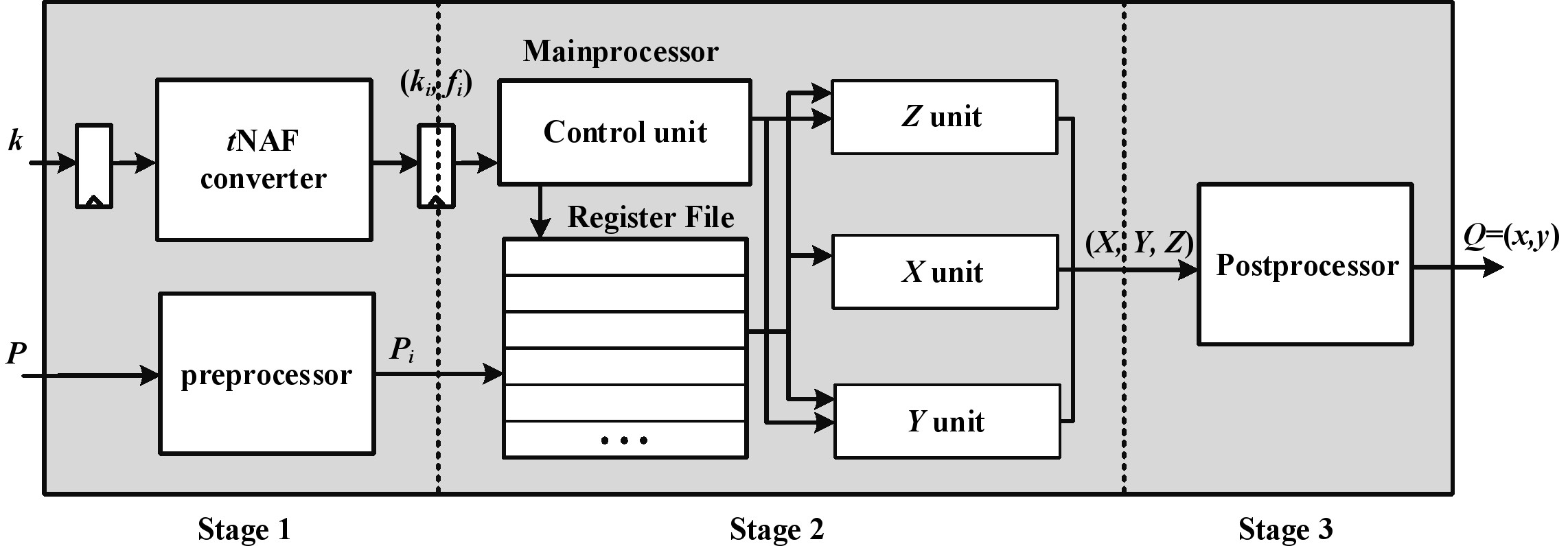}
	\caption{Structure of the ECC processor in \cite{K4}.}
	\label{fig:K4}
\end{figure}

A very high-speed FPGA-based ECC for Koblitz curves is described in \cite{K4}. It is based on a preliminary version that was presented in \cite{K1}. The implementation is optimized for both increased the performance and decreased the hardware resources for Koblitz curve K-163. In more details, the structure in \cite{K4} consists of four main components. The top level view of the structure is given in Fig.\ref{fig:K4}. The converter, converts the integer $k$ into width-4 $\tau$NAF and encodes it. The preprocessor computes the precomputed points, $P_1,... ,P_N$ for algorithm right to left point multiplication algorithm on Koblitz curves with precomputations \cite{K4}. These precomputations can be performed in parallel in the preprocessor concurrently with the converter. The loop iterations of the point multiplication algorithm is performed in the main processor after two previous computations. Finally, the result point $Q=(x,y)$ is maps from Lopez-Dahab coordinate to affine coordinate by the postprocessor.

The design of the ECC processors using two field multipliers over $\F_{2^{163}}$ with digit-serial processing is presented in \cite{K8}. The field operations are implemented using GNB representation over binary field. Also the point multiplication is computed using window-$\tau$NAF algorithm with $w$=2, 4, 8 and 16. In \cite{K12} a highly parallel structure to speed up the point multiplication for high-speed FPGA implementation on Koblitz curves is presented. The PA formulas are modified in order to employ 4 parallel field multipliers in the data-flow. Therefore, the number of the clock cycles of performing PA is reduced and speed of the point multiplication is increased.

In \cite{K14} the scalar conversion process in \cite{Brumley} is improved based on division by $\tau^2$. Two levels of optimizations are applied in the scalar conversion structure. First, the number of long integer subtractions during the scalar conversion is reduced. This optimization reduces the computation complexity and also simplifies the critical paths in the conversion structure. Then the architecture is pipelined. 

\begin{figure}[H]
	\centering
	\includegraphics[scale=0.35]{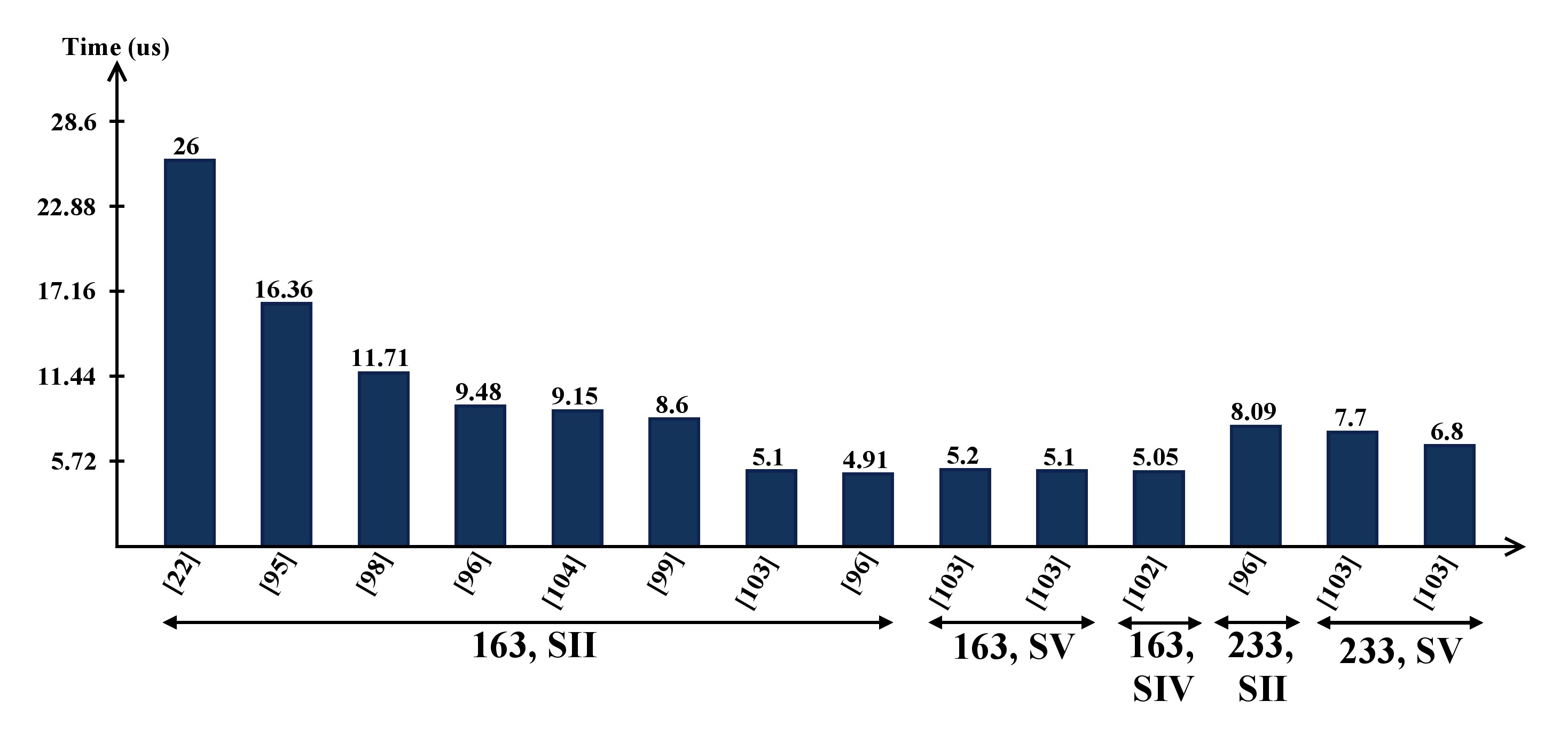}
	\caption{Graphical representation of the execution time for point multiplication on Koblitz curves.}
	\label{fig:K_1}
\end{figure}

\begin{figure}[H]
	\centering
	\includegraphics[scale=0.35]{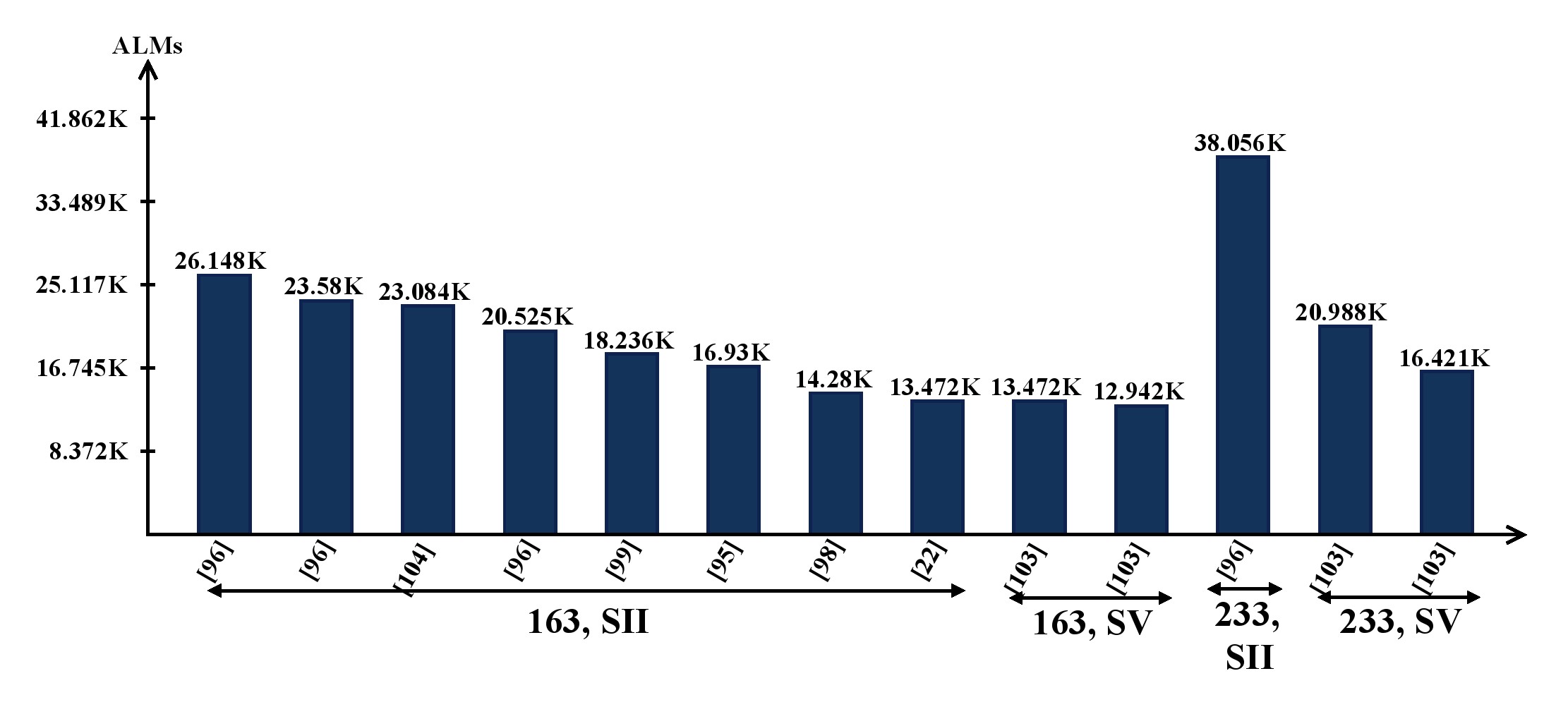}
	\caption{Graphical representation of the number of Slices for point multiplication on Koblitz curves.}
	\label{fig:K_2}
\end{figure}

Implementation results of the FPGA-based point multiplication on Koblitz curves are presented in Table \ref{table:K1}. The fastest implementations on K-163 in table compute point multiplication in 4.9l$\mu s$ \cite{K2} ($D$=55), 5.05$\mu s$ \cite{K8}, 5.1$\mu s$ \cite{K2} ($D$=41), 5.1$\mu s$ \cite{K11} ($D$=55) and 5.2$\mu s$ \cite{K11} ($D$=41). Also on K-233 computation times are 6.8$\mu s$ ($D$=78) and 7.7$\mu s$ ($D$=59) in \cite{K11}. The fastest implementation \cite{K2} require large amounts of ALMs on Stratix-II FPGA. The work \cite{K3} has the number 2431 Slices which is the lowest area in among of the K-163 curves. Fig.\ref{fig:K_1} and Fig.\ref{fig:K_2} show  graphical representation of the execution time and the number of Slices, respectively, based on type of FPGA and field size for different structures on Koblitz curves. In these figures implementation conditions for presented works are equal.

\begin{table}[H]
	\centering
	\captionsetup{justification=centering}
{\scriptsize	
\caption{Results of the FPGA-based point multiplication on Koblitz curves.}
	\centering 
	%\vspace*{-1}
	\label{table:K1}
	\begin{tabular}{|c|c|c|c|c|c|c|}
		\hline
		\bf Works(Year) & \bf Field & \bf Device & \bf Area & \bf F$_{max}$(MHz) & \bf Time($\mu s$)\\
		\hline
		\cite{K1}, PB, (2008) & 163 & SII (EP2S180F1020C3) & 16930 ALMs+21 M4Ks & 185 & 16.36 \\
		\hline
		\cite{K2}, $D$=41, PB, (2009) & 163 & SII (EP2S180F1020C3) & 20525 ALMs & 203.87 & 5.1 \\		
		\hline
		\cite{K2}, $D$=55, PB, (2009) & 163 & SII (EP2S180F1020C3) & 26148 ALMs & 187.48 & 4.91 \\		
		\hline
		\cite{K2}, $D$=59, NB, (2009) & 163 & SII (EP2S180F1020C3) & 23580 ALMs & 162.42 & 9.48 \\		
		\hline			
		\cite{K2}, $D$=59, PB, (2009) & 233 & SII (EP2S180F1020C3) & 38056 ALMs & 181.06 & 8.09  \\		
		\hline	
		\cite{K3}, PB, (2013) & 163 & V4 (XC4VFX12) & 2431 Slices & 155.376 & 273 \\
		\hline								
		\cite{K3}, PB, (2013) & 233 & V4 (XC4VFX12) & 2431 Slices & 155.376 & 604 \\
		\hline
		\cite{K4}, PB, (2011) & 163 & SII (EP2S180F1020C3) & 14280 ALMs+25 M4Ks & --- & 11.71 \\
		\hline
		\cite{K5}, $D$=41, GNB, (2016) & 163 & SII (EP2S180F1020C3) & 18236 ALMs & 187.9 & 8.6 \\
		\hline		
		\cite{K6}, GNB, (2006) & 163 & V2 (XC2V2000-6) & 6494 Slices+6 BRAMs & 128 & ---\\
		\hline			
		\cite{K8}, GNB, (2014) & 163 & SIV (EP4SGX180HF35C2) & 24270 ALUTs & 177.1 & 5.05 \\
		\hline			
		\cite{K10}, PB, (2008) & 163 & SII (EP2S180F1020C3) & 13472 ALMs & 155.5 & 26 \\
		\hline					
		\cite{K11}, $D$=41, GNB, (2015) & 163 & SV (5SGXMA3E2H29C2) & 12942 ALMs & 259.2 & 5.2 \\
		\hline
		\cite{K11}, $D$=55, GNB, (2015) & 163 & SV (5SGXMA3E2H29C2) & 13472 ALMs & 234.5 & 5.1 \\
		\hline		
		\cite{K11}, $D$=59, GNB, (2015) & 233 & SV (5SGXMA3E2H29C2) & 20988 ALMs & 245.7 & 7.7 \\
		\hline		
		\cite{K11}, $D$=78, GNB, (2015) & 233 & SV (5SGXMA3E2H29C2) & 16421 ALMs & 246.1 & 6.8 \\
		\hline		
		\cite{K12}, $D$=41, GNB, (2013) & 163 & SII (EP2S180F1020C3) & 23084 ALMs & 188.71 & 9.15 \\
		\hline		
		\cite{K13}, PB, (2016) & 163 & V5 (XC5LX110T) & 2708 Slices+5 BRAMs & 222.67 & 55 \\
		\hline
		\cite{K15}, ALU single, PB, (2008) & 233 & V2 (XC2V4000) & 14091 Slices & 51.7 & 8.72 \\
		\hline
		\cite{K15}, ALU parallel, PB, (2008) & 233 & V2 (XC2V4000) & 15916 Slices & 51.7 & 7.22\\
		\hline					
		\end{tabular}\\
		~\\ $D$: Digit Size; PB: Polynomial basis; GNB: Gaussian normal basis; SII:Stratix II; SV:Stratix V; SIV:Stratix IV; V5:Virtex-5; V7:Virtex-7.
}
\end{table}

\begin{itemize}
  \item \textbf{\textit{FPGA implementations of the binary Edwards, Generalized Hessian and Huff curves:}}
\end{itemize}

The hardware implementations of the point multiplication presented in \cite{E1}-\cite{E6} are based on binary Edwards and general Hessian curves. 
In \cite{E1} a design and implementation of the binary Edwards curves processor is explained. This work is the first FPGA-based unified processor in the literature. Furthermore, the structure is explored in terms of power analysis to make the design simple power attack preventive. In \cite{E2} parallelization in higher levels by full resource utilization of computing PA and PD formulas for both binary Edwards and general Hessian curves is performed. For computing of the point multiplication, $w$-coordinate differential formulations are used. The authors evaluate the LUT complexity and time-area tradeoffs of the processor on an FPGA by using a LUT-based pipelined and efficient digit-level GNB multiplier. 

\begin{figure}[H]
	\centering
	\includegraphics[scale=0.42]{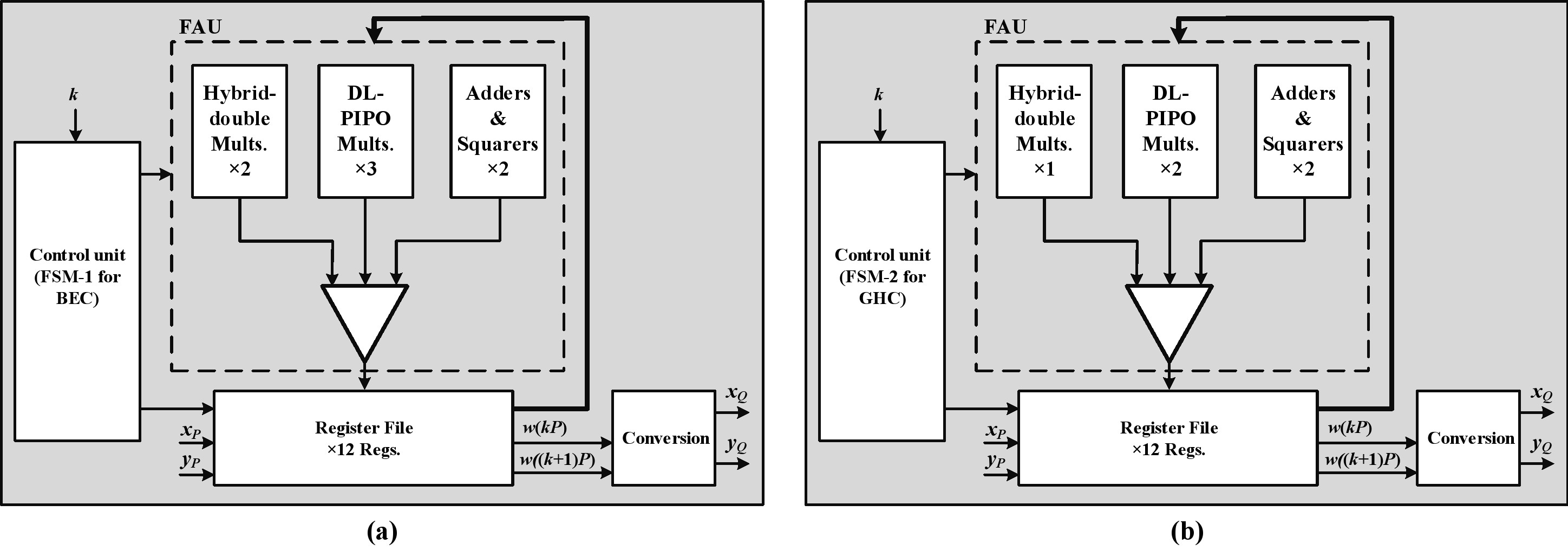}
	\caption{The architecture of crypto-processors in \cite{E3} for the point multiplication on (a) binary Edwards and (b)
	generalized Hessian curves.}
	\label{fig:E_1}
\end{figure}

To reduce the number of clock cycle in the point multiplication computation, an analysis of data-flow and maximum number of parallel field multipliers is used in \cite{E3}. Also the PA and PD formulas are modified. A digit-serial hybrid-double GNB multiplier is employed to reduce the data dependencies and the latency of the point multiplication. The architectures of the processors in \cite{E3} are shown in Fig.\ref{fig:E_1}. The point multiplication processors are composed of four main parts including field arithmetic part (FAU), register file, control part and conversion part to obtain the final affine coordinates. In FAU part of crypto-processors for binary Edwards curves, three single digit-level parallel-in parallel-out (DL-PIPO) GNB multipliers and two hybrid-double multipliers are employed. Also for Generalized Hessian curves, in FAU part two single DL-PIPO multipliers and a hybrid-double multiplier are employed. 

In \cite{E4} a high-speed binary Edwards curves point multiplication implementation is proposed based on a parallel design strategy. In this work, two field multipliers are employed and also power analysis attack resistance against a variety of attacks is provided. The hardware structure is implemented based on the parallelism layer concept. A parallelism unit (PU) realizes a single parallelism layer of the point multiplication computations. It is consists of two bit-parallel field multipliers, two field squarers and three field adders. These components operate in parallel form and output result of each component is ready with one clock cycle. Outputs of the PU components are stored in the register file (constructed by 23 registers) in order to be reused in future clock cycles. Generator irreducible polynomials of the field $\F_{2^{m}}$ are specific irreducible polynomial such as trinomial or pentanomial.

A high-speed hardware structures of the point multiplication based on Montgomery ladder algorithm for binary Edwards and generalized Hessian curves in GNB are presented in \cite{E6}. Computations of the PA and PD in the structure are simultaneously performed by pipelined digit-serial field multipliers. The field multipliers in parallel form are scheduled for reduce latency. The structure of digit-serial GNB multiplier is constructed based on regular and low-cost components of exponentiation by powers of 2 and multiplication by normal elements \cite{GNB_FPGA_B}. Therefore, the structures are area efficient and have low critical path delay. In these architectures, the point multiplication is implemented by using four and three field multipliers for $d_1 \neq d_2$. More details of the structure are presented in \cite{E6}.

\begin{figure}[H]
	\centering
	\includegraphics[scale=0.43]{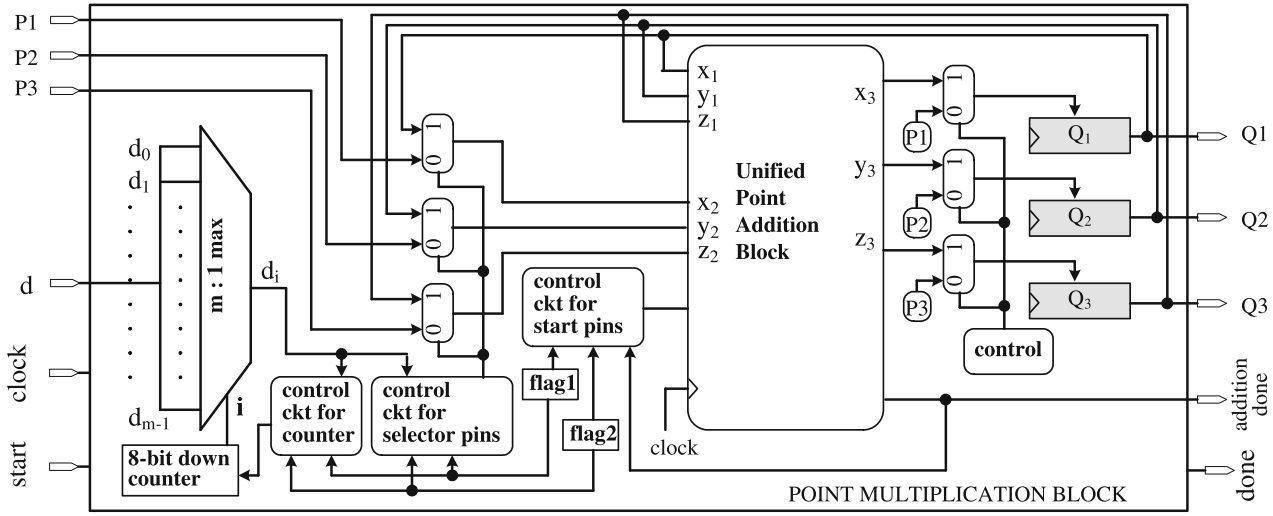}
	\caption{Architecture of Huff curve point multiplication in \cite{E8}.}
	\label{fig:E_6}
\end{figure}

In \cite{E7} the first hardware design of binary Huff curves is proposed, which also lead to unified point multiplication. To a faster circuit and better utilization of the FPGA resources, several optimized architectural features have been developed. In \cite{E8} provide an efficient hardware implementation of the unified Huff formula in projective coordinates on FPGA. Also side channel vulnerability is studied with simple power analysis. It is claimed that the formula is unified and there is not power consumption difference when computing PA and PD operations. The architecture of the point multiplication on Huff curve based on left-to-right binary algorithm is shown in Fig.\ref{fig:E_6}. The $Q$ registers are initialized with coordinates of input point $P$. There is a counter $i$ with counting range 0 to $m-2$. At each loop iteration the counter helps to select the corresponding bit of the scalar $d$. Two intermediate signals $flag_1$ and $flag_2$ are for detect the on-going point operation of either PA $(P+Q)$ or PD $(Q+Q)$. If PA operation is going on, $flag_2$ will be enabled and $flag_1$ will be disabled. Also if PD operation is going on, $flag_1$ will be enabled and $flag_2$ will be disabled. The PA and PD operations are computed using the same block implemented using unified addition formula. After the completion of one point operation, the $addition\_done$ signal will be enabled for one clock cycle during which the $Q$ registers are updated by the new intermediate result coming out from the unified point addition block. Finally, the $done$ signal will be enabled once the point multiplication is complete.

The results of the FPGA implementations of the point multiplication on BECs, GHCs and BHCs are shown in Table \ref{table:TE1}. In this table, we report Area$\times$Time metric because the FPGA platform in these work are the same.
Work \cite{E6} for BECs $(d_1\neq d_2)$ over $\F_{2^{233}}$ has better hardware consumption and execution time than that of work \cite{E1}. For Virtex-5 FPGA, BECs $(d_1\neq d_2)$ over $\F_{2^{163}}$, the execution time and maximum operation frequency of work \cite{E6} are 67\% and 25\% better than those of work in \cite{E2}, but hardware consumption in \cite{E2} are less than work \cite{E6}. Also, there are similar comparison results for BECs $(d_1=d_2)$ and GHCs over $\F_{2^{163}}$ between works \cite{E6} and \cite{E2}. For BECs $(d_1\neq d_2)$ over $\F_{2^{233}}$ and $\F_{2^{163}}$ implemented on Virtex-4 FPGA the work \cite{E6} has 40\% and 30\% less computation time compared to that of \cite{E3} respectively. Also for GHCs over $\F_{2^{163}}$ and $\F_{2^{233}}$ the work presented in \cite{E3} has 24\% and 35\% computation time more compared to that of the work \cite{E6} for the equal digit size. 

\begin{table}[H]
	\centering
	\captionsetup{justification=centering}
{\scriptsize	
\caption{Results of the FPGA implementations of the point multiplication on BECs, GHCs and BHCs.}
	\centering 
	\label{table:TE1}
	\begin{tabular}{|c|c|c|c|c|c|c|}
		\hline
		\bf Works(Year) & \bf Field & \bf Device & \bf Area & \bf F$_{max}$(MHz) & \bf Time($\mu s$) & \bf Area$\times$Time \\
		\hline
		\cite{E1}, BECs, $(d_1\neq d_2)$, PB, (2012) & 233 & V4 (XC4VLX140) & 21816 Slices & 47.384 & 190 & 4.145 \\
		\hline
		\cite{E2}, BECs, $(d_1\neq d_2)$, $D$=41, GNB, (2012) & 163 & V5 (XC5VLX110) & 5788 Slices  & 264.5  & 25.3 & 0.14644 \\		
		\hline		
		\cite{E2}, BECs, $(d_1 = d_2)$, $D$=41, GNB, (2012) & 163 & V5 (XC5VLX110) & 5788 Slices  & 264.5  & 19.8 & 0.1146 \\		
		\hline		
		\cite{E2}, GHCs, $D$=41, GNB, (2012) & 163 & V5 (XC5VLX110) & 5788 Slices  & 267.1  & 17.7 & 0.10245 \\		
		\hline		
		\cite{E3}, BECs, $(d_1 = d_2)$, $D$=33, GNB, (2014) & 163 & V4 (XC4VLX160) & 27778 Slices  & 217.2  & 17.5 & 0.48612 \\		
		\hline		
		\cite{E3}, BECs, $(d_1 = d_2)$, $D$=26, GNB, (2014) & 233 & V4 (XC4VLX160) & 29252 Slices  & 198.4  & 36.3 & 1.06185 \\		
		\hline		
		\cite{E3}, GHCs, $D$=33, GNB, (2014) & 163 & V4 (XC4VLX160) & 15992 Slices  & 218.2  & 15.9 & 0.2543 \\		
		\hline		
		\cite{E3}, GHCs, $D$=26, GNB, (2014) & 233 & V4 (XC4VLX160) & 16940 Slices  & 205.1  & 33.1 & 0.5607 \\		
		\hline				
		\cite{E4}, BECs, $(d_1 = d_2)$, PB, (2016) & 233 & V4 (XC4VFX140) & 40793 LUTs  & 67  & 49 & 1.9989 \\		
		\hline		
		\cite{E4}, BECs, $(d_1 = d_2)$, PB, (2016) & 233 & V5 (XC5VLX110) & 32874 LUTs  & 132  & 25 & 4.33937 \\		
		\hline					
		\cite{E6}, BECs, $(d_1 \neq d_2)$, $D$=41, GNB, (2016) & 163 & V4 (XC4VLX100) & 27365 Slices  & 247.396  & 10.52 & 0.28788 \\		
		\hline			
		\cite{E6}, BECs, $(d_1 = d_2)$, $D$=41, GNB, (2016) & 163 & V4 (XC4VLX100) & 20853 Slices  & 247.750  & 10.49 & 0.218748 \\		
		\hline		
		\cite{E6}, GHCs, $D$=41, GNB, (2016) & 163 & V4 (XC4VLX100) & 20752 Slices  & 247.037  & 10.54 & 0.21873 \\		
		\hline		
		\cite{E6}, BECs, $(d_1 \neq d_2)$, $D$=41, GNB, (2016) & 163 & V5 (XC5VLX110) & 11397 Slices  & 302.081  & 8.62 & 0.09824 \\		
		\hline			
		\cite{E6}, BECs, $(d_1 = d_2)$, $D$=41, GNB, (2016) & 163 & V5 (XC5VLX110) & 8645 Slices  & 302.093  & 8.6 & 0.074347 \\		
		\hline		
		\cite{E6}, GHCs, $D$=41, GNB, (2016) & 163 & V5 (XC5VLX110) & 8645 Slices  & 302.093  & 8.62 & 0.07452 \\		
		\hline			
		\cite{E6}, BECs, $(d_1 \neq d_2)$, $D$=26, GNB, (2016) & 233 & V4 (XC4VLX100) & 18278 Slices  & 333.970  & 21.6 & 0.394805 \\		
		\hline			
		\cite{E6}, BECs, $(d_1 = d_2)$, $D$=26, GNB, (2016) & 233 & V4 (XC4VLX100) & 13786 Slices  & 333.970  & 21.57 & 0.297364 \\		
		\hline		
		\cite{E6}, GHCs, $D$=26, GNB, (2016) & 233 & V4 (XC4VLX100) & 14052 Slices  & 333.970  & 21.6 & 0.30352 \\		
		\hline		
		\cite{E6}, BECs, $(d_1 \neq d_2)$, $D$=26, GNB, (2016) & 233 & V5 (XC5VLX110) & 6547 Slices  & 391.932  & 18.40 & 0.120465 \\		
		\hline			
		\cite{E6}, BECs, $(d_1 = d_2)$, $D$=26, GNB, (2016) & 233 & V5 (XC5VLX110) & 4987 Slices  & 391.932  & 18.38 & 0.09166 \\		
		\hline		
		\cite{E6}, GHCs, $D$=26, GNB, (2016) & 233 & V5 (XC5VLX110) & 5045 Slices  & 391.932  & 18.40 & 0.09283 \\		
		\hline	
		\cite{E6}, BECs, $(d_1 \neq d_2)$, $D$=59, GNB, (2016) & 233 & V5 (XC5VLX110) & 14343 Slices  & 337.603  & 11.03 & 0.1582 \\		
		\hline		
		\cite{E6}, GHCs, $D$=59, GNB, (2016) & 233 & V5 (XC5VLX110) & 8875 Slices  & 337.603  & 11.03 & 0.097891 \\		
		\hline					
		\cite{E7}, BHCs, PB, (2012) & 233 & V4 (XC4V140) & 20437 Slices  & 81  & 73 & 1.4919 \\		
		\hline	
		\cite{E8}, BHCs, PB, (2013) & 233 & V4 & 19352 Slices  & 134  & 55 & 1.06436 \\		
		\hline	
		\cite{E8}, BHCs, PB, (2013) & 233 & V6 & 7150 Slices  & 172  & 43 & 0.30745 \\		
		\hline			
		\cite{E8}, BHCs, PB, (2013) & 233 & V7 & 6032 Slices  & 183  & 40 & 0.24128 \\		
		\hline			
											
		\end{tabular}\\
		~\\ $D$: Digit Size; PB: Polynomial basis; GNB: Gaussian normal basis; V4:Virtex-4; V5:Virtex-5; V6:Virtex-6; V7:Virtex-7.
}
\end{table}

\subsubsection{FPGA implementations of the point multiplication on prime fields}

The ECC processors on prime fields utilize more hardware resources and are relatively slower than binary fields. The point multiplication implementation on prime fields can be categorized based on the modular reduction methods and modulus primes. Therefore, the prime field ECC hardware implementations can be split into three groups as follows: 

\begin{itemize}
  \item \textbf{Arbitrary prime field and curve parameters}
\end{itemize}
\begin{itemize}
  \item \textbf{Special curves or special modulus primes such as Mersenne and pseudo Mersenne prime numbers}
\end{itemize}
\begin{itemize}
  \item \textbf{Residue number systems (RNS) and redundant signed digits (RSD) based prime field ECC processors}
\end{itemize}

The special cases of implementations are efficient for the point multiplications, but for further applications such as digital signature generation cannot be applicable \cite{FP1}. For example, order of the base point in the ECDSA is not a special prime. Therefore, arbitrary prime fields are better for support the ECC such as curve transition, key agreement and signature generation which require the operations over another prime field. Modular multiplication is the most important operation in the elliptic curve point multiplication over $\F_p$. Two main methods are employed for implementation of modular multiplication. The first method is based on Montgomery method. It is widely used in implementations of arbitrary curves. The second method is multiply-then-reduce. It is used, with efficient modular reduction, for implementation of special curves over $\F_p$ where $p$ is the generalized/pseudo-Mersenne prime. The FPGA-based implementations of the point multiplication on prime fields are presented in \cite{FP1}-\cite{A43}. In this category, many of works are implemented based on the DSP blocks and embedded multipliers in FPGA. The works \cite{FP1}, \cite{FP2}, \cite{FP4}, \cite{FP9}, \cite{FP14}, \cite{FP21}-\cite{FP30}, \cite{FP32}-\cite{FP33} and \cite{FP35} utilized the inherent DSP blocks in FPGAs to optimize the area and performance. Scalable and flexible FPGA-based point multiplication implementations are proposed in \cite{FP31} and \cite{FP35} respectively. The hardware structure in these works support all five prime field elliptic curves recommended by NIST. The main special techniques for FPGA implementations of the point multiplication on prime fields are summarized as follows:  

\begin{enumerate}
\item in \cite{FP7} the \textbf{\textit{balanced ternary representation}} of the point multiplication based on multiple point tripling and point addition is presented. 
\item In \cite{FP26} a \textbf{\textit{single instruction based ultra-light ECC processor}} coupled with dedicated hard-IPs of the FPGAs is proposed.
\item In \cite{FP36} point multiplication algorithm is based on efficient \textbf{\textit{co-Z arithmetics}}, where addition of projective points share the same Z-coordinate. The algorithm is fast and secure against different attacks.
\end{enumerate}

In \cite{FP5} an application-specific instruction-set ECC processor based on redundant signed digit representation is proposed. The processor uses pipelining techniques for Karatsuba-Ofman multiplication algorithm. Also, an efficient modular adder without comparison and a high-throughput modular divider are implemented. The structure supports the NIST curve P-256. A hardware implementation of fast point multiplication using the balanced ternary representation is presented in \cite{FP7}. In this implementation, uses multiple point tripling and point addition. Here, 3P, 9P, 27P, etc. are precomputed by using fast tripling and use them in preforming of the final product over $\F_p$. This work is the first implementation of balanced ternary representation and pre-computation over $\F_p$ on FPGA platform. In \cite{FP8} a high-performance structure for the point multiplication over general prime field using Jacobian coordinates is presented. The structure is implemented based on a parallel field arithmetic unit. The field adder and subtractor are in parallel to four field multipliers. The field multiplier is optimized by radix-4 Booth encoding technique, while adder and subtractor are implemented by using available fast carry chains on FPGA. It is constructed based on the parallel arithmetic unit (PAU) unit, a register file, input/output multiplexing logic and a control unit. The PAU consists of 5 field arithmetic units. Add/Sub unit compute a single addition or subtraction operation in one clock cycle, also four multiplications can be computed in parallel by the four multipliers. The control signals are generated for execution of the respective field operation based on fetch and decode instructions. In this architecture, a division block is used for the final conversion from Jacobian to affine coordinates.

\begin{figure}[H]
	\centering
	%	\captionsetup{justification=centering}
	\includegraphics[scale=0.7]{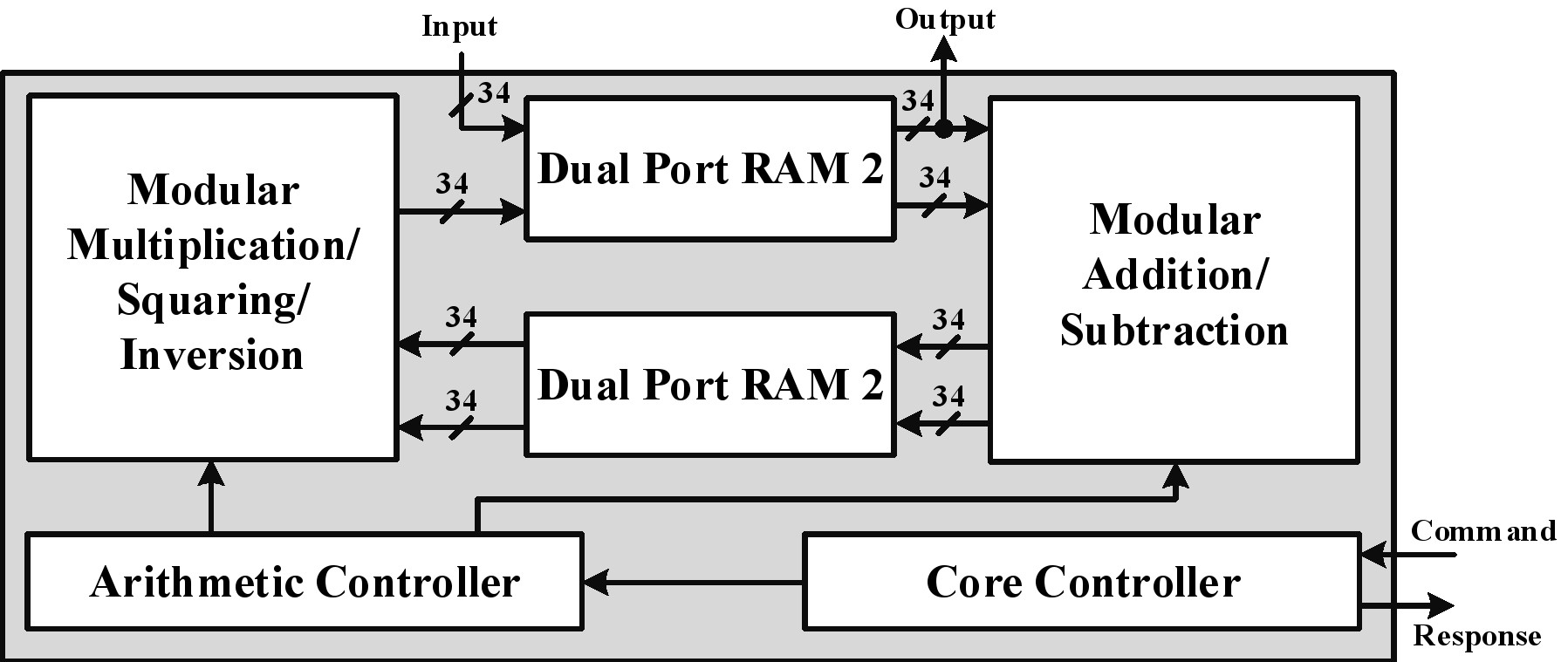}
	\caption{Architecture of the Curve25519 core in \cite{FP28}.}
	\label{fig:FP28_1}
\end{figure}

In \cite{FP25} a flexible hardware processor over five standard NIST prime fields P-192, P-224, P-256, P-384 and P-521 is proposed. The flexibility of the implementation is achieved through the software-controlled hardware programmability, which allows for different scenarios of computing atomic block sequences. A single instruction based lightweight ECC processor coupled with dedicated hard-IPs of the FPGAs is proposed in \cite{FP25}. This hardware structure is the first implementation of the point multiplication which requires less than 100 Slices on Virtex-5 and Spartan-6 FPGA. A secure and efficient implementation of a special ECC processor using the Curve25519 \cite{Curve25519} on FPGA is presented in \cite{FP28}. In the structure, the DSP blocks of FPGAs are used for field operations. Also, basic multi-core DSP-based architectures achieves a high-performance of more than 32000 point multiplications per second on a Xilinx Zynq 7020 FPGA. Architecture of the Curve25519 core in \cite{FP28} is shown in Fig.\ref{fig:FP28_1}. In this structure, two dual-ported BRAMs in butterfly configuration are used. In more details, the first BRAM only receives the results of the addition or subtraction unit and provides the input to the multiplication while the second BRAM stores the multiplication result and feeds the addition unit. Therefore, parallel operation is enabled and pipeline stalls through loading and write back can be avoided with only little overhead.

The field multiplier in the arithmetic unit is consists of 18 DSP blocks, 15 DSP blocks are used to compute partial products, one for a pre-reduction and two for the final modular reduction. Computation of partial products in the field multiplier can be interleaved with the reduction step in pipeline manner. In this work, the single core design with a dedicated inverter circuit and share it among several cores are augmented for an optimal area and performance trade-off.

In \cite{FP31} a high-performance scalable elliptic curve processor is presented (Fig.\ref{fig:FP31_1}). The double-and-add algorithm is selected using mixed affine and Jacobian coordinates for PAs and Jacobian coordinates for PDs. The processor is able to support all five NIST prime field elliptic curves. To achieve high speed and low hardware resource the structure takes advantage of the DSP48E blocks available in Virtex-5 FPGA. The parallelizes of the field operations reduce the number of clock cycle of the point multiplication. To better fit the structure of the addition/subtraction/reduction (AR) block into the reduction algorithms of the five NIST primes, the internal operation of the AR block uses a 32-bit data-path.

A detailed comparison for FPGA-based implementations of the elliptic curve point multiplication on prime fields are shown in Table \ref{table:T2} and Table \ref{table:T7}. Table \ref{table:T2} shows results of works which are implemented for arbitrary prime field or are scalable for support all NIST prime fields. Presented works in Table \ref{table:T7} are implemented over a special NIST prime field or are implemented on special curve.

\begin{table}[H]
	\centering
	\captionsetup{justification=centering}
{\scriptsize	
\caption{Results of the FPGA implementations of the elliptic curve point multiplication on arbitrary prime field and scalable works.}
	\centering 
	\label{table:T2}
	%\vspace*{-1}
	\begin{tabular}{|c|p{2.5cm}|c|c|c|c|c|}
		\hline
		\bf Works(Year) & \bf Prime Field & \bf Device & \bf Area & \bf F$_{max}$(MHz) & \bf Time($\mu s$) \\
		\hline
		\cite{FP1}, (2016) & Arbitrary 256 & Spartan-6 & 105 Slices + 2 DSPs + 2 BRAMs & 200.4 & 9200  \\
		\hline		
		\cite{FP2}, (2012) & Arbitrary 256 & V5 (XC5VLX110) & 3657 Slices + 10 DSPs & 263 & 860 \\
		\hline		
		\cite{FP2}, (2012) & Arbitrary 256 & V4 (XC4VFX12) & 2901 Slices + 14 DSPs & 227 & 1090 \\
		\hline		
		\cite{FP2}, (2012) & Arbitrary 256 & V2 (XC2VP30) & 3423 Slices + 14 18*18-bit MULs & 112 & 2240 \\
		\hline				
		\cite{FP6}, (2016) & Arbitrary 256 & V6 (XC6VLX130T) & 32.4K LUTs & 144 & 1430 \\
		\hline	
		\cite{FP6}, (2016) & Arbitrary 256 & V4 (XC4VFX140) & 35.7K Slices & 70 & 2960 \\
		\hline					
		\cite{FP8}, (2016) & Arbitrary 256 & V6 & 22151 LUTs & 95 & 2010 \\
		\hline			
		\cite{FP8}, (2016) & Arbitrary 256 & V5 & 31431 LUTs & 73 & 2620 \\
		\hline	
		\cite{FP8}, (2016) & Arbitrary 256 & V4 & 20579 Slices & 49 & 3910 \\
		\hline				
		\cite{FP11}, (2011) & Arbitrary 256 & V2 pro & 12K Slices & 36 & 9380 \\
		\hline			
		\cite{FP12}, (2004) & Arbitrary 256 & V2 (XC2V2000) & 3109 Slices & 44.42 & --- \\
		\hline								 
		\cite{FP14}, (2010) & Arbitrary 256 & SII (EP2S30F484C3)  & 9177 ALMs + 96 DSPs & 157.2 & 680  \\		
		\hline		
		\cite{FP19}, (2005) & Arbitrary 192 & V2 (XC2V1000) & 4729 LUTs + 1147 FFs + 2 BRAMs & 50 & 6000  \\		
		\hline												
		\cite{FP20}, (2006) & Arbitrary 256 & V2 (XC2VP30) & 15755 Slices + 256 18*18-bit MULs  & 40  & 3860  \\		
		\hline			
		\cite{FP22}, (2010) & Arbitrary 256 & V2 pro & 1832 Slices + 2 DSPs + 9 BRAMs  & 108.2  & 29830 \\		
		\hline			
		\cite{FP23}, (2009) & Arbitrary 256 & V5 (XC5VLX110) & 2025 Slices  & 100  & 9700 \\		
		\hline							
		\cite{FP25}, (2014) & NIST P-192, P-224, P-256, P-384 and P-521 & V6 (XCV6FX760)  & 32.9K LUTs + 289 DPSs + 128 BRAMs  & 100 & 300 to 3910 \\		
		\hline	
		\cite{FP27}, (2004) & Arbitrary 256 & V2 pro & 15755 Slices + 256 DPSs & 39.5 & 3840 \\		
		\hline			
		\cite{FP29}, (2006) & Arbitrary 256 & Zynq-7020 & 15755 Slices + 20 DSPs + 2 BRAMs & 39.46 & 3860 \\
		\hline	
		\cite{FP30}, (2010) & Arbitrary 256 & V5 (XC5VLX30)  & 20000 Slices & 200 & 1660 \\		
		\hline
		\cite{FP31}, (2014) & NIST P-192, P-224, P-256, P-384 and P-521 & V5 (XC5LX110T)  & 1980 Slices + 7 DPSs & 251.3 & 1709 to 28040 \\
		\hline			
		\cite{FP31}, (2014) & NIST P-192, P-224, P-256, P-384 and P-521 & V4 (XC4VFX100)  & 7020 Slices + 8 DPSs & 182 & 2361 to 38730 \\
		\hline	
		\cite{FP32}, (2016) & Arbitrary 192, 224, 256, 384 and 521 & V7 (XCVU440) & 6816 LUTs + 20 DSPs & 225 & 690 to 9700 \\	
		\hline	
		\cite{FP33}, (2015) & Arbitrary 192 & V5 (XC5VLX330T)  & 615 Slices & 191.42 & 675 \\		
		\hline						
		\cite{FP34}, (2013) & Arbitrary 256 & V5 & 1725 Slices + 37 DPSs + 10 BRAMs & 291 & 380 \\		
		\hline
		\cite{FP35}, (2009) & NIST P-192, P-224, P-256, P-384 and P-521  & V4 (XCV4FX100) & 20793 Slices + 32 DSPs & 43 & 6100 \\		
		\hline
		\cite{FP36}, (2012) & Arbitrary 256 & V5 (XUPV5LX110T)  & 41.6K Slices & 94.7 & 2660 \\		
		\hline													
		\cite{A22}, (2009) & Arbitrary 256 & V4 (XC4VLX200)  & 13661 Slices & 43 & 9200 \\		
		\hline															
		\end{tabular}\\
}
\end{table}

\begin{table}[H]
	\centering
	\captionsetup{justification=centering}
{\scriptsize	
\caption{Results of the FPGA implementations of the elliptic curve point multiplication on special NIST prime field and works on special curve.}
	\centering 
	\label{table:T7}
	%\vspace*{-1}
	\begin{tabular}{|c|p{2.5cm}|c|c|c|c|c|}
		\hline
		\bf Works(Year) & \bf Prime Field & \bf Device & \bf Area & \bf F$_{max}$(MHz) & \bf Time($\mu s$) \\
		\hline
		\cite{FP4}, (2015) & NIST P-384 & V4 (XC4VLX40) & 11883 Slices + 26 DSPs & 276 & 1030 \\
		\hline		
		\cite{FP5}, (2015) & NIST P-256 & V4 (XC4VLX160) & 50589 LUTs & 139 & 2600 \\
		\hline	
		\cite{FP5}, (2015) & NIST P-256 & V5 (XC5VLX110) & 34612 LUTs & 160 & 2260 \\
		\hline				
		\cite{FP7}, (2016) & NIST P-192 & V5 (XC5VLX110) & 2657 Slices & 48.147 & 11.05 \\
		\hline				
		\cite{FP9}, (2016) & FourQ, Mont 256 & Zynq-7020 & 565 LSs + 16 DSPs + 7 BRAMs & 175 & 310 \\
		\hline
		\cite{FP9}, (2016) & FourQ, End 256 & Zynq-7020 & 1691 LSs + 27 DSPs + 10 BRAMs & 175 & 157 \\
		\hline				
		\cite{FP14}, (2001) & NIST P-192 & VE (XCV1000E) & 11416 LUTs + 5735 FFs + 35 BRAMs & 40 & ---  \\		
		\hline	
		\cite{FP16}, (2007) & 160 & V2 pro & 1806 Slices + 3 BRAMs & 101 & 12716  \\		
		\hline	
		\cite{FP17}, (2011) & NIST P-256 & V2 pro & 1158 Slices + 3 BRAMs & 210 & 4520  \\		
		\hline	
		\cite{FP21}, (2008) & NIST P-256 & V4 (XC4VFX12) & 1715 Slices + 32 DPSs + 11 BRAMs  & 490 & 450 \\		
		\hline									
		\cite{FP17}, (2011) & NIST P-256 & V2 pro & 773 Slices + 1 DPSs + 9 BRAMs  & 210 & 10020 \\		
		\hline	
		\cite{FP26}, (2016) & NIST P-256 & V6 & 81 Slices + 8 DPSs + 22 BRAMs  & 171.5 & 11100 \\		
		\hline					
		\cite{FP26}, (2016) & NIST P-256 & Spartan-6 & 72 Slices + 8 DPSs + 24 BRAMs  & 156.25 & 12200 \\		
		\hline	
		\cite{FP28}, (2015) & Curve25519 & Zynq-7020 & 1029 LSs + 20 DSPs + 2 BRAMs & 100 & 397 \\
		\hline			
		\cite{FP37}, (2008) & NIST P-192 & V2 Pro (XC2VP30)  & 3173 Slices + 16 18*18-bit MULs + 6 BRAMs & 93 & 9900 \\		
		\hline															
		\cite{A43}, (2017) & NIST P-256 & V5 (XC5VLX330)  & 12300 Slices & 75.43 & 5260 \\		
		\hline															
		\end{tabular}\\
}
\end{table}

The works \cite{FP1}, \cite{FP26}, \cite{FP17} and \cite{FP22} focused on low-cost and compact implementations and other works \cite{FP5}, \cite{FP6}, \cite{FP9}, \cite{FP14}, \cite{FP21}, \cite{FP25}, \cite{FP27},  \cite{FP30}, \cite{FP33}-\cite{FP34} and \cite{A22}-\cite{A43} on high-speed implementations. Hardware resources in \cite{FP1} are 350 Slices with 2 MULTs and 2 BRAMs, for arbitrary prime field with 256-bit, which are the lowest in the comparison with other implementations. Therefore, \cite{FP1} achieves a good trade-off in the consumed slices and hardcores. The work \cite{FP26} for NIST P-256 only occupies 72 Slices, but the consumed 8 DSP blocks and 24 BRAMs on Spartan-6 FPGA. The work \cite{FP11} is the fastest FPGA-based implementation to date. It runs at 291 MHz on a Virtex-2 Pro FPGA and takes 380$\mu s$ per point multiplication. The designs \cite{FP9}, \cite{FP28} and \cite{FP29} are implemented on same FPGA Zynq-7020. The structure presented in \cite{FP9} has the lowest hardware consumption compared to \cite{FP28} and \cite{FP29}. The best computation time for performing of one point multiplication is 157$\mu s$ for FourQ, End 256 structure in \cite{FP9}. Also FourQ, End 256 structure in \cite{FP9} is 2.54 times faster in computation time than that of \cite{FP28}. The number of DSP blocks in \cite{FP9} is 27 and for \cite{FP28} is 20. Therefore, work \cite{FP9} has about 1.88 times better speed-area ratio than \cite{FP28}.
The implementations \cite{FP25}, \cite{FP31}, \cite{FP32} and \cite{FP35} are scalable FPGA-Based architectures and support five prime fields P-192, P-224, P-256, P-384 and P-521. The proposed ECC processor implemented on Virtex-7 in \cite{FP35} computes the point multiplication with size 192, 224, 256, 384 and 521 in 690$\mu s$, 1080$\mu s$, 1490$\mu s$, 4080$\mu s$ and 9700$\mu s$ respectively. The FPGA implementation of this work consumed 6818 LUTs and 20 DSP48E slices. It runs at a maximum clock frequency of 225 MHz. It also supports arbitrary curves in short Weierstrass form up to 1024-bit without the need to reconfigure the hardware. In this category of the implementations work \cite{FP25} has the best timing performance but hardware consumption in this work is 32.9K LUTs, 289 DPSs and 128 BRAMs.

\subsubsection{FPGA implementations of the point multiplication on dual-field}

General purpose ECC crypto-processors are implemented for both fields $\F_{2^m}$ and $\F_p$. These hardware implementations work in $\F_{2^m}$ as well as $\F_p$ which are categorized in dual-field implementations. They are usually slower than the two previous ECC hardware implementations. Important and desired factor in this group are flexibility and compatibility for support different standards, curve parameters, algorithms and security applications.

\begin{figure}[H]
	\centering
	\includegraphics[scale=0.45]{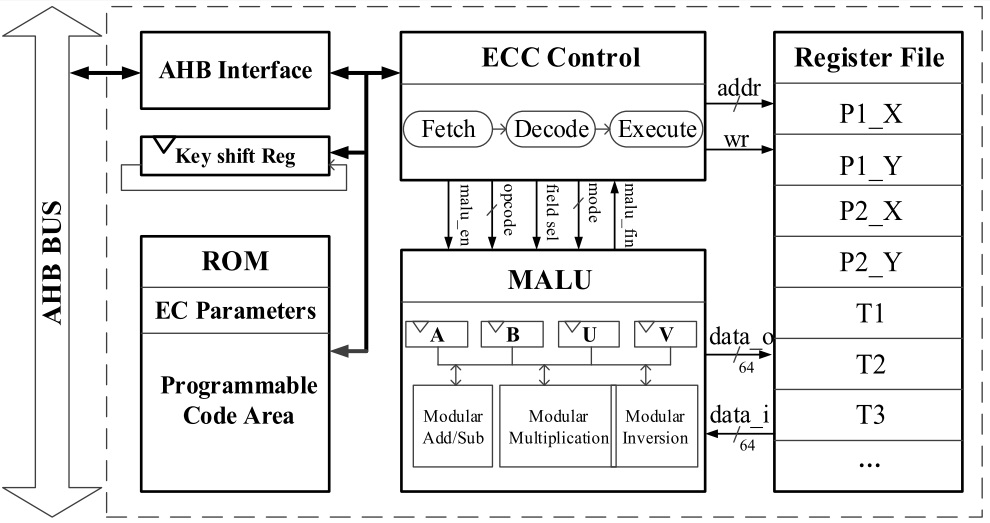}
	\caption{Architecture of the dual-field ECC processor in \cite{A48}.}
	\label{fig:A48_1}
\end{figure}

In \cite{A48} an efficient and flexible hardware implementation of dual-field ECC processor using the hardware-software approach is presented. The structure can support arbitrary elliptic curve based on Modular arithmetic logic unit (MALU). It can compute basic field operations and achieve high efficiency. The processor can be programmed by instruction set to compute different point operations and algorithms. The presented ECC processor in \cite{A48} is shown in Fig.\ref{fig:A48_1}. It is consists of a control unit, MALU, ROM memory, register file and AMBA-AHB interface. By initializing memory with curve parameters and instruction codes, the processor can flexibly perform arbitrary elliptic curve operations over dual-field and different point multiplication algorithms.

To achieve the flexibility and applicability for different elliptic curves in dual-field, in \cite{A48}, authors have integrated the multiple field operations into an MALU. In this circuit, adders are based on carry propagation adder and carry save adder. Modified Radix-4 Interleaved multiplication is used for field multiplier. Also, the plus-minus version of the Radix-4 binary GCD algorithm is used for field inversion and division operations.

In \cite{D1} to speed up point multiplication a processor based on parallel processing technique is presented. The processor consists of a controller that checks instruction-level parallelism (ILP) and multiple sets of modular arithmetic units accelerating field operations. The FPGA results of two dual-field works \cite{A48} and \cite{D1} are shown in Table \ref{table:FD_3}.

\begin{table}[H]
	\centering
	\captionsetup{justification=centering}
{\scriptsize	
\caption{FPGA results of two dual-field works \cite{A48} and \cite{D1}.}
	\centering 
	\label{table:FD_3}
	%\vspace*{-1}
	\begin{tabular}{|c|c|c|c|c|c|}
		\hline
		\bf Works(Year) & \bf Field size $\F_p$/$\F_{2^m}$ & \bf Device & \bf Area & \bf F$_{max}$(MHz) & \bf Time($\mu s$) $\F_p$/$\F_{2^m}$ \\
		\hline
		\cite{A48}, (2016) & 256/256 & V2 & 12425 LUTs & 55.7  & 8250\\
		\hline
		\cite{A48}, (2016) & 256/256 & V4 & 24003 LUTs & 36.5  & 12600\\		
		\hline
		\cite{D1}, (2006) & 160/163 & V2 pro & 8954 Slices+6 BRAMs & 100 & 1040/840 \\		
		\hline		
		\end{tabular}\\
}

\end{table}

\subsection{\textbf{ASIC Hardware Implementations of the Elliptic Curve Cryptosystems}}

In this section, we present a review of different ASIC hardware implementations for the ECC processor. In general, three different types of the ASIC implementations in elliptic curve cryptosystems are over binary fields, prime fields and dual-fields. Many of the works in the ASIC implementations have been focused on the applications which have limited hardware resources with low-power such as smart cards, Wireless Sensor Networks (WSN) and Radio Frequency Identification (RFID) tags. Therefore, in recent years much of the efforts have been confined to designing the lightweight ECC processors. The ASIC hardware implementations of the ECC are reported in \cite{L_BECs}, \cite{A10}, \cite{A21}, \cite{A22}, \cite{A43}, \cite{A48} and \cite{A1}-\cite{A57}. The many of implementations \cite{A10}, \cite{A21}, \cite{A22}, \cite{A43}, \cite{A48}, \cite{A1}-\cite{A5}, \cite{A7}-\cite{A8}, \cite{A11}, \cite{A13}-\cite{A14}, \cite{A16}, \cite{A18}, \cite{A20}, \cite{A24}-\cite{A30}, \cite{A32}-\cite{A33}, \cite{A37}, \cite{A39}-\cite{A42}, \cite{A49}-\cite{A50} and \cite{A54}, in this category, are only synthesized by the Synopsis Design Compiler (Design Vision tool) with CMOS technology. These works are not implemented in layout level. The works \cite{L_BECs}, \cite{A6}, \cite{A9}, \cite{A12}, \cite{A15}, \cite{A17}, \cite{A19}, \cite{A23}, \cite{A31}, \cite{A34}-\cite{A36}, \cite{A38}, \cite{A44}-\cite{A47}, \cite{A51}-\cite{A53} and \cite{A55}-\cite{A57} are implemented in layout level. In following subsections we present three different types of the ASIC implementations in elliptic curve cryptosystems in more details.
 
\subsubsection{\textit{ASIC implementations of the ECC on binary fields}} 

The works \cite{L_BECs}, \cite{A10}, \cite{A21}, \cite{A1}, \cite{A4}, \cite{A6}-\cite{A11}, \cite{A13}-\cite{A14}, \cite{A16}, \cite{A18}, \cite{A20}, \cite{A23}-\cite{A27}, \cite{A29}-\cite{A33}, \cite{A37}-\cite{A40} and \cite{A54}-\cite{A55} are ASIC implementation of the ECC on binary fields $\F_{2^m}$. 

The main special techniques for FPGA implementations of the point multiplication on prime fields are summarized as follows:  

\begin{enumerate}

\item In \cite{L_BECs} by using the \textbf{\textit{logical effort technique}} the delay is optimally decreased and the drive ability of the structure in the point multiplication is increased.

\item In \cite{A4} the ECC processor is designed based on \textbf{\textit{programmable cellular automata}}. 

\item In \cite{A29} an \textbf{\textit{optimized RAM-macro block}} is used and the design allows reduces the complexity by sharing different resources of the controller and the data-path.

\item In \cite{A32} a \textbf{\textit{new technique to compute point additions in affine coordinates}} on Koblitz curves is proposed. This technique is based on applying a efficient inversion algorithm, which is implemented by fewer registers than the traditional schemes.

\end{enumerate}

In \cite{L_BECs} an efficient ASIC implementation  of  point multiplication on binary Edwards curves with GNB representation. The implementation is a low-cost structure constructed by one digit-serial field multiplier. The  field multiplier is busy during PA and PD computations. In this work, by using the logical effort technique the delay is optimally decreased and the drive ability of the structure in the point multiplication is increased.

\begin{figure}[H]
	\centering
	%	\captionsetup{justification=centering}
	\includegraphics[scale=0.6]{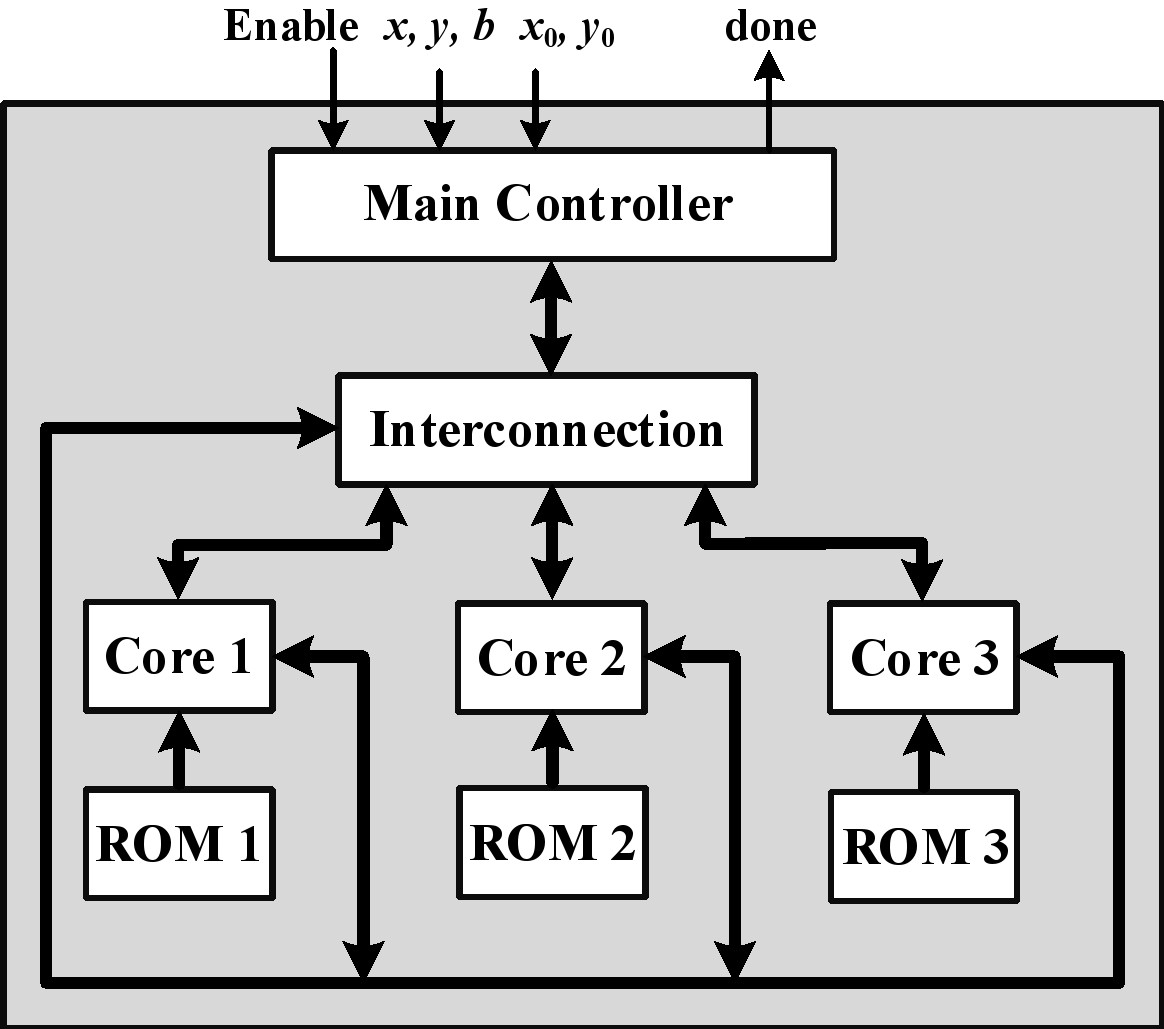}
	\caption{Structure of pseudo-multi-core ECC processor in \cite{A21}.}
	\label{fig:F3}
\end{figure}

In \cite{A21} a ECC processor over $\F_{2^{163}}$ for the cryptographic applications that require high-performance is proposed. It has three 5-stage pipelined field RISC cores and a control unit to achieve instruction-level parallelism for the point multiplication. To decrease the latency customized instructions are proposed. The internal connections among three finite field cores and the main controller is obtained based on the analysis of both data dependency and critical path. The structure is illustrated in Fig.\ref{fig:F3}. This structure is called pseudo-multi-core because this implementation achieves parallelism in instruction-level, not process level.

The instruction set, $AB$, $A+B$, $(A+B)^2$ and $A^4$, for the parallelized Lopez-Dahab algorithm, in each core is obtained by analyzing the algorithm level and the hardware level.

\begin{figure}[H]
	\centering
	\includegraphics[scale=0.5]{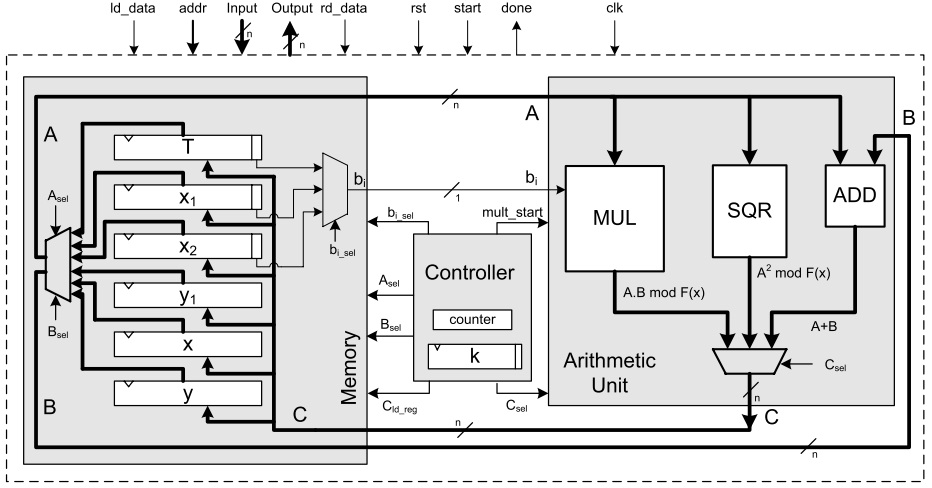}
	\caption{Low-area ECC processor over $\F_{2^n}$ in \cite{A1}.}
	\label{fig:F111}
\end{figure}

In \cite{A1} a highly area optimized ECC processor for binary field is designed. The fast squarer circuit is used to construct an addition chain for efficient hardware implementation of the inversion. Therefore, an ASIC implementation of the processor using a modified Montgomery ladder point multiplication based on affine coordinate is presented. The design is for binary elliptic curves ranging from 113 to 193 bits. Area consumed is between 10k and 18k gates on a 0.35$\mu m$ CMOS process for the different curves. Fig.\ref{fig:F111} shows ECC processor presented in \cite{A1}. The three units: field addition (ADD), field multiplication (MUL) and field squaring (SQR) are connected inside a single arithmetic unit sharing the common input data-bus $A$. The output results of the three previous units are selected at the output data-bus $C$ by the control signal $C_{sel}$ and one 3 to 1 multiplexer. The field adder needs an additional data-bus $B$ for the second input operand and the field multiplier requires a serial bit $b_i$ for the multiplicand. The operands are stored in the registers with the output being selected for $A$, $B$ and $b_i$ using multiplexers with control signals $A_{sel}$, $B_{sel}$ and $b_{i\_sel}$. All registers are connected in parallel to the data-bus $C$ based on load signal $C_{ld\_reg}$.

An architecture of a elliptic curve processor for RFID tags over $\F_{2^{163}}$ is proposed in \cite{A8}. The processor is able to perform the point multiplications as well as general field operation such as additions and multiplications which are required for the different cryptographic protocols. By applying several techniques, the number of registers in register file are reduced from 9 to 6. A redundant field operation is introduced to obtain an efficient field arithmetic. Furthermore, the structure can support several cryptographic protocols. Elliptic curve PA and PD circuit (EC Add/Doubler) consists of control unit-1 (Control1), the modular arithmetic logic unit (MALU) and a register file. Control unit-1 receives the curve parameters and gives the result of the point multiplication via control unit-2 (Control2). Also control unit-2 reads in bytes a scalar (key) via the bus manager and controls the elliptic curve Add/Doubler based on the Montgomery ladder algorithm.

In \cite{A30} a lightweight coprocessor based on 283-bit Koblitz curve that implements high security ECC is presented. For the fast point multiplication the scalars are given as specific $\tau$-adic expansions. This work is the first lightweight different of the conversion algorithm from integers to $\tau$-adic. Therefore, the first lightweight implementation of Koblitz curves that includes the scalar conversion is introduced in \cite{A30}. Also the structure is the first lightweight ASIC multiplication for Koblitz curves that includes a set of countermeasures against simple power analysis, differential power analysis, timing attacks and safe error fault attacks. The processor consists of an ALU, an address generation unit, a shared memory and a control unit based on FSM. The ALU is connected with the memory block using an input register pair and an 2 to 1 multiplexer at output. The central part of the ALU consists of a 16 bits integer adder/subtracter circuit, a 16 bits binary field multiplier and two binary field adders. The some constant parameters are stored in a small ROM called Reduction-ROM which are used during modular reductions and multiplications.

In \cite{A32} an efficient implementation of the point multiplication on Koblitz curves for extremely-constrained applications in term of area is proposed. The field multiplication is designed by an efficient bit-serial multiplier with GNB representation. The addition and accumulation of this GNB multiplier is shared with other field additions.

A processor for ECC over $\F_{2^{163}}$ in \cite{A38} is presented. It is flexible enough to support several cryptographic protocols. The chip for hardware realization processor is fabricated using UMC 130$nm$ 1P8M process, resulting in a core area of 0.54 $mm^2$. The energy consumption to perform one point multiplication is 5.1$\mu J$.

\begin{figure}[H]
	\centering
	%	\captionsetup{justification=centering}
	\includegraphics[scale=0.55]{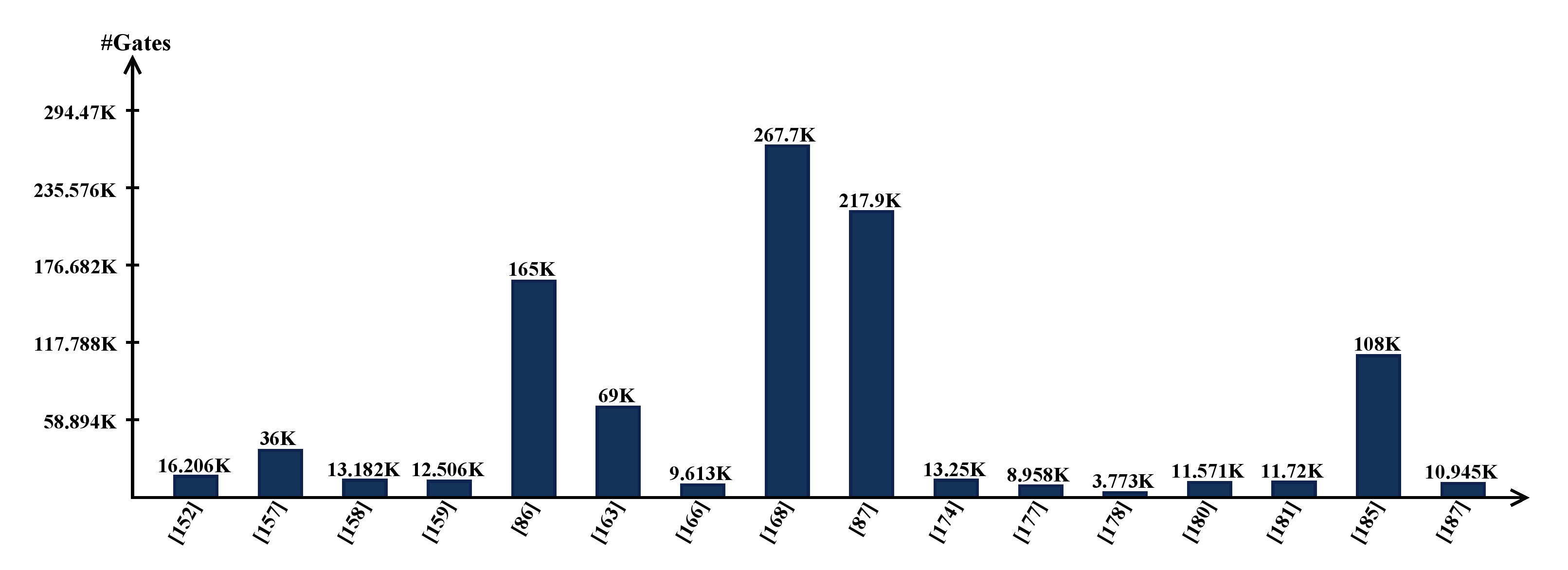}
	\caption{The number of gates for ECC ASIC implementations over $\F_{2^{163}}$.}
	\label{fig:AB1}
\end{figure}

In Table \ref{table:T_AB}, we present the timing characteristics, area and power consumption of the previous ASIC hardware implementations of the ECC on $\F_{2^{m}}$. The works  \cite{A21}, \cite{A1}, \cite{A4}, \cite{A6}-\cite{A11}, \cite{A13}-\cite{A14}, \cite{A16}, \cite{A18}, \cite{A23}-\cite{A27}, \cite{A29}-\cite{A30}, \cite{A40}, \cite{A54} and \cite{A56} are implemented based on binary Weierstrass curves. The fastest design in this category is presented work \cite{A21} with execution time for one point multiplication equal 5.4$\mu s$ in 180$nm$ CMOS technology. Hardware resources in this work is equivalent 217.9K gates. The fully programmable processor in \cite{A54} can handle various curve parameters and an arbitrary irreducible polynomial. In addition, a wide range of the field size can be supported by changing the program and reconfiguring the data-path in the MALU cores. The type of the elliptic curve in the works \cite{A30}, \cite{A32}, \cite{A37} and \cite{A38} is Koblitz curves. For Koblitz curves the proposed structure in \cite{A30} has minimum area consumption compared to other works, it is equal to 4323 gates and also work \cite{A37} has minimum computation time with hardware resources equal to 108K gates. The work \cite{A32} is aimed at the low-area constrained application, such as RFID. So it consumed 11571 logic gates with power consumption 0.66$\mu$W at 106 KHz. Fig.\ref{fig:AB1} shows the number of gates for works which are implemented over $\F_{2^{163}}$.

\begin{table}[H]
	\centering
	\captionsetup{justification=centering}
{\scriptsize	
\caption{Results of the timing characteristics, area and power consumption of the previous ASIC implementations of the ECC on $\F_{2^{m}}$}
	\centering 
	\label{table:T_AB}
	%\vspace*{-1}
	\begin{tabular}{|p{2.5cm}|p{1cm}|c|c|c|c|c|}
		\hline
		\bf Works(Year) & \bf Field size & \bf Technology & \bf Area & \bf F$\bf_{max}$(MHz) & \bf Time($\bf\mu s$) & \bf \shortstack{Power \\ consumption} ($\bf\mu$W)\\
		\hline
		 \cite{A1}, BWCs, PB, (2013) & 163 &  350$nm$ AMI & 16.206K gates & 13.56 & 27900 & ---\\
		\hline
	     \cite{A4}, BWCs PB, (2011)  & 256 & 120$nm$ & 1.29 $mm^2$ & 312 & 850 & 23100\\		
		\hline
		 \cite{A6}, BWCs, PB, (2007) & 163 & 180$nm$ TSMC & 36K gates+1KB RAM & 125 & 62 & ---\\ 
		\hline
	     \cite{A7}, BWCs, PB, (2007) & 163 & 180$nm$ & 13.182K gates & --- & --- & ---\\
		\hline
		 \cite{A8}, BWCs, PB, (2008) & 163 & 130$nm$ UMC & 12.506K gates & 1.13 & 244080 & 36.63\\
		\hline
		 \cite{A9}, BWCs, PB, (2004) & 191 & 130$nm$ & 0.16 $mm^2$ & 10 & 34143 & ---\\
		\hline
		 \cite{A10}, BWCs, PB, (2000) & 163 & 250$nm$ CE71 & 165K gates & 66 & 1100 & ---\\
		\hline
	     \cite{A11}, BWCs, PB, (2003) & 178 & 500$nm$ & 112K gates & 20.83 & --- & 150000 @ 20MHz\\
		\hline
	     \cite{A13}, BWCs, PB, (2009) & 163 & 180$nm$ TSMC & 1.92$mm$*1.92$mm$, 69K gates & 181 & 1260 & 136000 \\
		\hline
	     \cite{A14}, BWCs, PB, (2005) & 191 & 350$nm$ & 68K gates & 125 & 590 & --- \\
		\hline
	     \cite{A16}, BWCs, PB, (2009) & 163 & 130$nm$ IBM & 9.613K gates & --- & --- & --- \\
		\hline
	     \cite{A18}, BWCs, PB, (2010) & 163 & 130$nm$ & 267.7K gates & 199 & 11.1 & --- \\
		\hline
	     \cite{A21}, BWCs, PB, (2010) & 163 & 180$nm$ TSMC & 217.9K gates & 263 & 5.4 & --- \\
		\hline
		 \cite{A23}, BWCs, PB, (2007) & 283 & 250$nm$ & 1.9$mm^2$ & --- & 175 & 50.6 \\
		\hline
		 \cite{A24}, BWCs, PB, (2003) & 251 & 350$nm$ & 2.75$mm^2$ & 100 & 5500 & 13600\\
		\hline
		 \cite{A25}, BWCs, PB, (2002) & 192 & 350$nm$ & 16.847K gates & 10 & 126 & ---\\
		\hline		
		 \cite{A26}, BWCs, PB, (2009) & 163 & 180$nm$ UMC & 13.25K gates & 46 & 2792000 & 8.57 @ 106KHz\\
		\hline			
		 \cite{A27}, BWCs, PB, (2006) & 131 & 130$nm$ & 6.718K gates & --- & 115000 & 30 @ 500KHz\\
		\hline					
		 \cite{A29}, BWCs, PB, (2011) & 163 & 130$nm$ UMC & 8.958K gates & --- & --- & 32.34 @ 1MHz\\
		\hline	
		 \cite{A30}, BWCs, PB, (2015) & 163 & 130$nm$ UMC & 3.773K gates & --- & 30310 & 6.11 @ 1MHz \\
		\hline	
		 \cite{A30}, BKCs, PB, (2015) & 163 & 130$nm$ UMC & 4.323K gates & --- & 26300 & 6.11 @ 1MHz \\
		\hline	
		 \cite{A32}, BKCs, GNB, (2014) & 163 & 65$nm$ & 11.571K gates & --- & 1006600 & 0.66 @ 106KHz\\
		\hline
		 \cite{A33}, BECs, PB, (2010) & 163 & 130$nm$ & 11.72K gates & --- & 547870 & 7.27 @ 400KHz\\
		\hline
		 \cite{A37}, BKCs, PB, (2007) & 163 & 130$nm$ & 108K gates & 555.6 & 27 & ---\\
		\hline	
	    \cite{A38}, BKCs, PB, (2016) & 163 & 130$nm$ UMC & 735$\mu m$*735$\mu m$ & 555.6 & 102000 & 50.4 @ 847.5KHz\\
		\hline	
		 \cite{A39}, BECs, PB, (2015) & 163 & 65$nm$ TSMC & 10.945K gates & --- & --- & ---\\
		\hline			
		 \cite{A40}, BWCs, PB, (2014) & 160 & 130$nm$ & 12.448K gates & --- & --- & 42.42 @ 1MHz\\
		\hline			
		 \cite{A54}, CONFIG-I, BWCs, PB, (2007) & 163, 193, 283 and 571 & 130$nm$ & 393K gates & 292 & 54 to 1349 & ---\\
		\hline														
		 \cite{A54}, CONFIG-II, BWCs, PB, (2007) & 163, 193, 283 and 571 & 130$nm$ & 244K gates & 292 & 54 to 1349 & ---\\
		\hline
			
		 \cite{L_BECs}, BECs, GNB, (2017) & 233 & 180$nm$ & 29.524K gates+10*233 Regs & 1070.66 & 118.6 & ---\\
		\hline			
					
		\end{tabular}\\
}

\end{table}

\subsubsection{\textit{ASIC implementations of the ECC on prime fields}}

The works \cite{A22}, \cite{A43}, \cite{A2}, \cite{A3}, \cite{A5}, \cite{A12}, \cite{A28}, \cite{A41}, \cite{A42}, \cite{A53} and \cite{A57} are ASIC implementations of the ECC on prime fields. In \cite{A22} two different parallelization techniques to accelerate the point multiplication over $\F_{p}$ in affine coordinates are presented. The proposed implementations are resist against different side channel attacks based on power and time analysis. The both architectures are synthesized for 160, 192, 224 and 256 bits on FPGA and also ASIC implementation in 130$nm$ CMOS technology is performed. In \cite{A43} to achieve a high-speed and low-area hardware structure of elliptic curve point multiplication over a prime field, a combined PA and PD architecture is presented by using efficient modular arithmetic in Jacobian coordinates. In \cite{A5} a high-performance ECC processor for general prime curves is presented. By using a unified systolic array the field addition, field subtraction, field multiplication and field division are efficiently implemented. The structure is pipelined and pipeline stall problems are successfully solved by using two optimization methods. The processor, is synthesized in 130$nm$ standard cell technology, it takes 1.01$ms$ to compute a 256 bits point multiplication for general curves. The proposed ASIC implementation in \cite{A12} is includes a 3-stage pipelined full-word modular Montgomery multiplier which needs a few clock cycles. The precomputation steps of field multiplication based on Montgomery method are implemented by hardware. The ECC field arithmetic unit has programmable data-path, so, arbitrary field lengths are supported for implementation.

A high-performance implementation of ECC over SCA-256 prime field by considering an all-new isochronous architecture is proposed in \cite{A41}. It is resist against simple power analysis and double attack with minimum time cost. Also random cycles are inserted in the structure to differential power analysis. By modifying Montgomery ladder point multiplication algorithm the PA and PD can operate synchronously. The processor achieves 211$\mu s$ and 8.5$\mu J$ for one point multiplication with 208k gates using CMOS standard cell library of 130$nm$.

The timing characteristics, area and power consumption of the previous ASIC hardware implementations of the ECC on $\F_p$ are illustrated in Table \ref{table:T_2}. Compared to other related designs, the work \cite{A42} outperforms other implementations in terms of execution time and area/time product. Hardware consumed in \cite{A28} is equal to 30.3K gates in 130$nm$ CMOS technology which is the lowest compared to other designs.  

\begin{table}[H]
	\centering
	\captionsetup{justification=centering}
{\scriptsize	
\caption{Results of the timing characteristics, area and power consumption of the previous ASIC Implementations of the ECC on $\F_p$.}
	\centering 
	\label{table:T_2}
	%\vspace*{-1}
	\begin{tabular}{|c|c|c|c|c|c|c|}
		\hline
		\bf Works & \bf Field size & \bf Technology & \bf Area & \bf F$\bf_{max}$(MHz) & \bf Time ($\bf \mu s$) & \bf \shortstack{Power \\ consumption} ($\bf\mu$W)\\
		\hline
		 \cite{A3}, (2010)& 256 &  180$nm$ & 132K gates & 671 & 850 & ---\\
		\hline
		 \cite{A5}, (2007) & 256 & 130$nm$ & 122K gates & 556 & 1010 & ---\\		
		\hline
		 \cite{A12}, (2012) & 256 & 90$nm$ & 540K gates, 2.72$mm^2$ & 185 & 120 & ---\\ 
		\hline
		  \cite{A22}, Design 1, (2009) & 256 & 130$nm$ & 106.7K gates & 137.7 & 2680 & ---\\
		\hline
		 \cite{A22}, Design 2, (2009) & 256 & 130$nm$ & 109.2K gates & 110 & 3610 & ---\\
		\hline
		 \cite{A28}, (2004) & 167 & 130$nm$ & 30.3K gates & --- & 34143 & 990 @ 20MHz\\
		\hline
		 \cite{A41}, (2015) & 256 & 130$nm$ & 208K gates & 215 & 211 & ---\\ 
		\hline
		 \cite{A42}, (2014) & 256 & 130$nm$ & 659K gates & 163.7 & 20.36 & ---\\ 
		\hline		
		 \cite{A43}, (2016) & 256 & 90$nm$ & 447K gates, 0.93$mm^2$ & 546.5 & 730 & ---\\ 
		\hline				
		\end{tabular}\\
}

\end{table}

\subsubsection{\textit{ASIC implementations of the ECC on dual-fields}}

\begin{figure}[H]
	\centering
	\includegraphics[scale=0.5]{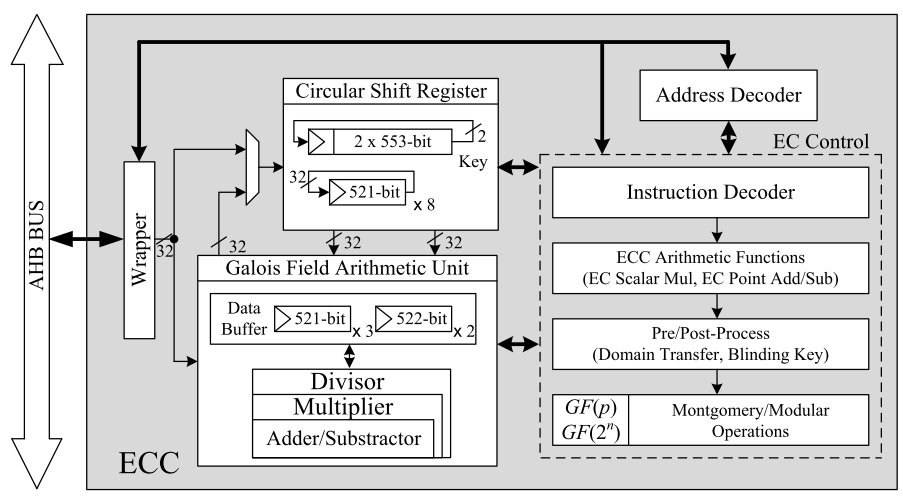}
	\caption{Architecture of the dual-field ECC processor in \cite{A17}.}
	\label{fig:F6}
\end{figure}

Flexibility and scalability for ASIC implementation of elliptic curve applications is important and interesting subject in literature. The works \cite{A48}, \cite{A15}, \cite{A17}, \cite{A19}, \cite{A34}-\cite{A36} and \cite{A44}-\cite{A52} are ASIC implementations of the ECC over both prime fields $\F_p$ and binary fields $\F_{2^m}$ to support a wide range of elliptic curves and applications. In \cite{A17} a unified division algorithm and a free precomputation structure are proposed to speedup the $\F_p/\F_{2^n}$ elliptic curve arithmetic operations. The structure is optimized by a very compact field arithmetic unit with the fully pipelined technique. Also, a key-blinded technique with regular computation is implemented against the power analysis attacks without time cost. After fabricated in 90$nm$ CMOS 1P9M process, area of ECC processor is 0.55$mm^2$. It can perform the point multiplication in 19.2$ms$ over $\F_{p^{521}}$ and 8.2$ms$ over $\F_{2^{409}}$, respectively. Fig.\ref{fig:F6} shows the ECC architecture with a standard AMBA AHB bus interface. The inputs are consist of user public/private-key, elliptic curve coordinates, elliptic curve parameters and protocol instructions. The instruction decoder and pre-/post-process are combined in the processor. After the instruction decoding, the pre-process stage is to convert the coordinates and parameters into the Montgomery domain and blind the key value to avoid power analysis attacks. All dual-field modular and Montgomery operations are integrated into the pipelined Galois field arithmetic unit with circuit sharing.

In \cite{A19} presents a parallel, scalable and high-throughput dual-field ECC architecture. This processor has features all ECC functions with the programmable field and curve parameters over both the prime and binary fields. Using 130$nm$ CMOS technology, the core size of the processor is 1.44$mm^2$. The results show that the ECC processor can perform one 160-bit point multiplication with coordinate conversion over $\F_p$ in 608$\mu s$ at 121MHz with only 70$m$W and the field $\F_{2^m}$ in 372$\mu s$ at 146MHz with 82.1$m$W. The ECC instructions and data are fed into the input buffer through the AMBA AHB interface. The main controller decodes the instructions that support comprehensive cryptographic functions, including the coordinate conversion, PA, PD, point multiplication, Montgomery pre-/postprocessing, modular exponentiation and common field operations.
The ECC processor has been fabricated using TSMC 130$nm$ 1.2-$V$ 1P8M CMOS technology. The size of the ECC processor chip is 5.15$mm^2$, where the core size is only 1.44$mm^2$ (1.2 $ \times $ 1.2 $ mm $). Table \ref{table:T3} summarizes the previously published results of the ASIC implementations of the dual-field ECC.

\begin{table}[H]
	\centering
	\captionsetup{justification=centering}
{\scriptsize	
\caption{Results of the previously published ASIC implementations of the dual-field ECC.}
	\centering 
	\label{table:T3}
	%\vspace*{-1}
	\begin{tabular}{|c|c|c|c|c|c|c|}
		\hline
		\bf Works(Year) & \bf \shortstack{Field size \\ $\F_p$/$\F_{2^m}$} & \bf Technology & \bf Area & \bf \shortstack{F$_{max}$(MHz) \\ $\F_p$/$\F_{2^m}$} & \bf \shortstack{Time($\mu s$) \\ $\F_p$/$\F_{2^m}$} & \bf \shortstack{Power consumption ($\mu$W) \\ $\F_p$/$\F_{2^m}$}\\
		\hline
		\cite{A15}, (2010) & 256/256 & 130$nm$ TMSC & 184K gates, 1.30$mm^2$ & 75/114  & 368/252 &68400/58200 \\
		\hline
		\cite{A17}, (2010) & 256/163 & 90$nm$ & 170K gates, 0.55$mm^2$  & 147/188 & 4400/1150 & 67600/72500\\		
		\hline
		\cite{A19}, (2009) & 160/160 & 130$nm$ TSMC & 169K gates, 1.44$mm^2$ & 121/146 & 608/372 & 70000/82000\\ 
		\hline
		\cite{A34}, (2008) & 160/160 & 130$nm$ & 150.5K gates, 1.06$mm^2$ & 217/350 & 340/155 & ---/---\\
		\hline
		\cite{A35}, (2010) & 163/163 & 130$nm$ & 331.7K gates, 2.34$mm^2$ & 415/415 & 440/440 & ---/---\\
		\hline
		\cite{A36}, (2011) & 160/160 & 130$nm$ & 179K gates, 1.35$mm^2$ & 141.3/158.1 & 385/272 & 32300/46200 \\
		\hline		
		\cite{A44}, (2015) & 163/256 & 65$nm$ & 179K gates, 1.10$mm^2$ & 500/500 & 320/80 & 120000/125000 \\
		\hline		
		\cite{A45}, (2008) & 160/160 & 180$nm$ & 17.81$mm^2$ & 233/233 & ---/--- & 10000 @ 1MHz \\
		\hline		
		\cite{A46}, (2011) & 256/256 & 90$nm$ & 122K gates, 0.45$mm^2$ & 250/277.8 & 770/590 & 31000/35600 \\
		\hline		
		\cite{A47}, (2014) & 160/160 & 90$nm$ UMC & 96K gates, 0.41$mm^2$ & 256/260 & 250/220 & ---/--- \\
		\hline		
		\cite{A48}, (2016) & 256/256 & 55$nm$ & 189K gates, 0.35$mm^2$ & 316/316 & 1450/1450 & ---/--- \\
		\hline	
		\cite{A50}, (2003) & 192/160 & 130$nm$ & 117.5K gates & 137.7/510.2 & 1440/190 & ---/--- \\
		\hline	
		\cite{A57}, Radix-2, (2012) & 160/160 & 90$nm$ & 61.3K gates, 0.21$mm^2$ & 277/277 & 710/610 & ---/--- \\
		\hline
		\cite{A57}, Radix-4, (2012) & 160/160 & 90$nm$ & 83.2K gates, 0.29$mm^2$ & 238/238 & 430/390 & ---/--- \\
		\hline	
		\cite{A57}, Radix-2, (2012) & 521/409 & 90$nm$ & 168K gates, 0.58$mm^2$ & 250/263 & 8080/4650 & ---/--- \\
		\hline
		\cite{A57}, Radix-4, (2012) & 521/409 & 90$nm$ & 265K gates, 0.93$mm^2$ & 232/238 & 4570/2770 & ---/--- \\
		\hline										
		\end{tabular}\\
}

\end{table}

The present work in \cite{A47} needs a preprocessing to convert the operands into the Montgomery domain. But in recent work \cite{A48} no domain conversion is needed, it is implemented in the ordinary prime/binary finite fields. Overall, ECC processor in \cite{A48} achieves high efficiency and flexibility due to the elaborate MALU structure and soft-hardware approach respectively. It can be used for different ECC standards, different elliptic curves and different point multiplication algorithms. In \cite{A19} for design-for-testability (DFT), six scan chains were inserted with a fault coverage of 99.77\%. The best work in term of computation time is \cite{A44} for 163-bit binary field and 256-bit for prime field.

\section{Conclusion}
\label{sec-end}

In this paper, a comprehensive study of hardware implementations of elliptic curve cryptography is presented. For fair comparison and better analysis, the implementations are categorized and presented based on used finite field, type of elliptic curves, representation basis and implementation platforms. In the survey, different elliptic curves, point multiplication algorithms and finite field arithmetics and also their effect on implementations are defined and discussed. The implementations are compared in terms of hardware consumption which is important for any cost-sensitive application and execution time which is important for many applications especially in high-speed application. The study shows that FPGAs are suitable for reconfigurable applications and ASIC implementations are suitable for Lightweight implementations of the ECC which is one of the attractive fields in ECC implementation. The most previous works have similar selection in the implementation, i.e., in selecting type of finite field, elliptic curve, point multiplication algorithm and algorithm of field operations. For example, the Montgomery ladder and the Itoh-Tsujii algorithm are widely used for point multiplication and field inversion respectively. Several proposed solutions and important issues that can be helpful in hardware implementation of the ECC are as follows:

\begin{itemize}
  \item The number of researches concerning the implementation of ECC in low-cost microcontroller-based devices is increasing. It is important and feasible to implement cryptographic applications in constrained environments and be able to achieve acceptable performance. Therefore, implementation of the ECC on microcontrollers and microprocessors is attractive subject.
 
\end{itemize}

\begin{itemize}
   \item Binary Edwards curves (BECs) are complete and without exception points, the point addition law which makes them attractive for implementation and intrinsically resistant to Simple power analysis. The design and implementation of the hardware structures for the point multiplication on binary Edwards curves can be a popular research topic and more work could be done to take the best advantage of these curves.
\end{itemize}

\begin{itemize}
\item The logical effort technique is a procedure for achieving the least delay for a given load in a logic circuit. In the design of the field multiplier in the ECC structure, to balance the delay among the stages and to obtain a minimum over all delay, the logical effort technique can be applied. This technique is suitable for high-speed hardware implementation of the ECC. We can design an algorithmic and automatics approach based on logical effort for compute size of transistors in the critical path delay for different loads. Also, in this case the best trade-offs between area and speed can be achieved.
\end{itemize}

\begin{itemize}
\item Lightweight hardware implementation of the ECC for FPGAs and ASIC design has been a popular research topic due to the development low-cost hardware embedded applications. 
\end{itemize}

\begin{itemize}
\item The field addition and multiplication operations in $\F_{3^m}$ are comparable in performance to a space equivalent characteristic two alternative. So, implementation of the ECC on $\F_{3^m}$ can be an important issue in future. 
\end{itemize}

\begin{itemize}
\item To design ECC processor based on large field multipliers for high-level security application in a significant speedup, the Fine-Grain pipelining and retiming technique (minimize the clock  period and the number of registers in the circuit) should be applied.
\end{itemize}

\begin{itemize}
\item For flexible and scalable ECC processors, more work could be done to take the best advantage of full size hardware for different applications and concurrent multi-point multiplication algorithms.
\end{itemize}

\begin{itemize}
\item For high-speed and low-area ECC processors, more work could be done to take the best advantage of different scheduling algorithms, for example, List scheduling, Force directed and Iterative refinement. The aim of the high-speed implementation is reduce the number of field operation especially field multiplier. Therefore, the new scheduling methods can be useful for computation of the point addition and point doubling, in differential addition coordinate, in binary Edwards and generalized Hessian curves.
\end{itemize}


\begin{thebibliography}{1}

\bibitem{tth}
The package is TTH, available at http://hutchinson.belmont.ma.us/tth/ .

\bibitem{use2e}
As the mark-up of the \TeX\ source for this document makes clear, your file
  should be coded in \LaTeX 2${\varepsilon}$, not \LaTeX\ 2.09 or an earlier
  release. Also, please use the \texttt{article} document class.

\bibitem{inclme}
Among whom are the author of this document. The ``real'' references and notes
  contained herein were compiled using B{\small{IB}}\TeX\ from the sample .bib
  file \texttt{scibib.bib}, the style package \texttt{scicite.sty}, and the
  bibliography style file \texttt{Science.bst}.

\bibitem{nattex}
One of the equation editors we use, Equation Magic (MicroPress Inc., Forest
  Hills, NY; http://www.micropress-inc.com/), interprets native \TeX\ source
  code and generates an equation as an OLE picture object that can then be cut
  and pasted directly into Word. This editor, however, does not handle \LaTeX\
  environments (such as \texttt{\{array\}} or \texttt{\{eqnarray\}}); it can
  interpret only \TeX\ codes. Thus, when there's a choice, we ask that you
  avoid these \LaTeX\ calls in displayed math --- for example, that you use the
  \TeX\ \verb+\matrix+ command for ordinary matrices, rather than the \LaTeX\
  \texttt{\{array\}} environment.

\end{thebibliography}


\begin{thebibliography}{10}
	
	%%%%%%%%%%%%%%%%%%%%%%%%%%%%%%%%%%%%%%
	\bibitem{Miller}
	{Miller, V.S.},
	{\it Use of elliptic curve in cryptography}, { Advances in Cryptology, in: Proceedings of the Crypto'85, 1986, pp. 417-426.}
	
	\bibitem{Koblitz}
	{Koblitz, N.},
	{\it Elliptic curve cryptosystems}, {Math. Comput. Vol. 48, 1987, pp. 203-209.}
	
	\bibitem{Guide}
	{Hankerson, D., Menezes, A., Vanstone, S.},
	{\it Guide to Elliptic Curve Cryptography} {1st ed., Springer-Verlag, New York, 2003.}
	
	\bibitem{TLS}
	{Dierks, T., Rescorla, E.},
	{\it The Transport Layer Security (TLS) Protocol} {Version 1.2, August 2008.}
	
	\bibitem{WTLS}
	{WAP WTLS,},
	{\it Wireless Application Protocol Wireless Transport Layer Security Specification} {Wireless Application Protocol Forum, February 1999. Drafts available at http://www.wapforum.org}

	\bibitem{ANSI}
	{ANSI X9.62-1999},
	{\it The Elliptic Curve Digital Signature Algorithm, ANSI, Washington, D.C., USA, 1999.}

	\bibitem{IEEE}
	{IEEE P1363},
	{\it Editorial Contribution to standard for Public Key Cryptography, 2000.}
	
	\bibitem{ISO}
	{ISO/IEC  14888-3},
	{\it Information technology Security techniques  Digital signatures with appendix Part 3: Discrete logarithm based mechanisms, 2006.}

	\bibitem{NIST}
	{FIPS},
	{\it Federal Information Processing Standards Publications (FIPS)186-2, U.S.	Department of Commerce/NIST: Digital Signature Standard (DSS), 2000.}

	\bibitem{SCA}
	{State Cryptography Administration of China},
	{\it Public Key Cryptographic Algorithm SM2 Based on Elliptic Curves, 2010.}

	\bibitem{Koblitz2}
	{Koblitz, N.},
	{\it CM-curves with good cryptographic properties}, {in Proceedings of the Annual International Cryptology Conference (Advances in Cryptology-CRYPTO), Lecture Notes in Computer Science, New York: Springer, Vol.576, 1991, pp. 279-287.}	 
	
	\bibitem{BLF8}
	{Bernstein, D., Lange, T. and Rezaeian Farashahi, R.},
	{\it Binary Edwards Curves}, {in Proceedings of the International Workshop on Cryptographic Hardware and Embedded Systems (CHES), Worcester, Lecture Notes in Computer Science, Vol. 5154, 2008, pp. 244-265.}

			 
	\bibitem{FJ}
	{Rezaeian Farashahi, R., Joye,  M.},
	{\it Efficient arithmetic on Hessian curves}, {in Proceedings of the 13th Int. Conf. Practice Theory of Public Key Cryptography (PKC), 2010, pp. 243-260.}	
	
    \bibitem{Huff}
	{Devigne, J., Joye, M.},
	{\it Binary huff curves}, {in Proceedings of the Cryptographers'Track at the RSA Conference (CT-RSA), LNCS, Springer-Heidelberg, Vol. 6558, 2011, pp. 340-355.}

    \bibitem{Curve25519}
	{Bernstein, D.},
	{\it Curve25519: New Diffie-Hellman Speed Records}, {Berlin, Heidelberg: Springer-Heidelberg, 2006, pp. 207-228. Available: http://dx.doi.org/10. 1007/11745853 14}

    \bibitem{MontCur}
	{Montgomery, P.L.},
	{\it Speeding the Pollard and Elliptic Curve Methods of Factorization}, {Mathematics of Computation. Vol. 48 (177), 1987, pp. 243-264.}

    \bibitem{GLV}
	{Gallant, R.P., Lambert, R.J., and Vanstone, S.A.},
	{\it Faster Point Multiplication on Elliptic Curves with Efficient Endomorphisms}, {in Proceedings of the Annual International Cryptology Conference (Advances in Cryptology-CRYPTO) 2001, LNCS 2139, 2001, pp. 190-200.}

    \bibitem{GLS}
	{Galbraith, S.D., Lin, X., Scott, M.},
	{\it Endomorphisms for faster elliptic curve cryptography on a large class of curves}, {J. Cryptology, Vol. 24, No. 3, 2011, pp. 446-469.}

    \bibitem{Hankerson}
	{Hankerson, D., Karabina, K., Menezes, A.},
	{\it Analyzing the Galbraith-Lin-Scott point multiplication method for elliptic curves over binary fields}, {IEEE Trans. Comput., Vol. 58, No. 10, 2009, pp. 1411-1420.}
	
    \bibitem{Jacobian}
	{Billet, O., Joye, M.},
	{\it The Jacobi model of an Elliptic Curve and the Side-channel Analysis}, {Applied Algebra, Algebraic Algorithms and Error-Correcting Codes, Vol. 2643, 2003, pp 34-42.}
  
    \bibitem{LD}
  	{Lopez, J., Dahab, R.},
  	{\it Improved algorithms for elliptic curve arithmetic in $GF_{2^n}$}, {in Proceedings of the Sel. Areas Cryptography, 1999, pp. 201-212.}
  	
	\bibitem{K10}
	{Jarvinen, K. and Skytta, J.},
	{\it On Parallelization of High-Speed Processors for Elliptic Curve Cryptography}, {IEEE Trans. Very Large Scale Integr. Syst., Vol. 16, No. 9, 2008, pp. 1162-1175.}	
 
	\bibitem{Solinas}
	{Solinas, J.A.},
	{\it Efficient arithmetic on Koblitz curves}, {Des. Codes Cryptogr., Vol. 19, 2000, pp. 195-249.} 
		
	\bibitem{MLP}
	{Montgomery, P.L.,},
	{\it Speeding the Pollard and elliptic curve methods of factorization}, {Mathematics of Computation, Vol. 48, 1987, pp. 243-264.} 		

	\bibitem{W18}
	{Rashidi, B., Sayedi, S.M., Rezaeian Farashahi, R.,},
	{\it High-speed Hardware Architecture of Scalar Multiplication for Binary Elliptic Curve Cryptosystems}, {Microelectronics Journal, Vol. 52, 2016, pp. 49-65.} 	

	\bibitem{S1}
	{San, C.V.},
	{\it A Survey of Elliptic Curve Cryptosystems, Part I: Introductory}, {NAS Technical Report-NAS-03-012, August 2003.} 

	\bibitem{Ash}
	{Ash, D.W., Blake, I.F., and Vanstone, S.A.},
	{\it Low Complexity Normal Bases}, {Discrete Applied Mathematics, Vol. 25, 1989, pp. 191-210.} 		

	\bibitem{Mullin}
	{Mullin, R.C., Onyszchuk, I.M., Vanstone, S.A., and Wilson, R.M.},
	{\it Optimal normal bases in $GF{p^m}$}, {Discrete Applied Mathematics, Vol. 22, No. 2, Feb 1989, pp. 149-161.} 	

	\bibitem{Huapeng_Wu}
	{Wu, H.},
	{\it Efficient Computations in Finite Fields with Cryptographie Significance}, {PhD thesis, University of Waterloo, Ontario, Canada, 1998.} 
	
	\bibitem{Inv_B}
	{Rashidi,B., Rezaeian Farashahi, R., Sayedi, S.M.},
	{\it High-performance and high-speed implementation of polynomial basis Itoh–Tsujii inversion algorithm over $GF(2^m)$} {IET Information Security, 2016, Vol. 11, Iss. 2,  pp. 66-77.}
	

	\bibitem{Fault_B}
	{Rashidi, B., Sayedi, S.M., Rezaeian Farashahi, R.},
	{\it Efficient implementation of bit-parallel fault tolerant polynomial basis multiplication and squaring over $GF(2^m)$}, {IET Comput. Digit. Tech., Vol. 10, Iss. 1, 2016, pp. 18-29.} 
	
	\bibitem{FF_Rodrigez}
	{Rodriguez-Henriquez, F., Saqib, N.A., Diaz-Perez, A.},
	{\it Cryptographic algorithms on reconfigurable hardware}, {Springer US, New York, 2006, 1st edn.}
	
	\bibitem{FF_Book}
	{Deschamps, J.P., Imana, J.L., Sutter, G.D.},
	{\it Hardware implementation of finite-field arithmetic}, {McGraw-Hill, New York, 2009, 1st edn.}
	
	\bibitem{Gorge}
	{Guajardo, J., Guneysu, T., Kumar, S.S., Paar, C., Pelzl,J.},
	{\it Efficient Hardware Implementation of Finite Fields with Applications to Cryptography}, {Acta Appl. Math, Vol. 93, 2006, pp. 75-118.}

	\bibitem{Conf_B}
	{Rashidi, B., Rezaeian Farashahi, R., Sayedi, S.M.},
	{\it High-speed and Pipelined Finite Field Bit-Parallel 
	Multiplier over $GF(2^m)$ for Elliptic Curve 
	Cryptosystems}, {in Proceedings of the 11th International ISC Conference on Information Security and Cryptology (ISCISC), 3-4 Sept. 2014, pp. 15-20.}			

	\bibitem{Isecure_B}
	{Rashidi, B., Rezaeian Farashahi, R. and Sayedi, S.M.},
	{\it Efficient Implementation of Low Time Complexity and Pipelined Bit-Parallel Polynomial Basis Multiplier over Binary Finite Fields}, {The ISC Int'l Journal of
	Information Security, Vol. 7, No. 2, 2015, pp. 101-114.}
	
	\bibitem{Mont_B}
	{Rashidi, B., Rezaeian Farashahi, R. and Sayedi, S.M.},
	{\it Fast and pipelined bit-parallel Montgomery multiplication and squaring over $GF(2^m)$}, {12$^{th}$ International Iranian Society of Cryptology Information Security and Cryptology (ISCISC), 2015, pp. 17-22.}
	
	\bibitem{HandBook}
	{Cohen, H., Frey, G., Avanzi, R., Doche, C., Lange, T., Nguyen, K., Vercauteren, F.},
	{\it Handbook of Elliptic and Hyperelliptic Curve Cryptography}, {first edn., CRC Press, Boca Raton, 2006.}
	
	\bibitem{VLSIONB_B}
	{Rashidi, B., Sayedi, S.M. and Rezaeian Farashahi, R.},
	{\it An efficient and high-speed VLSI implementation of optimal normal basis multiplication over $GF(2^m)$}, {Integration, the VLSI journal, Vol. 55, 2016, pp. 138-154.}
	
	\bibitem{GNB_FPGA_B}
	{Rashidi, B., Sayedi, S.M. and Rezaeian Farashahi, R.},
	{\it Efficient and low-complexity hardware architecture of Gaussian normal basis multiplication over $GF(2^m)$ for elliptic curve cryptosystems}, {IET Circuits Devices Syst., Vol. 10, 2016, pp. 1-10.}
	
	\bibitem{GNB_VLSI_B}
	{Rashidi, B., Sayedi, S.M. and Rezaeian Farashahi, R.},
	{\it High-speed VLSI implementation of digit-serial Gaussian normal basis multiplication over $GF(2^m)$}, {eprint.iacr.org/2016/966, 2016.}

	\bibitem{SM}
	{Sen, J.},
	{\it Cryptography and Security in Computing}, {first edn., Publisher InTech, 2012.}

	\bibitem{ITA}
	{Itoh, T. and Tsujii, S.},
	{\it A fast algorithm for computing multiplicative inverses in $GF(2^m)$ using normal bases}, {Inf. Comput., Vol. 78, No. 3, 1988, pp. 171-177.}

	\bibitem{Inv_GNB1}
	{Jarvinen, P., Dimitrov, S., and Azarderakhsh, R.},
	{\it  A generalization of addition chains and fast inversions in binary fields}, {IEEE Trans. Comput., 2015, Vol. 64, No. 9, pp. 2421-2432.}	

	\bibitem{Inv_GNB2}
	{Rashidi, B.},
	{\it  High-speed hardware implementation of Gaussian normal basis inversion algorithm over $\F_{2^m}$}, {Microelectronics Journal, 2017, Vol. 63, pp. 138-147.}

 	\bibitem{SoftPM}
 	{Hankerson, D., Hernandez, J.L., and Menezes, A.},
 	{\it Software Implementation of Elliptic Curve Cryptography over Binary Fields}, {in Proceedings of the International Workshop on Cryptographic Hardware and Embedded Systems (CHES), Worcester, Lecture Notes in Computer Science, LNCS 1965, 2000, pp. 1-24.}

 	\bibitem{smart}
 	{Smart, N.P.},
 	{\it A comparison of different finite fields for elliptic curve cryptosystems}, {Computers and Mathematics with Applications Vol. 42, 2001, pp. 91-100.}

 	\bibitem{MM}
 	{Montgomery, P.L.},
 	{\it Modular multiplication without trial division}, {Math. Comp., Vol. 44 , 1985, pp. 519-521.}

 	\bibitem{Crandall}
 	{Crandall, R.},
 	{\it Method and apparatus for public key exchange in a cryptographic system}, {U.S. Patent Number 5159632, 1992.}

 	\bibitem{Omura}
 	{Omura., J.K.},
 	{\it A Public Key Cell Design for Smart Card Chips}, {in Proceedings of the International Symposium on Information Theory and its Applications, 1990, pp. 27-30.}

 	\bibitem{Marcelo}
 	{Kaihara, M.E. and Takagi, N.},
 	{\it Bipartite Modular Multiplication}, {in Proceedings of the International Symposium on Information Theory and its Applications, 1990, pp. 27-30.}
 	
 	\bibitem{Kazuo}
 	{Sakiyama, K., Knez, M., Fan, J., Preneel, B., Verbauwhede, I.},
 	{\it Tripartite modular multiplication}, {Integration, the VLSI journal, Vol. 44, 2011, pp. 259-269.}

 	\bibitem{Re}
 	{Shenoy, N.},
 	{\it Retiming: Theory and practice}, {Integration, the VLSI journal, Vol. 22, 1997, pp. 1-21.}

	\bibitem{Sch}
	{Brucker, P.},
	{\it Scheduling Algorithms}, {Fifth edn., Springer-Verlag Berlin Heidelberg, 2007.}

	\bibitem{HD}
	{Azarderakhsh, R. and Reyhani-Masoleh, A.},
	{\it Low-complexity multiplier architectures for single and hybrid-double multiplications in Gaussian normal bases}, {IEEE Trans. Comput., Vol. 62 , NO. 4, 2013, pp. 744-757.}

	\bibitem{LE}
	{Sutherland, I., Sproull, R.F.},
	{\it Logical Effort: Designing for Speed on the Back of an Envelope}, {IEEE Advanced Research in VLSI, MIT Press, 1991.}

	\bibitem{L_BECs}
	{Rashidi, B., Sayedi, S.M. and Rezaeian Farashahi, R.},
	{\it Full-Custom Hardware Implementation of Point Multiplication on Binary Edwards Curves for ASIC Elliptic Curve Cryptosystem Applications}, {IET Circuits Devices Syst., accepted, 2017.}

	\bibitem{AES}
	{Rezaeian Farashahi, R., Rashidi, B. and Sayedi, S.M.},
	{\it FPGA based fast and high-throughput 2-slow retiming 128-bit AES encryption algorithm}, {Microelectronics Journal, Vol. 45, 2014, pp. 1014-1025}

    \bibitem{GLS1}
	{Govem, B., Jarvinen, K., Aerts, K., Verbauwhede, I. and Mentens, N.},
	{\it A Fast and Compact FPGA Implementation of Elliptic Curve Cryptography Using Lambda Coordinates}, {in Proceedings of the International Conference on Cryptology in Africa (AFRICACRYPT), LNCS 9646, 2016, pp. 63-83.}

    \bibitem{GLS2}
	{Azarderakhsh, R. and Karabina, K.},
	{\it A New Double Point Multiplication Method and its Implementation on Binary Elliptic Curves with Endomorphisms}, {IEEE Trans. Comput., Vol. 63, Iss. 10, October 2014, pp. 2614-2619.}


	\bibitem{W1}
	{Hernandez-Rodriguez, S.M., Rodriguez-Henriquez, F.},
	{\it An FPGA Arithmetic Logic Unit for Computing Scalar Multiplication using the Half-and-Add Method}, {in Proceedings of the International Conference on Reconfigurable Computing and FPGAs (ReConFig), 2005, pp. 1-7.}

	\bibitem{W2}
	{Hossain, M.S., Saeedi, E, and Kong, Y.},
	{\it High-Speed, Area-Efficient, FPGA-Based Elliptic Curve Cryptographic Processor over NIST Binary Fields}, {in Proceedings of the IEEE International Conference on Data Science and Data Intensive Systems, 2015, pp. 175-181.}
	
	\bibitem{W3}
	{Leong, P.H.W., and Leung, I.K.H.},
	{\it A Microcoded Elliptic Curve Processor Using FPGA Technology}, {IEEE Trans. Very Large Scale Integr. Syst., Vol. 10, No. 5, 2002, pp. 550-559.}
	
	\bibitem{W4}
	{Rodriguez-Henriquez, F., Saqib, N.A., Diaz-Perez, A.},
	{\it A fast parallel implementation of elliptic curve point multiplication over $GF(2^m)$}, {IEEE Trans. Very Large Scale Integr. Syst., Microprocess. Microsyst., Vol. 28, 2004, pp. 329-339.}
	
	\bibitem{W5}
	{Goodman, J., Chandrakasan, A.},
	{\it An Energy Efficient Reconfigurable Public-Key
	Cryptography Processor Architecture}, {in Proceedings of the International Workshop on Cryptographic Hardware and Embedded Systems (CHES), Worcester, Lecture Notes in Computer Science, Vol. 1965, 2000, pp. 175-190.}		

	\bibitem{W6}
	{Antao, S., Chaves, R., and Sousa, L.},
	{\it Efficient FPGA Elliptic Curve Cryptographic Processor over $GF(2^m)$}, {in Proceedings of the International Conference on ICECE Technology (FPT), 2008, pp. 357-360.}	

	\bibitem{W7}
	{Gura, N., Shantz, S.C., Eberle, H., Gupta, S., Gupta, V., Finchelstein, D., Goupy, E. and Stebila, D.},
	{\it An End-to-End Systems Approach to Elliptic
	Curve Cryptography}, {in Proceedings of the International Workshop on Cryptographic Hardware and Embedded Systems (CHES), Worcester, Lecture Notes in Computer Science, Vol. 2523, 2003, pp. 349-365.}

	\bibitem{W8}
	{Orlando, G., Paar, C.},
	{\it A high-performance reconfigurable elliptic curve processor for $GF(2^m)$}, {in Proceedings of the International Workshop on Cryptographic Hardware and Embedded Systems (CHES), Worcester, Lecture Notes in Computer Science, Vol. 1965, (Springer-Verlag), 2000, pp. 41-56.}
	
	\bibitem{W9}
	{Cohen, A.E., Parhi, K.K.},
	{\it Implementation of Scalable Elliptic Curve Cryptosystem Crypto-Accelerators for $GF(2^m)$}, {in Proceedings of the Thirty-Eighth Asilomar Conference on Signals, Systems and Computers, 2004, pp. 471-477.}	

	\bibitem{W10}
	{Yong-ping, D., Xue-cheng, Z., Zheng-lin, L., Yu, H., Li-hua, Y.I.},
	{\it Design of highly efficient elliptic curve crypto-processor with two multiplications over $GF(2^{163})$}, {J. Chin. Univ. Posts Telecommun., Vol. 16, No. 2, 2009, pp. 72-79.}

	\bibitem{W11}
	{Fayed, M.A., Watheq El-Kharashi, M., Gebali, F.},
	{\it A high-speed, high-radix, processor array architecture for real-time elliptic curve cryptography over $GF(2^m)$}, {in Proceedings of the IEEE International Symposium on Signal Processing and Information Technology, 2007, pp. 56-61.}

	\bibitem{W12}
	{Fournaris, A.P. and Koufopavlou, O.},
	{\it Low Area Elliptic Curve Arithmetic Unit}, {in Proceedings of the IEEE International Symposium on Circuits and Systems (ISCAS), Taipei, 24-27 May 2009, pp. 1397-1400.}

	\bibitem{W13}
	{Morales-Sandoval, M. and Feregrino-Uribe, C.},
	{\it A Hardware Architecture for Elliptic Curve Cryptography and Lossless Data Compression}, {in Proceedings of the 15th International Conference on Electronics, Communications and Computers (CONIELECOMP), 2005, pp. 1-6.}


	\bibitem{W14}
	{U. A. Khan, Z., and  Benaissa, M.},
	{\it High-Speed and Low-Latency ECC Processor
	Implementation over $GF(2^m)$ on FPGA}, {IEEE Trans. Very Large Scale Integr. Syst., Vol. , No. , 2016, pp. 1-12.}
	
	\bibitem{W15}
	{Liu, S., Ju, L, Cai, X., Jia, Z., Zhang, Z.},
	{\it High Performance FPGA Implementation of Elliptic
	Curve Cryptography over Binary Fields}, {in Proceedings of the IEEE 13th International Conference on Trust, Security and Privacy in Computing and Communications, 2014, pp. 148-155.}

	\bibitem{W16}
	{Li, L. and Li, S.},
	{\it High-Performance Pipelined Architecture of Elliptic Curve Scalar Multiplication over $GF(2^m)$}, {IEEE Trans. Very Large Scale Integr. Syst., Vol. 24, Iss. 4, 2016, pp. 1223-1232.}
	
	\bibitem{W17}
	{Mahdizadeh, H. and Masoumi, M.},
	{\it Novel architecture for efficient FPGA implementation of elliptic curve cryptographic processor over $GF(2^{163})$}, {IEEE Trans. Very Large Scale Integr. Syst., Vol. 21, No. 12, 2013, pp. 2330-2333.}	

	\bibitem{W19}
	{Morales-Sandoval, M., and Feregrino-Uribe, C.},
	{\it $GF(2^m)$ Arithmetic Modules for Elliptic Curve Cryptography}, {in Proceedings of the IEEE International Conference on Reconfigurable Computing and FPGA's(ReConFig), 2006, pp. 1-8.}	

	\bibitem{W20}
	{Nguyen, N., Gaj, K., Caliga, D. and El-Ghazawi, T.},
	{\it Implementation of Elliptic Curve Cryptosystems on a Reconfigurable Computer }, {in Proceedings of the IEEE International Conference on Field-Programmable Technology (FPT), 2003, pp. 60-67.}


	\bibitem{W21}
	{Roy, S.S., Rebeiro, C. and Mukhopadhyay, D.},
	{\it Theoretical modeling of elliptic curve scalar multiplier on LUT-based FPGAs for area and speed}, {IEEE Trans. Very Large Scale Integr. Syst., Vol. 21, No. 5, 2013, pp. 901-909.}	

	\bibitem{W22}
	{Saqib, N.A., Rodriguez-Henriquez, F. and Diaz-Perez, A.},
	{\it A Parallel Architecture for Fast Computation of Elliptic Curve Scalar Multiplication over $GF(2^m)$}, {in Proceedings of the 18th IEEE International Parallel and Distributed Processing Symposium (IPDPS’04), doi:10.1109/IPDPS.2004.1303124, 2004, pp. 1-8}	

	\bibitem{W23}
	{Schmalisch, M. and Timmermann, D.},
	{\it A Reconfigurable Arithmetic Logic Unit for Elliptic Curve Cryptosystems over $GF(2^m)$}, {in Proceedings of the IEEE 46th Midwest Symposium on Circuits and Systems, 2004, pp. 831-834.}
	
	\bibitem{W24}
	{Shohdy, S., El-sisi, A., Ismail, N.},
	{\it FPGA Implementation of elliptic curve point multiplication over $GF(2^{191})$}, { Advances in Information Security and Its Applications (ISA), Lecture Notes in Computer Science, Vol. 5576, Springer-Verlag, Germany, 2009, pp. 619-634.}	
	
 	\bibitem{W25}
 	{Bartolini, S., Branovic, I., Giorgi, R. and Martinelli, E.},
 	{\it Effects of Instruction-Set Extensions on an
 	Embedded Processor: A Case Study on Elliptic-Curve Cryptography over $GF(2^m)$}, {IEEE Trans. Comput., Vol. 57, No. 5, May 2008, pp. 672-685.}

 	\bibitem{W26}
 	{Ansari, B. and Hasan, M.A.},
 	{\it High-performance architecture of elliptic curve scalar multiplication}, {IEEE Trans. Comput., Vol. 57, No. 11, 2008, pp. 1443-1453.} 	
 	
 	\bibitem{W27}
 	{Sutter, G.D., Deschamps, J.P. and Imana, J.L.},
 	{\it Efficient elliptic curve point multiplication using digit-serial binary field operations}, {IEEE Trans. Ind. Electron., Vol. 60, No. 1, 2013, pp. 217-225.} 
 	
	\bibitem{W28}
	{Chelton, W.N. and Benaissa, M.},
	{\it Fast Elliptic Curve Cryptography on FPGA}, {IEEE Trans. Very Large Scale Integr. Syst., Vol. 16, No. 2, 2008, pp. 198-205.}			
 	
	\bibitem{W29}
	{Kim, C.H., Kwon, S. and Hong, C.P.},
	{\it FPGA implementation of high performance elliptic curve cryptographic processor over $GF(2^{163})$}, {J. Syst. Archit., Vol. 54, No. 10, 2008, pp. 893-900.}	 	
 	
	\bibitem{W30}
	{Gao, L., Shrivastava, S., and Sobelman, G.E.},
	{\it Elliptic Curve Scalar Multiplier Design Using FPGAs}, {in Proceedings of the International Workshop on Cryptographic Hardware and Embedded Systems (CHES), Worcester, Lecture Notes in Computer Science, Vol. 1717, (Springer-Verlag), 1999, pp. 257-268.}	

	\bibitem{W31}
	{C.C.Cheung, R., Jean-baptiste Telle, N., Luk, W. and Y.K.Cheung, P.},
	{\it Customizable Elliptic Curve Cryptosystems}, {IEEE Trans. Very Large Scale Integr. Syst., Vol. 13, No. 9, 2005, pp. 1048-1059.}
 
	\bibitem{W32}
	{Khan, Z.U.A. and Benaissa, M.},
	{\it Throughput/Area Efficient ECC Processor using 
	Montgomery Point Multiplication on FPGA}, {IEEE Trans. circuits and systems-II express briefs, Vol. 62, Iss. 11, 2015, pp. 1078-1082.} 
 
	\bibitem{A10}
	{Okada, S., Torii, N., Itoh, K. and Takenaka, M.},
	{\it Implementation of Elliptic Curve Cryptographic Coprocessor over $GF(2^m)$ on an FPGA}, {in Proceedings of the International Workshop on Cryptographic Hardware and Embedded Systems (CHES), Worcester, Lecture Notes in Computer Science, Vol. 1965, (Springer-Verlag), 2000, pp. 25-40.} 
	
	\bibitem{A21}
	{Zhang, Y., Chen, D., Choi, Y., Chen, L. and Ko, S.-B.},
	{\it A high performance ECC hardware implementation with instruction-level parallelism over $GF(2^{163})$}, {Microprocess. Microsyst., Vol. 34, No. 6, 2010, pp. 228-236.} 

	\bibitem{W33}
	{Rebeiro, C., Roy, S.S. and Mukhopadhyay, D.},
	{\it Pushing the Limits of High-Speed $GF(2^m)$
	Elliptic Curve Scalar Multiplication on FPGAs}, {in Proceedings of the International Workshop on Cryptographic Hardware and Embedded Systems (CHES), Worcester, Lecture Notes in Computer Science, Vol. 7428, (Springer-Verlag), 2012, pp. 494-511.} 
	
	\bibitem{K1}
	{Jarvinen, K. and Skytta, J.},
	{\it High-Speed Elliptic Curve Cryptography Accelerator for Koblitz Curves}, {in Proceedings of the 16th International Symposium on Field-Programmable Custom Computing Machines, 2008, pp. 109-118.} 

	\bibitem{K2}
	{Jarvinen, K. and Skytta, J.},
	{\it Fast point multiplication on Koblitz curves: Parallelization method and implementations}, {Microprocess. Microsyst., Vol. 33, 2009, pp. 106-116.} 

	\bibitem{K3}
	{Cinnati Loi, K.C. and Ko, S.B.},
	{\it High performance scalable elliptic curve cryptosystem processor for Koblitz curves}, {Microprocess. Microsyst., Vol. 37, 2013, pp. 394-406.} 

	\bibitem{K4}	
	{Jarvinen, K.},
	{\it Optimized FPGA-based elliptic curve cryptography processor for high-speed applications}, {Integration, the VLSI journal, Vol. 44, 2011, pp. 270-279.} 
		
	\bibitem{K5}
	{Al-Somani, T.},
	{\it Very efficient point multiplication on Koblitz curves}, {IEICE Electronics Express, Vol. 13, No. 9, 2016, pp. 1-6.} 
	
	\bibitem{K6}
	{Dimitrov, V.S., Jarvinen, K., Jacobson, M.J., Chan, W.F. and Huang, Z.},
	{\it FPGA Implementation of Point Multiplication on Koblitz Curves Using Kleinian Integers}, {in Proceedings of the International Workshop on Cryptographic Hardware and Embedded Systems (CHES), Worcester, Lecture Notes in Computer Science, Vol. 1965, (Springer-Verlag), 2006, pp. 445-459.}

	\bibitem{K7}
	{Dimitrov, V.S. and Jarvinen, K.},
	{\it Provably Sublinear Point Multiplication on Koblitz Curves and Its Hardware Implementation}, {IEEE Trans. Comput., Vol. 57, No. 11, 2008, pp. 1469-1481.}
	
	\bibitem{K8}
	{Realpe-Munoz, P., Trujillo-Olaya, V., and Velasco-Medina, J.},
	{\it Design of Elliptic Curve Cryptoprocessors over $GF(2^{163})$ on Koblitz Curves}, {in Proceedings of the Latin American Symposium On Circuits And Systems, 2014, pp. 1-4.}


	\bibitem{K11}
	{C. Realpe-Munoz, P. and Velasco-Medina, J.},
	{\it High-performance elliptic curve cryptoprocessors over $GF(2^m)$ on Koblitz curves}, {Analog Integr. Circ. Sig. Process, Vol. 85, 2015, pp. 129-138.}

	\bibitem{K12}
	{Azarderakhsh, R. and Reyhani-Masoleh, A.},
	{\it High-Performance Implementation of Point Multiplication on Koblitz Curves}, {IEEE Trans. circuits and systems-II express briefs, Vol. 60, Iss. 1, 2015, pp. 41-45.} 

	\bibitem{K13}
	{Cinnati Loi, K.C. and Ko, S.B.},
	{\it Parallelization of Scalable Elliptic Curve Cryptosystem Processors in $GF(2^m)$}, {Microprocess. Microsyst., Vol. 45, 2016, pp. 10-22.} 

	\bibitem{K14}
	{Sinha Roy, S., Fan, J. and Verbauwhede, I.},
	{\it Accelerating Scalar Conversion for Koblitz Curve Cryptoprocessors on Hardware Platforms}, {IEEE Trans. Very Large Scale Integr. Syst., Vol. 23, Iss. 5, 2015, pp. 810-818}

	\bibitem{K15}
	{Ahmadi, O., Hankerson, D. and Rodriguez-Henriquez, F.},
	{\it Parallel Formulations of Scalar Multiplication on Koblitz Curves}, {Journal of Universal Computer Science, Vol. 14, No. 3, 2008, pp. 481-504.}
	
	\bibitem{Brumley}
	{Brumley, B.B. and Jarvinen, K.},
	{\it Conversion algorithms and implementations for Koblitz curve cryptography}, {IEEE Trans. Comput., Vol. 59, No. 1, 2010, pp. 81-92.}	
	
	\bibitem{E1}
	{Chatterjee, A. and Sengupta, I.},
	{\it Design of a high performance Binary Edwards Curve based processor secured against side channel analysis}, {Integration, the VLSI Journal, Vol. 45, No. 3, 2012, pp. 331-340.}	
	
	\bibitem{E2}
	{Azarderakhsh, R. and Reyhani-Masoleh, A.},
	{\it Efficient FPGA Implementations of Point Multiplication on Binary Edwards and Generalized Hessian Curves Using Gaussian Normal Basis}, {IEEE Trans. on VLSI Systems, Vol. 20, No. 8, 2012, pp. 1453-1466.}
	
	\bibitem{E3}
	{Azarderakhsh, R. and Reyhani-Masoleh, A.},
	{\it Parallel and High-Speed Computations of Elliptic Curve Cryptography Using Hybrid-Double Multipliers}, {IEEE Trans. on Parallel and Distributed Systems, Vol. 26, Iss. 6, 2015, pp. 1668-1677.}
		
	\bibitem{E4}
	{Fournaris, AP., Sklavos, N. and Koulamas, C.},
	{\it A High Speed Scalar Multiplier for Binary Edwards Curves}, {in Proceedings of the Third Workshop on Cryptography and Security in Computing Systems, ACM, 2016, pp. 41-44.}
	
	\bibitem{E5}
	{Batina, L., Hogenboom, J., Mentens, N., Moelans, J. and Vliegen, J.},
	{\it Side-channel evaluation of FPGA implementations
	of binary Edwards curves}, {in Proceedings of the 17th IEEE International Conference on Electronics, Circuits, and Systems (ICECS), 2010, pp. 1255-1258.}	

	\bibitem{E6}
	{Rashidi, B., Farashahi, R.R. and Sayedi, S.M.},
	{\it High-speed Hardware Implementations of Point Multiplication for Binary Edwards and Generalized Hessian Curves}, {eprint.iacr.org/2017/005, 2017.}
	
	\bibitem{E7}
	{Chatterjee, A. and Sengupta, I.},
	{\it High-Speed Unified Elliptic Curve Cryptosystem on FPGAs Using Binary Huff Curves}, {in Proceedings of the Progress in VLSI Design and Test (VDAT), LNCS 7373, 2012, pp. 243-251.}	

	\bibitem{E8}
	{Ghosh, S., Kumar, A., Das, A. and Verbauwhede, I.},
	{\it On the Implementation of Unified Arithmetic
	on Binary Huff Curves}, {in Proceedings of the International Workshop on Cryptographic Hardware and Embedded Systems (CHES), Worcester, Lecture Notes in Computer Science, Vol. 8086, (Springer-Verlag), 2013, pp. 349-364.}	
	
	\bibitem{FP1}
	{Ma, Y., Zhang, Q., Liu, Z., Tu, C. and Lin, J.},
	{\it Low-Cost Hardware Implementation of Elliptic Curve Cryptography for General Prime Fields}, {in Proceedings of the International Conference on Information and Communications Security (ICICS), LNCS 9977, 2016, pp. 292-306.}
	
	\bibitem{FP0}
	{Hamilton, M. and P. Marnane, W.},
	{\it FPGA Implementation of an Elliptic Curve Processor using the GLV Method}, {in Proceedings of the International Conference on Reconfigurable Computing and FPGAs, 2009, pp. 249-254.}
	
	\bibitem{FP2}
	{Lai, J.Y., Wang, Y.S. and Huang, C.T.},
	{\it High-Performance Architecture for Elliptic Curve Cryptography over Prime Fields on FPGAs}, {Interdisciplinary Information Sciences, Vol. 18, No. 2, 2012. pp. 167-173.}
	
	\bibitem{FP3}
	{Baldwin, B., Moloney, R., Byrne, A., McGuire, G. and
	P. Marnane, W.},
	{\it A Hardware Analysis of Twisted Edwards Curves for an Elliptic Curve Cryptosystem}, {in Proceedings of the 5th International Workshop on Applied Reconfigurable Computing, 2009, pp. 355-361}
	
	\bibitem{FP4}
	{Wu, T.},
	{\it Elliptic Curve $ GF(p) $ Point Multiplier by Dual Arithmetic Cores}, {in Proceedings of the 11th International Conference on  ASIC (ASICON), 2015, DOI:10.1109/ASICON.2015.7516997, pp. 1-4}	
	
	\bibitem{FP5}
	{Marzouqi, H., Al-Qutayri, M., Salah, K., Schinianakis, D. and Stouraitis, T.},
	{\it A High-Speed FPGA Implementation of an RSD-Based ECC Processor}, {IEEE Trans. on VLSI Systems, Vol. 24, Iss. 1, 2016, pp. 151-164.}	
	
	\bibitem{FP6}
	{Javeed K. and Wang, X.},
	{\it Low latency flexible FPGA implementation of point multiplication on elliptic curves over $ GF(p) $}, {Int. J. Circ. Theor. Appl., Vol. 45, Iss. 2, 2016, pp. 214-228.}
	
	\bibitem{FP7}
	{Shylashree, N. and Sridhar, V.},
	{\it Hardware realization of fast elliptic curve point multiplication using balanced ternary representation and pre-computation over $GF(p)$}, {Journal of Discrete Mathematical Sciences and Cryptography, Vol. 19, No. 1, 2016, pp. 141-161.}	
	
	\bibitem{FP8}
	{Javeed K., Wang, X. and Scott, M.},
	{\it High performance hardware support for elliptic curve cryptography over general prime field}, {Microprocess. Microsyst., In Press, 2016, DOI:10.1016/j.micpro.2016.12.005}	
	
	\bibitem{FP9}
	{Jarvinen, K., Miele, A., Azarderakhsh, R. and Longa, P.},
	{\it FourQ on FPGA: New Hardware Speed Records for Elliptic Curve Cryptography over Large Prime Characteristic Fields}, {in Proceedings of the International Workshop on Cryptographic Hardware and Embedded Systems (CHES), Worcester, Lecture Notes in Computer Science, Vol. 9813, (Springer-Verlag), 2016, pp. 517-537.}
	
	\bibitem{FP10}
	{Al-Khaleel, O., Papachristou, C., Wolff, F. and Pekmestzi, K.},
	{\it An Elliptic Curve Cryptosystem Design Based on FPGA Pipeline Folding}, {in Proceedings of the 13th IEEE International On-Line Testing Symposium (IOLTS), 2007, DOI: 10.1109/IOLTS.2007.15., pp. 1-6.}	

	\bibitem{FP11}
	{Ghosh, S., Mukhopadhyay, D., and Roychowdhury, D.},
	{\it Petrel: Power and Timing Attack Resistant Elliptic Curve Scalar Multiplier Based on Programmable $GF(p)$ Arithmetic Unit}, {IEEE Trans. circuits and systems-I: Regular Paper, Vol. 58, No. 8, 2011, pp. 1798-1812.} 
	
	\bibitem{FP12}
	{Daly, A., Marnane, W., Kerins, T. and Popovici, E.},
	{\it An FPGA implementation of a $ GF(p) $ ALU for encryption processors}, {Microprocess. Microsyst., Vol. 28, 2004, pp. 253-260.} 	

	\bibitem{FP13}
	{Berna Ors, S., Batina, L., Preneel, B., Vandewalle, J.},
	{\it Hardware Implementation of an Elliptic Curve Processor over $GF(p)$}, {in Proceedings of the Application-Specific Systems, Architectures, and Processors (ASAP'03) , 2003, pp. 1-11.}	
	
	\bibitem{FP14}
	{Guillermin, N.},
	{\it A High Speed Coprocessor for Elliptic Curve Scalar Multiplications over $\F_p$}, {in Proceedings of the International Workshop on Cryptographic Hardware and Embedded Systems (CHES), Worcester, Lecture Notes in Computer Science, Vol. 6225, (Springer-Verlag), 2010, pp. 48-64.}

	\bibitem{FP15}
	{Orlando, G. and Paar, C.},
	{\it A Scalable $ GF(p) $ Elliptic Curve Processor Architecture for Programmable Hardware}, {in Proceedings of the International Workshop on Cryptographic Hardware and Embedded Systems (CHES), Worcester, Lecture Notes in Computer Science, Vol. 2162, (Springer-Verlag), 2001, pp. 348-363.}
		
	\bibitem{FP16}
	{Laue, R. AND A.Huss, S.},
	{\it Parallel Memory Architecture for Elliptic Curve Cryptography over $ GF(p) $ Aimed at Efficient FPGA Implementation}, {Journal of VLSI Signal Processing, Vol. 51, Iss. 1, 2007, pp. 39-55.} 

	\bibitem{FP17}
	{Varchola, M., Guneysu, T. and Mischke, O.},
	{\it MicroECC: A Lightweight Reconfigurable Elliptic Curve Crypto-Processor}, {in Proceedings of the  International Conference on Reconfigurable Computing and FPGAs, 2011, pp. 204-210.}	
	
	\bibitem{FP18}
	{M. Schinianakis, D., P. Fournaris, A., E. Michail, H., P. Kakarountas, A. and Stouraitis, T.},
	{\it An RNS Implementation of an $\F_p$ Elliptic Curve Point Multiplier}, {IEEE Trans. circuits and systems-I: Regular Paper, Vol. 56, No. 6, 2009, pp. 1202-1213.} 	

	\bibitem{FP19}
	{Shuhua, W., and Yuefei, Z.},
	{\it A Timing-and-Area Trade off $ GF(p) $ Elliptic Curve Processor Architecture  for FPGA}, {in Proceedings of the International Conference on Communications, Circuits and Systems, 27-30 May 2005, pp. 1308-1312.}	
	
	\bibitem{FP20}
	{J. McIvor, C., McLoone, M. and V. McCanny, J.},
	{\it Hardware Elliptic Curve Cryptographic Processor Over $ GF(p) $}, {IEEE Trans. circuits and systems-I: Regular Paper, Vol. 53, No. 9, 2006, pp. 1946-1957.} 	
	
	\bibitem{FP21}
	{Guneysu, T. and Paar, C.},
	{\it Ultra High Performance ECC over NIST Primes on Commercial FPGAs}, {in Proceedings of the International Workshop on Cryptographic Hardware and Embedded Systems (CHES), Worcester, Lecture Notes in Computer Science, Vol. 5154, (Springer-Verlag), 2008, pp. 62-78.}	

	\bibitem{FP22}
	{Vliegen, j., Mentens, N., Genoe, J., Braeken, A., Kubera, S., Touhafi, A. and Verbauwhede, I.},
	{\it A compact FPGA-based architecture for elliptic curve cryptography over prime fields}, {in Proceedings of the Application-specific Systems Architectures and Processors (ASAP), 2010, pp. 313-316.}
	
	\bibitem{FP23}
	{Hamilton, M. and P. Marnane, W.},
	{\it FPGA Implementation of an Elliptic Curve Processor using the GLV Method}, {in Proceedings of the International Conference on Reconfigurable Computing and FPGAs, 2009, pp. 249-254.}	

	
	\bibitem{FP25}
	{Alrimeih, H. and Rakhmatov, D.},
	{\it Fast and Flexible Hardware Support for ECC Over Multiple Standard Prime Fields}, {IEEE Trans. on VLSI Systems, Vol. 22, Iss. 12, 2014, pp. 2661-2674.}


	\bibitem{FP26}
	{Vliegen, J., Mentens, N., Genoe, J., Braeken, A., Kubera, S., Touhafi, A. and Verbauwhede, I.},
	{\it ECC on Your Fingertips: A Single Instruction Approach for Lightweight ECC Design in $ GF(p) $}, {in Proceedings of the International Conference on Selected Areas in Cryptography, 2015, pp. 161-177.}	
	
	\bibitem{FP27}
	{McIvor, C., McLoone, M.,  McCanny, J.V.},
	{\it ECC on Your Fingertips: A Single Instruction Approach for Lightweight ECC Design in $ GF(p) $}, {in Proceedings of the Irish Signals and Systems Conference, 2004, pp. 589-594.}		
	
	\bibitem{FP28}
	{Sasdrich, P. and Guneysu, T.},
	{\it Implementing Curve25519 for Side-Channel–Protected Elliptic Curve Cryptography}, {ACM Trans. on Reconfigurable Technology and Systems, Vol. 9, No. 1, 2015, pp. 1-15.}		
	
	\bibitem{FP29}
	{Koppermann, P., De Santis, D., Heyszl, J. and Sigl, G.},
	{\it X25519 Hardware Implementation for Low-Latency Applications}, {in Proceedings of the Euromicro Conference on Digital System Design, 2016, pp. 99-106.}		
	
	\bibitem{FP30}
	{Ali Tawalbeh, L., Mohammad, A., Abdul-Aziz Gutub, A.},
	{\it Efficient FPGA Implementation of a Programmable Architecture for $ GF(p) $ Elliptic Curve Crypto Computations}, {J Sign Process Syst., Vol. 59, 2010, pp. 233-244.}	
	
	\bibitem{FP31}
	{Chi Cinnati Loi, K. and Ko, S.B.},
	{\it Scalable Elliptic Curve Cryptosystem FPGA Processor for NIST Prime Curves}, {IEEE Trans. on VLSI Systems, Vol. 23, Iss. 11, 2015, pp. 2753-2756.}
	
	\bibitem{FP32}
	{Amiet, D., Curiger, A. and Zbinden, P.},
	{\it Flexible FPGA-Based Architectures for Curve Point Multiplication over $ GF(p) $}, {in Proceedings of the Euromicro Conference on Digital System Design, 2016, pp. 107-114.}	

	\bibitem{FP33}
	{Shylashree , N. and Sridhar, V.},
	{\it FPGA Implementation of High Speed Scalar Multiplication for ECC in $ GF(p) $}, {in Proceedings of the IEEE Region 10 Conference (TENCON), 2015, DOI: 10.1109/TENCON.2015.7373070., pp. 1-7.}

	\bibitem{FP34}
	{Ma, Y., Liu, Z., PAN, W., STATE, J.J.},
	{\it FPGA Implementation of High Speed Scalar Multiplication for ECC in $ GF(p) $}, {in Proceedings of the International Conference on Selected Areas in Cryptography, 2013, pp. 421-437.}
	
	\bibitem{FP35}
	{Ananyi, K., Alrimeih, H. and Rakhmatov, D.},
	{\it Flexible Hardware Processor for Elliptic Curve Cryptography Over NIST Prime Fields}, {IEEE Trans. on VLSI Systems, Vol. 17, No. 9, 2009, pp. 1099-1112.}
	
	\bibitem{FP36}
	{Baldwin, B., R.Goundar, R., Hamilton, M. and P. Marnane, W.},
	{\it Co-Z ECC scalar multiplications for hardware, software and hardware–software co-design on embedded systems}, {J Cryptogr Eng , Vol. 2, 2012, pp. 221-240.}
	
	\bibitem{FP37}
	{Fan, J., Sakiyama, K., and Verbauwhede, I.},
	{\it Elliptic curve cryptography on embedded multicore systems}, {J Cryptogr Eng , Vol. 2, 2012, pp. 221-240.}	

	\bibitem{A22}
	{Ghosh, S., Alam, M., Roy Chowdhury, D., and Sen Gupta, I.},
	{\it Parallel crypto-devices for $GF(p)$ elliptic curve multiplication resistant against side channel attacks}, {Computers and Electrical Engineering, Vol. 35, 2009, pp. 329-338.}

	\bibitem{A43}
	{Selim Hossain, M., Kong, Y., Saeedi, E. and C. Vayalil, N.},
	{\it High-performance elliptic curve cryptography processor over NIST prime fields}, {IET Computers and Digital Techniques, Vol. 11, Iss. 1, 2017, pp. 33-42.}
															
 	\bibitem{A48}
 	{Liu, Z., Liu, D. and Zou, X.},
 	{\it An Efficient and Flexible Hardware Implementation of the Dual-Field Elliptic Curve Cryptographic Processor}, {IEEE Trans. Ind. Electron., Vol. 64, Iss. 3, 2017, pp. 2353-2362.} 
 
	\bibitem{D1}
	{Sakiyama, K., De Mulder, E., Preneel, B. and Verbauwhede, I.},
	{\it A Parallel Processing Hardware Architecture for Elliptic Curve Cryptosystems}, {in Proceedings of the International Conference on Acoustics, Speech and Signal Processing, 2006, pp. 904-907.}
	
	\bibitem{A1}
	{S. Kumar, S. and Paar, C.}
	{\it Are standards compliant Elliptic Curve Cryptosystems feasible on RFID?}, {in Proceedings of the in Workshop on RFID Security and Light-Weight Cryptography, July 2006.}
	
 	\bibitem{A2}
 	{Ahmadi, H.R., Afzali-Kusha, A. and Pedram, M.},
 	{\it A power-optimized low-energy elliptic-curve crypto-processor}, {IEICE Electronics Express, Vol. 7, No. 23, 2010, pp. 1752-1759.} 
 	
	\bibitem{A3}
	{Chen, C. and Qin, Z.}
	{\it Improved Elliptic Curve Cryptographic Processor for General Curves over $ GF(p) $}, {in Proceedings of the 10th International Conference on Signal Processing (ICSP), 2010, pp. 1849-1852.}
	
	\bibitem{A4}
	{Guitouni, Z., Chotin-Avot, R., Machhout, M.,  Mehrez, H. and Tourki, R.}
	{\it High Performances ASIC based Elliptic Curve Cryptographic Processor over $ GF(2^m) $}, {IJCA Special Issue on Network Security and Cryptography, 2011, pp. 1-10.}
	
	\bibitem{A5}
	{Chen, G., Bai, G. and Chen, H.},
	{\it A High-Performance Elliptic Curve Cryptographic Processor for General Curves Over $ GF(p) $ Based on a Systolic Arithmetic Unit}, {IEEE Trans. circuits and systems-II express briefs, Vol. 54, No. 5, 2007, pp. 412-416.} 	
	
	\bibitem{A6}
	{Ansari, B. and Wu, H.}
	{\it Efficient Finite Field Processor for $ GF(2^{163}) $ and its VLSI Implementation}, {in Proceedings of the International Conference on Information Technology (ITNG'07), 2007, pp. 1-6.}	
	
	\bibitem{A7}
	{Ki Lee, Y. and Verbauwhede, I.}
	{\it A Compact Architecture for Montgomery Elliptic Curve Scalar Multiplication Processor}, {in Proceedings of the International Workshop on Information Security Applications, 2007, pp. 115-127.}		
	
	\bibitem{A8}
	{Ki Lee, Y., Sakiyama, K., Batina, L., Verbauwhede, I.},
	{\it Elliptic Curve Based Security Processor for RFID}, {IEEE Trans. Comput., Vol. 57, Iss. 11, 2009, pp. 1514-1527.}		
	
	\bibitem{A9}
	{Aigner, H., Bock, H., Hutter, M. and Wolkerstorfer, J.},
	{\it A Low-Cost ECC Coprocessor for Smartcards}, {in Proceedings of the International Workshop on Cryptographic Hardware and Embedded Systems (CHES), Worcester, Lecture Notes in Computer Science, Vol. 3156, (Springer-Verlag), 2004, pp. 107-118.}		

	\bibitem{A11}
	{Schroeppel, R., Beaver, C., Gonzales, R., Miller, R. and Draelos, T.},
	{\it A Low-Power Design for an Elliptic Curve Digital Signature Chip}, {In Proceedings of the International Workshop on Cryptographic Hardware and Embedded Systems (CHES), Worcester, Lecture Notes in Computer Science, Vol. 2523, (Springer-Verlag), 2002, pp. 366-380.}		
	
	\bibitem{A12}
	{Chung, S.C., Lee, J.W., Chang,, H.C. and Lee, C.Y.},
	{\it A High-Performance Elliptic Curve Cryptographic Processor over $ GF(p) $ with SPA Resistance}, {in Proceedings of the International Symposium on Circuits and Systems (ISCAS), 2012, pp. 1456-1459.}		

	\bibitem{A13}
	{Hong, J.H. and Wu, W.C.},
	{\it The Design of High Performance Elliptic Curve Cryptographic}, {in Proceedings of the International Midwest Symposium on Circuits and Systems, 2009, pp. 527-530.}

	\bibitem{A14}
	{Hal, C.S., Kim, J.H., Choi, B.Y., Lee, J.H. and Kim, H.W.},
	{\it $ GF(2^{191}) $ Elliptic Curve Processor using Montgomery Ladder and High Speed Finite Field Arithmetic Unit}, {in Proceedings of the IEEE Region 10 Conference (TENCON), 2005, DOI:10.1109/TENCON.2005.301250.}
	
	\bibitem{A15}
	{MuthuKumar, B., and Jeevananthan, S.},
	{\it High Speed Hardware Implementation of an Elliptic Curve Cryptography (ECC) Co-Processor}, {in Proceedings of the Trendz in Information Sciences and Computing (TISC), 2010, pp. 176-180.}	

	\bibitem{A16}
	{Leinweber, L., Papachristou, C., and G. Wolff, F.},
	{\it Efficient Architectures for Elliptic Curve Cryptography Processors for RFID}, {in Proceedings of the International Conference on Computer Design (ICCD), 2009, pp. 372-377.}	

	\bibitem{A17}
	{Lee, J.W., Chen, Y.L., Tseng, C.Y., Chang, H.C. and Lee, C.Y.},
	{\it A 521-bit Dual-Field Elliptic Curve Cryptographic Processor with Power Analysis Resistance}, {in Proceedings of the 36th European Solid State Circuits Conference, 2010, pp. 206-209.}	
	
	\bibitem{A18}
	{Lai, J.Y., Hung, T.Y., Yang. K.H. and Huang, C.T.},
	{\it High-Performance Architecture for Elliptic Curve Cryptography over Binary Field}, {in Proceedings of the 36th European Solid State Circuits Conference, 2010, pp. 3933-3936.}
	
	\bibitem{A19}
	{Lai, J.Y. and Huang, C.T.},
	{\it A Highly Efficient Cipher Processor for Dual-Field Elliptic Curve Cryptography}, {IEEE Trans. circuits and systems-II express briefs, Vol. 56, No. 5, 2009, pp. 394-398.} 	

	\bibitem{A20}
	{Leung, P.K., Choy, C.S., Chan, C.F. and Pun, K.P.},
	{\it A Low Power Asynchronous $ GF(2^{173}) $ ALU for Elliptic Curve Crypto-processor}, {in Proceedings of the 3rd International IEEE-NEWCAS Conference, 2003, pp. 337-340.}

	\bibitem{A23}
	{Peter, S., Langendorfer, P. and Piotrowski, K.},
	{\it Flexible Hardware Reduction for Elliptic Curve Cryptography in $ GF(2^m) $}, {in Proceedings of the  Design, Automation and Test in Europe Conference and Exhibition, 2007, pp. 1-6.}
	
	\bibitem{A24}
	{Huang, C. Lai, J., Ren, I. and Zhang, Q.},
	{\it Scalable Elliptic Curve Encryption Processor for Portable Application}, {in Proceedings of the IEEE ICASIC, 2003,  DOI:10.1109/ICASIC.2003.1277458. pp. 1312-1316.}							 			 														

	\bibitem{A25}
	{Kim, J.H. and Lee, D.H.},
	{\it A compact finite field processor over $ GF(2^m) $ for elliptic curve cryptography}, {in Proceedings of the International Symposium on Circuits and Systems, 2002, pp. 340-343.}	
	

	\bibitem{A26}
	{Hein, D., Wolkerstorfer, J. and Felber, N.},
	{\it ECC Is Ready for RFID-A Proof in Silicon}, {in Proceedings of the International Workshop on Selected Areas in Cryptography, LNCS 5381, 2008, pp. 401-413.}		
					
	\bibitem{A27}
	{Batina, L., Mentens, N., Sakiyama, K., Preneel, B. and Verbauwhede, I.},
	{\it Low-Cost Elliptic Curve Cryptography for Wireless Sensor Networks}, {in Proceedings of the European Workshop on Security in Ad-hoc and Sensor Networks, LNCS 4357, 2006, pp. 6-17.}
	
	\bibitem{A28}
	{Ozturk, E., Sunar, B. and Savas, E.},
	{\it Low-Power Elliptic Curve Cryptography Using Scaled Modular Arithmetic}, {in Proceedings of the International Workshop on Cryptographic Hardware and Embedded Systems (CHES), Worcester, Lecture Notes in Computer Science, Vol. 3156, (Springer-Verlag), 2004, pp. 92-106.}		

	\bibitem{A29}
	{Wenger, E. and Hutter, M.},
	{\it A Hardware Processor Supporting Elliptic Curve Cryptography for Less than 9 kGEs}, {in Proceedings of the International Conference on Smart Card Research and Advanced Applications, LNCS 7079, 2011, pp. 182-198.}

	\bibitem{A30}
	{Roy, S.S., Jarvinen, K. and Verbauwhede, I.},
	{\it Lightweight Coprocessor for Koblitz Curves: 283-Bit ECC Including Scalar Conversion with only 4300 Gates}, {in Proceedings of the International Workshop on Cryptographic Hardware and Embedded Systems (CHES), Worcester, Lecture Notes in Computer Science, Vol. 9293, (Springer-Verlag), 2015, pp. 102-122.}	

	\bibitem{A31}
	{Goodman, J. and P. Chandrakasan, A.},
	{\it An Energy-Efficient Reconfigurable Public-Key Cryptography Processor}, {IEEE Journal of Solid-State Circuits, Vol. 36, No. 11, 2001, pp. 1808-1820.} 

	\bibitem{A32}
	{Azarderakhsh, R., Jarvinen, K. and Mozaffari-Kermani, M.},
	{\it Efﬁcient Algorithm and Architecture for Elliptic Curve Cryptography for Extremely Constrained Secure Applications}, {IEEE Trans. circuits and systems-I: Regular Paper, Vol. 61, No. 4, 2014, pp. 1144-1155.}
	
	\bibitem{A33}
	{Kocabas, U., Fan, J. and Verbauwhede, I.},
	{\it Implementation of Binary Edwards Curves for very-constrained devices}, {in Proceedings of the 21st IEEE International Conference on Application-specific Systems Architectures and Processors (ASAP), 2010, pp. 185-191.}

	\bibitem{A34}
	{Lai, J.Y. and Huang, C.T.},
	{\it Elixir: High-Throughput Cost-Effective Dual-Field Processors and the Design Framework for Elliptic Curve Cryptography}, {IEEE Trans. on VLSI Systems, Vol. 16, No. 11, 2008, pp. 1567-1580.}

	\bibitem{A35}
	{Chen, J.H., Shieh, M.D. and Lin, W.C.},
	{\it A High-Performance Unified-Field Reconfigurable Cryptographic Processor}, {IEEE Trans. on VLSI Systems, Vol. 18, No. 8, 2010, pp. 1145-1158.}

	\bibitem{A36}
	{Lai, J.Y. and Huang, C.T.},
	{\it Energy-Adaptive Dual-Field Processor for High-Performance Elliptic Curve Cryptographic Applications}, {IEEE Trans. on VLSI Systems, Vol. 19, No. 8, 2011, pp. 1512-1517.}
	
	\bibitem{A37}
	{Sakiyama, K., Batina, L., Preneel, B. and Verbauwhede, I.},
	{\it High-performance Public-key Cryptoprocessor for Wireless Mobile Applications}, {Mobile Netw Appl, Vol. 12, 2007, pp. 245-258.}	

	\bibitem{A38}
	{Rozic, V., Reparaz, O. and Verbauwhede, I.},
	{\it A 5.1uJ per point-multiplication elliptic curve cryptographic processor}, {Int. J. Circ. Theor. Appl., Vol. 45, Iss. 2, 2016, pp. 245-258.}	
	
	\bibitem{A39}
	{Koziel, B., Azarderakhsh, R. and Mozaffari-Kermani, M.},
	{\it Low-Resource and Fast Binary Edwards Curves Cryptography}, {in Proceedings of the International Conference in Cryptology in India (INDOCRYPT), LNCS 9462, 2015, pp. 347-369.}	

	\bibitem{A40}
	{Pessl, P. and Hutter, M.},
	{\it Curved Tags-A Low-Resource ECDSA Implementation tailored for RFID}, {in Proceedings of the International Workshop on Radio Frequency Identification: Security and Privacy Issues (RFIDSec), LNCS 8651, 2014, pp. 156-172.}

	\bibitem{A41}
	{Zhang, D. and Bai, G.},
	{\it Ultra High-Performance ASIC Implementation of SM2 with Power-Analysis Resistance}, {in Proceedings of the International Conference on Electron Devices and Solid-State Circuits (EDSSC), 2015, pp. 523-526.}	

	\bibitem{A42}
	{Zhao, Z. and Bai, G.},
	{\it Ultra High-Speed SM2 ASIC Implementation}, {in Proceedings of the IEEE 13th International Conference on Trust, Security and Privacy in Computing and Communications, 2014, pp. 182-188.}

	\bibitem{A44}
	{Li, W., Yi, W., Ma, C., Yi, S., Yang, X., Dai, Z.},
	{\it A high-throughput processor for dual-field elliptic curve cryptography}, {in Proceedings of the International Conference on  Information and Communications Technologies (ICT 2015), 2015, pp. 1-5.}	

	\bibitem{A45}
	{Chen, G., Bai, G. and Chen, H.},
	{\it A Dual-Field Elliptic Curve Cryptographic Processor Based on a Systolic Arithmetic Unit}, {in Proceedings of the International Symposium on Circuits and Systems (ISCAS), 2008, pp. 3298-3301.}	

	\bibitem{A46}
	{Chen, Y.L. Lee, J.W. Liu, P.C. Chang, H.C., and Lee, C.Y.},
	{\it A Dual-Field Elliptic Curve Cryptographic Processor with a Radix-4 Unified Division Unit}, {in Proceedings of the International Symposium on Circuits and Systems (ISCAS), 2011, pp. 713-716.}	
	
	\bibitem{A47}
	{Lee, J.W., Chung, S.C., Chang, H.C. and Lee, C.Y.},
	{\it Efficient Power-Analysis-Resistant Dual-Field Elliptic Curve Cryptographic Processor Using Heterogeneous Dual-Processing-Element Architecture}, {IEEE Trans. on VLSI Systems, Vol. 22, No. 1, 2014, pp. 49-61.}	

	\bibitem{A49}
	{D. Targhetta, A., E. Owen, D., L. Israel, F. and V. Gratz, P.},
	{\it Energy-Efficient Implementations of $ GF(p) $ and $ GF(2^m) $ Elliptic Curve Cryptography}, {in Proceedings of the 33rd IEEE International Conference on Computer Design (ICCD), 2015, pp. 704-711.}	
	
	\bibitem{A50}
	{Satoh, A. and Takano, K.},
	{\it A Scalable Dual-Field Elliptic Curve Cryptographic Processor}, {IEEE Trans. Comput., 2003, Vol. 52, No. 4, pp. 449-460.}		

	\bibitem{A51}
	{Wolkerstorfer, J.},
	{\it Dual-Field Arithmetic Unit for $ GF(p) $ and $ GF(2^m) $}, {in Proceedings of the International Workshop on Cryptographic Hardware and Embedded Systems (CHES), Worcester, Lecture Notes in Computer Science, Vol. 2523, (Springer-Verlag), 2003, pp. 500-514.}

	\bibitem{A52}
	{Wu, Y. and Zeng, X.},
	{\it A New Dual-Field Elliptic Curve Cryptography Processor}, {in Proceedings of the International Symposium on Circuits and Systems (ISCAS), 2006, pp. 305-308.}	

	\bibitem{A53}
	{Ahmadi, H. and Afzali-Kusha, A.},
	{\it Very Low-Power Flexible $ GF(p) $ Elliptic-Curve Crypto-Processor for Non-Time-Critical Applications }, {in Proceedings of the International Symposium on Circuits and Systems, (ISCAS), 2009, pp. 904-907.}	

	\bibitem{A54}
	{Sakiyama, K., Batina, L., Preneel, B. and Verbauwhede, I.},
	{\it Multicore Curve-Based Cryptoprocessor with Reconfigurable Modular Arithmetic Logic Units over $ GF(2^n) $}, {IEEE Trans. Comput., Vol. 56, No. 9, 2007, pp. 1269-1282.}	

	\bibitem{A55}
	{Liu, Z., Liu, D., Zou, X., Lin. H. and Cheng, J.},
	{\it Design of an Elliptic Curve Cryptography Processor for RFID Tag Chips}, {Sensors, Vol. 14, No. 10, 2014, pp. 17883-17904.}	

	\bibitem{A56}
	{Marzouqi, H., Al-Qutayri, M., Salah, K. and Saleh, H.},
	{\it Very Low-Power Flexible $ GF(p) $ Elliptic-Curve Crypto-Processor for Non-Time-Critical Applications }, {in Proceedings of the IEEE 59th International Midwest Symposium on Circuits and Systems (MWSCAS), 16-19 October 2016, Abu Dhabi, UAE, 2016, pp. 1-4.}

	\bibitem{A57}
	{Lee, J.W., Chung, S.C., Chang, H.C. and Lee, C.Y.},
	{\it An Efficient Countermeasure against Correlation Power-Analysis Attacks with Randomized Montgomery Operations for DF-ECC Processor}, {in Proceedings of the International Workshop on Cryptographic Hardware and Embedded Systems (CHES), Worcester, Lecture Notes in Computer Science, Vol. 7428, (Springer-Verlag), 2012, pp. 548-564.}
	
										
\end{thebibliography}
\end{document}